\documentclass[pre,showpacs,preprintnumbers,twocolumn,amsmath,amssymb,superscriptaddress]{revtex4-2}
\usepackage{bm}
\usepackage{graphicx}% Include figure files
\usepackage{dcolumn}% Align table columns on decimal point
\usepackage{times}
\usepackage{color}
\usepackage[utf8]{inputenc}
\usepackage[T1]{fontenc}
\usepackage{mathptmx}
\usepackage{etoolbox}
\usepackage{subcaption}

\makeatletter
\def\@email#1#2{%
 \endgroup
 \patchcmd{\titleblock@produce}
  {\frontmatter@RRAPformat}
  {\frontmatter@RRAPformat{\produce@RRAP{*#1\href{mailto:#2}{#2}}}\frontmatter@RRAPformat}
  {}{}
}%
\makeatother

\begin{document}

%\preprint{AIP/123-QED}

\title{Scale free chaos in the confined Vicsek flocking model}
% Force line breaks with \\

\author{R. Gonz\'alez-Albaladejo}
\affiliation{Departamento de Matem\'atica Aplicada, Universidad Complutense de Madrid, 28040 Madrid, Spain}
\affiliation{Gregorio Mill\'an Institute for Fluid Dynamics, Nanoscience and Industrial Mathematics, Universidad Carlos III de Madrid, 28911 Legan\'{e}s, Spain}
\author{A. Carpio}
\affiliation{Departamento de Matem\'atica Aplicada, Universidad Complutense de Madrid, 28040 Madrid, Spain}
\affiliation{Gregorio Mill\'an Institute for Fluid Dynamics, Nanoscience and Industrial Mathematics, Universidad Carlos III de Madrid, 28911 Legan\'{e}s, Spain}
\author{L. L. Bonilla$^*$}
\affiliation{Gregorio Mill\'an Institute for Fluid Dynamics, Nanoscience and Industrial Mathematics, Universidad Carlos III de Madrid, 28911 Legan\'{e}s, Spain}
\affiliation{Department of Mathematics, Universidad Carlos III de Madrid, 28911 Legan\'{e}s, Spain. 
$^*$Corresponding author. E-mail: bonilla@ing.uc3m.es}

%%%%%%%%%%%%%%%%% END OF PREAMBLE %%%%%%%%%%%%%%%%

 \date{\today}

\begin{abstract}
The Vicsek model encompasses the paradigm of active dry matter. Motivated by collective behavior of insects in swarms, we have studied finite size effects and criticality in the three dimensional, harmonically confined Vicsek model. We have discovered a phase transition that exists for appropriate noise and small confinement strength. On the critical line of confinement versus noise, swarms are in a state of scale-free chaos characterized by minimal correlation time, correlation length proportional to swarm size and topological data analysis. The critical line separates dispersed single clusters from confined multicluster swarms. Scale-free chaotic swarms occupy a compact region of space and comprise a recognizable `condensed' nucleus and particles leaving and entering it. Susceptibility, correlation length, dynamic correlation function and largest Lyapunov exponent obey power laws. The critical line and a narrow criticality region close to it move simultaneously to zero confinement strength for infinitely many particles. At the end of the first chaotic window of confinement, there is another phase transition to infinitely dense clusters of finite size that may be termed flocking black holes.
\end{abstract} 
\maketitle

\section{Introduction}\label{sec:1}
The Vicsek model (VM) deals with the paradigm of dry active matter whose components influence their collective behavior without being immersed in a fluid or other medium \cite{cha20,mar13}. In the VM, $N$ self-propelled particles move with constant speed in a box with periodic boundary conditions, follow discrete time dynamics, and their velocities are directed to the average velocity of their neighbors plus an alignment noise \cite{vic95,vic12}. The VM undergoes an ordering transition for noise below a critical value (or particle density above a critical value) in which a gas of disordered particles filling the box evolves to ordered patterns such as traveling bands, or to an ordered ``liquid'' within boxes that are not sufficiently large \cite{cha20}. This flocking transition has attracted immense interest, for it seems to describe similar phenomena in very many physical and biological systems exhibiting critical behavior.

For reasons that are not completely understood, many biological systems live close to criticality, which induces power law behavior of macroscopic variables \cite{mor11,bia12,non17}. Examples include networks of neurons in vertebrate retinas \cite{mor11} (see Ref.~\onlinecite{non17} for the influence of random subsampling of large retinal datasets in signatures of criticality), amino acid frequencies in proteins \cite{tan17}, or flocking phenomena in animals \cite{sum10,aza18,cav18}. In many such systems, their components interact only with neighboring ones, defined either metrically or topologically \cite{cav18}. However, the maximum correlation length separating two mutually influencing components is proportional to the system size and not to any intrinsic length associated to individuals. This is a manifestation of scale free behavior and power laws obeyed by macroscopic variables such as the correlation length, susceptibility to changes in the polarization order parameter and dynamic correlation \cite{cav18}.

 Insect swarms provide particularly rich empirical data and peculiar critical behavior \cite{oku86,oku01,kel13,puc14,att14plos,att14,cav17,gor16,sin17}. Male midges and other diptera form swarms near visual markers to attract females for reproductive purposes \cite{oku86,oku01,puc14}. While small swarms track the marker shape, larger swarms are more isotropic and shape independent \cite{puc14}. In laboratory experiments, swarms form far from the walls of the enclosure that contains them \cite{kel13}. Topological data analysis of experiments shows that the swarm can be thought of as a condensed phase (the swarm nucleus) surrounded by a vapor phase (insects leaving and entering the nucleus \cite{sin17}). Isotropic swarms lack translation invariance and their cohesion may be explained by a confining harmonic potential \cite{gor16}. 

Natural swarms present collective behavior and strong correlations but not global order. The polarization order parameter (the mean of the directions of insect velocities) is quite small but the correlation length (measuring the  largest distance between two insects whose velocity fluctuations still influence each other) is proportional to the swarm size \cite{cav17}. It is also much larger than all other length scales, such as insect size, average separation between insects, etc. Macroscopic variables, such as the correlation length, the susceptibility to polarization changes and the dynamic correlation, follow power laws as functions of the distance to criticality, with critical exponents that differ from those of equilibrium and many nonequilibrium phase transitions \cite{hua87,att14plos,att14,cav17}. Cavagna and coworkers have shown that the characteristic timescale, static and dynamic connected correlation functions depend on the control parameters (density, noise, \ldots) only through the correlation length. This is the finite-size scaling hypothesis, which is similar to that found in second-order equilibrium phase transitions \cite{hua87}. Finite-size scaling allows us to extrapolate power laws of macroscopic variables obtained for finite $N$ to the case of infinitely many particles, which characterize phase transitions \cite{ami05}. Attempts to use the ordering transition of the three-dimensional (3D) VM to explain observed critical exponents produced exponent values quite different from experimental ones \cite{att14,cav17}. 

In this paper, we study the 3D VM confined by a harmonic potential \cite{att14plos,att14,cav17}. Confining harmonic potentials have long been proposed as models for swarm behavior \cite{oku86,oku01,gor16}. We have made a number of discoveries. As the confinement parameter $\beta$ decreases, the VM changes from a period 2 attractor at large confinement values to other periodic, quasiperiodic and chaotic attractors. Using $\beta$ as control parameter, there are windows of chaotic solutions followed by intervals of non-chaotic behavior. As the number of particles $N$ increases, the first chaotic window begins at smaller values of $\beta$. Inside this window, we have found scale free chaos, for which the correlation length is proportional to the size of the swarm for increasing $N$, the polar order parameter is small and macroscopic quantities such as correlation length, susceptibility, dynamic correlation and the largest Lyapunov exponent exhibit power laws. At the beginning of the critical region, a line in the noise-$\beta$ plane separates chaotic single cluster  from multicluster swarms. Similar to observations \cite{sin17}, the single cluster swarm consists of a `condensed' nucleus and particles leaving and entering it. The single-to-multicluster chaos line and critical region move to $\beta=0$ as $N\to\infty$. Thus, we have found a scale-free-chaos phase transition. At the end of the first chaotic window, we have  found a different phase transition to infinitely dense clusters of finite size that is reminiscent of gravitational collapse \cite{gun07,alb20,chavanis20}. The finite size clusters of infinitely many particles may be termed {\em flocking black holes}. As $N\to\infty$, the critical line for collapse to them occurs for $\beta\to\infty$.

The rest of the paper is as follows. We present the confined Vicsek model in Section \ref{sec:2} and find different attractors as the confinement parameter decreases from a large value. Among them, period 2, period 4 periodic solutions, quasiperiodic solutions, large period solutions, and chaotic attractors with positive Lyapunov exponents. Section \ref{sec:3} discusses algorithms to calculate the largest Lyapunov exponent (LLE) and how to distinguish deterministic and noisy chaos from noise, using Gao {\em et al}'s scale dependent Lyapunov exponents \cite{gao06}. Section \ref{sec:4} uses ideas from statistical physics, modified correlation functions, and finite-size scaling to obtain the main results of the paper: the existence of a line of phase transitions within the noisy chaos region of the parameter space. Section \ref{sec:5} describes a different phase transition from multicluster chaos to the formation of clusters of finite size and infinite particle density reminiscent of gravitational collapse \cite{gun07,alb20,chavanis20}. Section \ref{sec:6} uses topological data analysis to characterize the phase transitions from single cluster to multicluster chaotic attractors. Section \ref{sec:7} discusses our findings and it contains our conclusions whereas the Appendices are devoted to technical matters. Appendix \ref{ap:a} describes our nondimensionalization of the confined Vicsek model. Appendix \ref{ap:b} describes the Benettin algorithm \cite{ben80,ott93,cen10}, scale dependent Lyapunov exponents \cite{gao06} and the Gao-Zheng algorithm to extract the largest Lyapunov exponent from high dimensional reconstructions of the chaotic attractor using lagged coordinates \cite{gao94}. Appendix \ref{ap:c} discusses dynamic and static correlation functions, how to calculate them and different definitions of critical lines at finite number of particles. Appendix \ref{ap:d} discusses two solvable examples illustrating the relation between susceptibility and correlation time.

\section{Confined Vicsek model} \label{sec:2}
We have nondimensionalized the VM governing equations using data from natural swarms (see Appendix \ref{ap:a}):
\begin{eqnarray}
&&\mathbf{x}_i(t+1)=\mathbf{x}_i(t)+ \mathbf{v}_i(t+1),\nonumber\\
&&\mathbf{v}_i(t+1)=v_0  \mathcal{R}_\eta\!\left[\Theta\!\left(\sum_{|\mathbf{x}_j-\mathbf{x}_i|<R_0}\mathbf{v}_j(t)-\beta\mathbf{x}_i(t)\right)\!\right]\!, \label{eq1}
\end{eqnarray}
where $i=1,\ldots,N$, $v_0=1$ is a constant speed, $R_0=1$, $\beta$ is the confinement strength. The position and velocity of the $i$th particle at time $t$ are $\mathbf{x}_i(t)$ and $\mathbf{v}_i(t)$, respectively. In \eqref{eq1}, $\Theta(\mathbf{x})= \mathbf{x}/|\mathbf{x}|$ and $\mathcal{R}_\eta(\mathbf{w})$ rotates the unit vector $\mathbf{w}$ randomly within a spherical cone  centered at its origin and spanning a solid angle in $(-\frac{\eta}{2},\frac{\eta}{2})$. The 2D VM is defined similarly. Initially, the particles are randomly placed within a sphere with unit radius and the particle velocities point outwards. 

\begin{widetext}
\begin{center}
\begin{figure}[ht]
\begin{center}
\includegraphics[trim={0.2cm 0.1cm 0.5cm 0.5cm},clip,width=5.9cm]{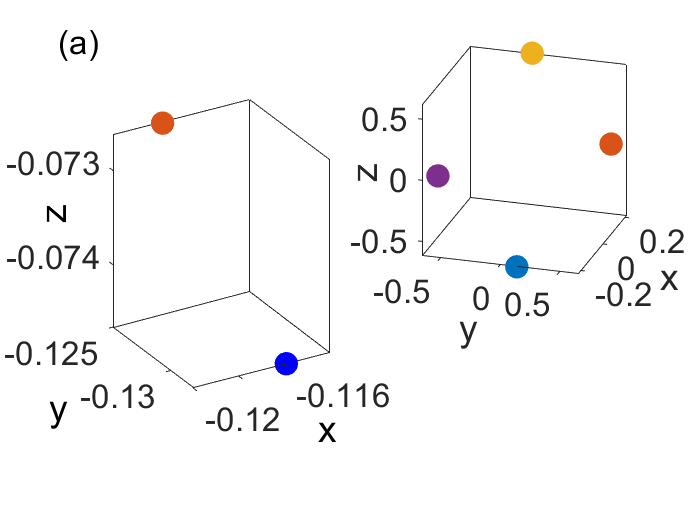}
\includegraphics[trim={0.2cm 0.1cm 0.5cm 0.5cm},clip,width=5.9cm]{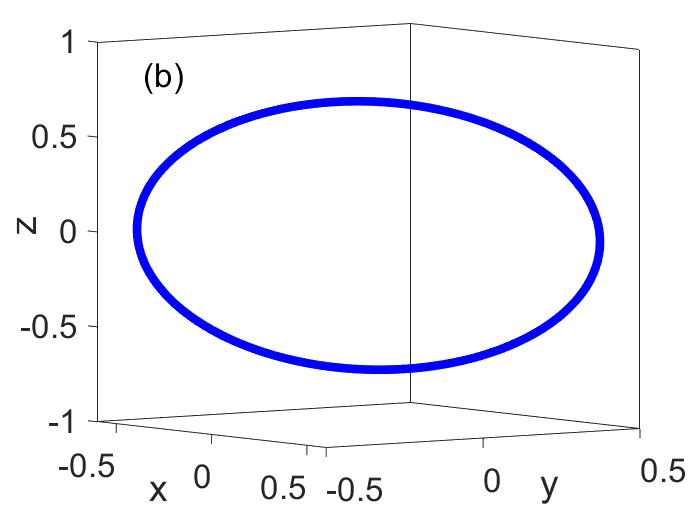}
\includegraphics[trim={0.2cm 0.1cm 0.5cm 0.5cm},clip,width=5.9cm]{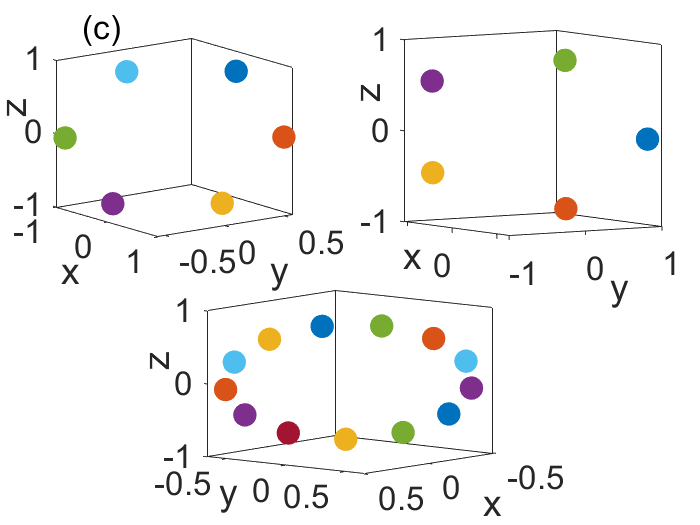}\\
\includegraphics[trim={0.2cm 0.1cm 0.5cm 0.5cm},clip,width=5.9cm]{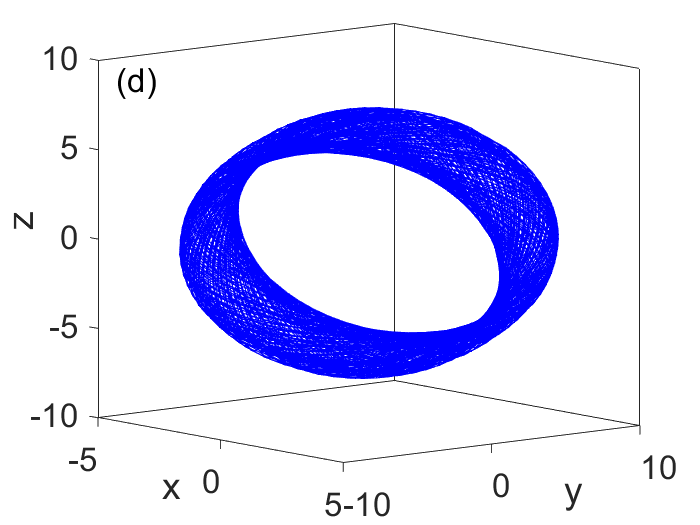}
\includegraphics[trim={0.2cm 0.1cm 0.5cm 0.5cm},clip,width=5.9cm]{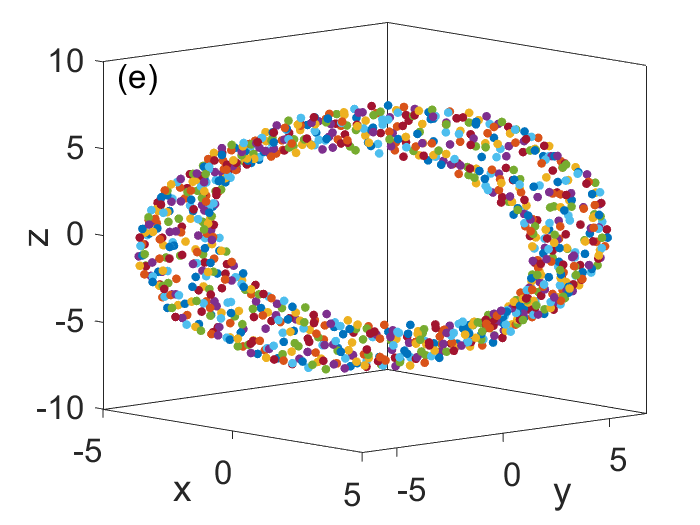}
\includegraphics[trim={0.2cm 0.1cm 0.5cm 0.5cm},clip,width=5.9cm]{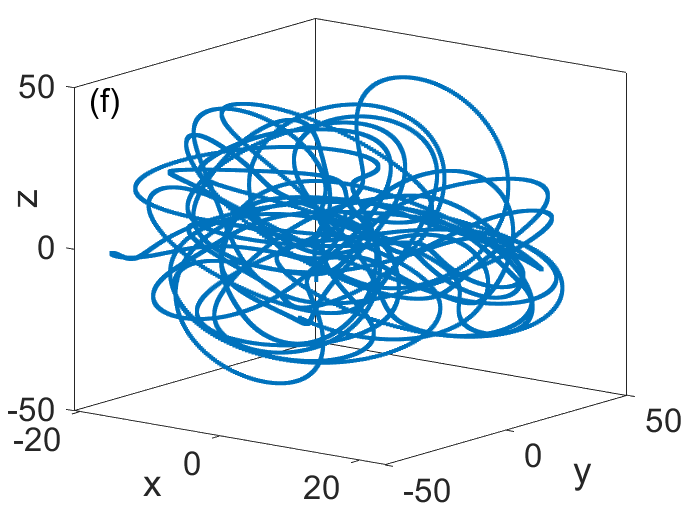}
\\
\end{center}
\caption{Center of mass trajectories of different attractors for $N=128$, $R_0=v_0=1$, $\eta=0$ and different $\beta$. {\bf (a)} Period 2 ($\beta=60000$) and period 4 ($\beta=300$) attractors. {\bf (b)} Quasiperiodic attractor that appears at $\beta=2N=256$. {\bf (c)} Periodic solutions with larger periods:  6 ($\beta= N=128$), 5 ($\beta\approx 177$),and 13 ($\beta\approx 225$). {\bf (d)-(e)} Torus-like chaotic attractor for $\beta=1$ depicted for a long and a shorter time interval. {\bf (f)} Chaotic attractor for $\beta=0.01$: the center-of-mass trajectory will fill a sphere-like body if depicted for much longer times. Note that increasing $\beta$ confines the motion to smaller volumes. }
 \label{fig1}
\end{figure}
\end{center}
\end{widetext}

The VM exhibits a variety of attractors for different values of confinement $\beta$ and alignment noise $\eta$, as depicted in Fig.~\ref{fig1} for $\eta=0$ and $N=128$. For large $\beta$, the swarm occupies the unit sphere and it is pulsating with period 2: all particles reverse their velocities at each time step. The center of mass (CM) of the swarm occupies two positions ($\beta=60000$) or four ($\beta=300$, period 4 attractor) as shown in Fig.~\ref{fig1}(a). As $\beta$ decreases, there appear quasiperiodic attractors interspersed with periodic attractors with higher periods, and chaotic attractors, cf. Fig.~\ref{fig1}(b)-(f). 

\begin{widetext}\begin{center}
\begin{figure}[ht!]
\includegraphics[trim={0.2cm 0.1cm 0.5cm 0.5cm},clip,width=5.9cm]{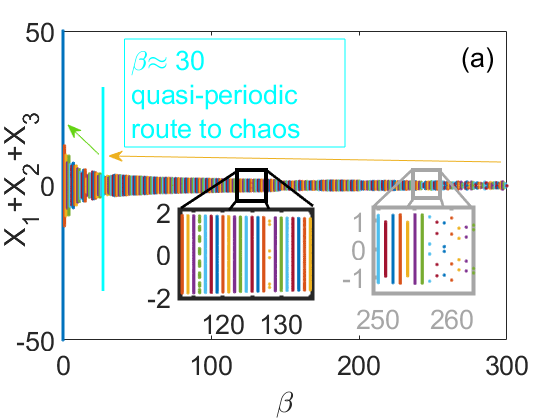}
\includegraphics[trim={0.2cm 0.1cm 0.5cm 0.5cm},clip,width=5.9cm]{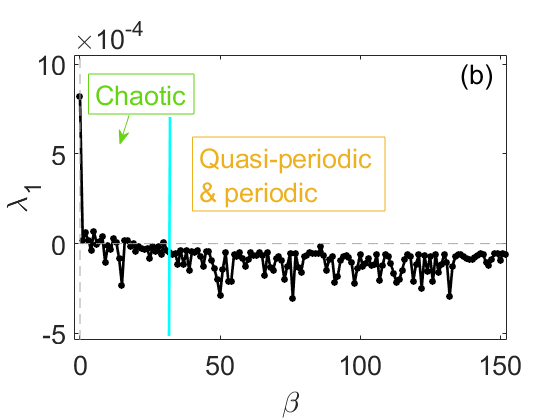}
\includegraphics[trim={0.2cm 0.1cm 0.5cm 0.5cm},clip,width=5.9cm]{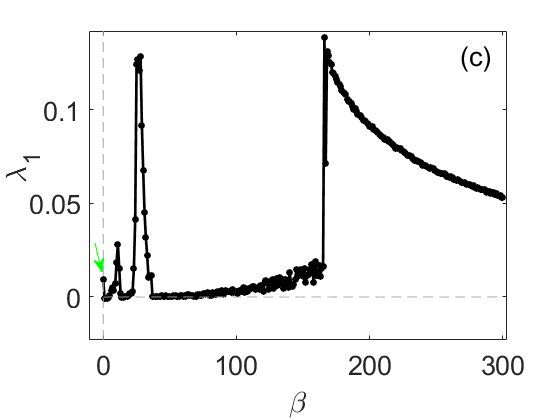}
\caption{ {\bf (a)} Bifurcation diagram of the sum of CM coordinates in nondimensional units and {\bf (b)} largest Lyapunov exponent (LLE) versus $\beta$ for $\eta=0$. The boxes in Panel (a) about $\beta=N$ and $\beta=2N$ correspond to period 6 solutions and others interspersed with quasiperiodic solutions, and a period 4 to quasiperiodic transition, respectively. Chaotic solutions appear following the quasiperiodic route to chaos. In panel (a), note the large increase of the range of CM values as $\beta$ decreases. {\bf (c)} Same as Panel (b) for $\eta=0.5$. Note how noise increases the value of the LLE and induces chaos for confinement values that correspond to quasiperiodic attractors for $\eta=0$. The area marked by green arrow in Panel (c) corresponds to the scale-free chaos transitions discussed in the present paper. Other parameters are as in Fig.~\ref{fig1}. 
}
\label{fig2}
\end{figure}
\end{center}
\end{widetext}

Fig.~\ref{fig2}(a)-(b) show how different attractors in Fig.~\ref{fig1} appear as the parameter $\beta$ changes. Periodic and quasiperiodic attractors exist for large confinement values and quasiperiodicity turns into chaos below $\beta\approx 30$. The chaotic attractor first looks like a torus and its central hole is filled as $\beta$ decreases, cf. Figs.~\ref{fig1}(d)-(f). As shown in cf. Fig.~\ref{fig2}(c), the alignment noise increases LLE values, there are parameter regions where noise induces chaos and there are scale-free chaos transitions, which will be discussed later.

\begin{widetext}
\begin{center}
\begin{figure}[ht]
\begin{center}
\includegraphics[width=5.8cm]{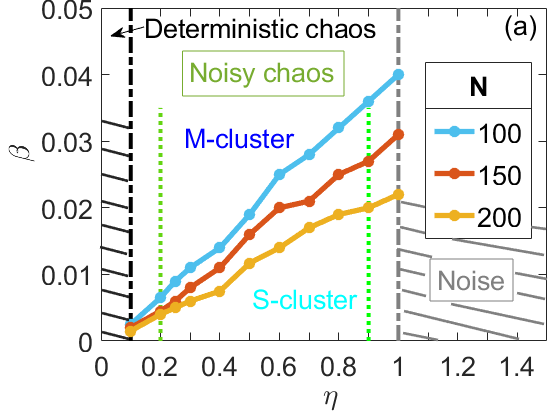}
\includegraphics[width=5.8cm]{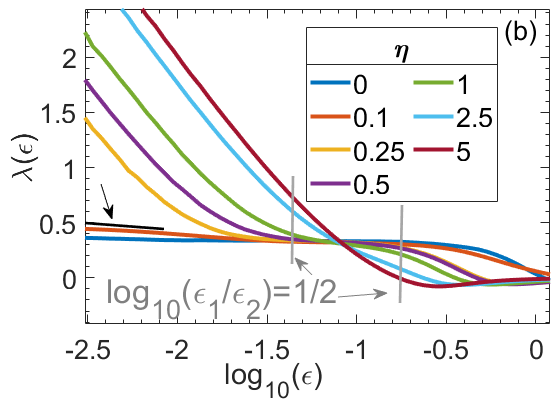}
\includegraphics[width=6.1cm]{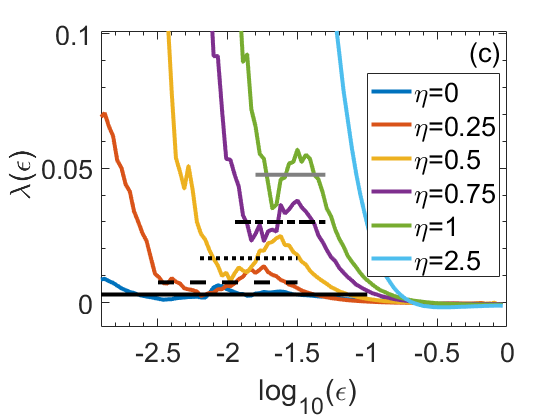}
\end{center}
\caption{{\bf Scale-free chaos. (a)} Phase diagram $\beta$ vs $\eta$ exhibiting regions of deterministic and noisy chaos, and of noisy disorder. The vertical lines at $\eta=0.2$ and 0.9 correspond to the maximum correlation length observed in experiments and to the noise for which the dynamic correlation function ceases to be flat near $t=0$, respectively. Noise swamps chaos for $\eta\geq1$. The three lines of critical points in the noisy chaos region correspond to critical confinement $\beta_c(N,\eta)$ for $N=100, 150, 200$.  They separate multicluster (M-cluster) from single cluster (S-cluster) chaos. {\bf (b)} Largest scale-dependent Lyapunov exponent as a function of the scale parameter $\epsilon$ for $N=100$, different values of $\eta$, two lagged coordinates $m=2$ and $\beta=\beta_c(N,\eta)$, see Appendix \ref{ap:b}. The LLE is the value of $\lambda(\varepsilon)$ at a plateau $(\varepsilon_1,\varepsilon_2)$ whose width satisfies $\log_{10}(\varepsilon_2/\varepsilon_1)\geq 1/2$ (Appendix \ref{ap:b}). The vertical lines mark the width of the critical plateau at which: $\log_{10}(\varepsilon_2/\varepsilon_1)= 1/2$ and correspond to the grey vertical dot-dashed line in Panel (a). The black line and arrow mark the very small slope of the SDLE for noise values close to deterministic chaos. By convention \cite{gao06}, noise swamps chaos when $\log_{10}(\varepsilon_2/\varepsilon_1)< 1/2$. {\bf (c)}  Largest scale-dependent Lyapunov exponent as a function of the scale parameter $\varepsilon$ for $N=100$, different values of $\eta$, and $\beta=\beta_c(N,\eta)$ with $m=6$, instead of $m=2$ as in Panel (b). The averages of the oscillations corresponding to the plateau region in Panel (b) increase with the noise $\eta$ indicating that so does the LLE: $\lambda_1(0)\sim 0.003, \lambda_1(0.25)\sim 0.0075, \lambda_1(0.5)\sim 0.0165, \lambda_1(0.75)\sim 0.03, \lambda_1(1)\sim 0.0476$. } 
 \label{fig3}
\end{figure}
\end{center}
\end{widetext}

\section{Deterministic and noisy chaos} \label{sec:3} 
For small confinement values and appropriate noise, the VM exhibits chaotic attractors characterized by positive values of the largest Lyapunov exponent (LLE). It is important to assess the role of noise. As explained in Appendix \ref{ap:b}, three methods to calculate the LLE produce the same values and yield complementary information: (i) applying the Benettin algorithm to Eq.~\eqref{eq1} \cite{ben80,ott93,cen10}; (ii) using the Gao-Zheng test \cite{gao94} on time traces of the swarm center-of-mass (CM) motion $\mathbf{X}(t)$ (which could be acquired from measurements of natural swarms); (iii) scale-dependent Lyapunov exponents from time traces, which discriminate between chaos and noise \cite{gao06}. 

Fig.~\ref{fig3}(a) is the phase diagram $(\eta,\beta)$ displaying phases of deterministic, noisy chaos and noise. Inside the noisy chaos phase we have indicated the critical lines of scale-free chaos for different $N$. Chaos is scale free on those lines because the correlation length defined in the next section is proportional to the swarm size for all values of $N$. To distinguish chaotic and noise phases, we have plotted in Fig.~\ref{fig3}(b) the scale-dependent Lyapunov exponent (SDLE) $\lambda(\varepsilon)$. For $\varepsilon_t=\varepsilon$, the SDLE satisfies $\ln\lambda(\varepsilon_t)= (\ln\varepsilon_{t+\Delta t}-\ln\varepsilon_t)/\Delta t$, where $\varepsilon_t$ and $\varepsilon_{t+\Delta t}$ are the average separation between nearby trajectories at times $t$ and $t+\Delta t$. Appendix \ref{ap:b} explains how to calculate $\lambda(\varepsilon)$ from time traces of center-of-mass motion with $m$-dimensional lagged vectors \cite{gao06}. If $m=2$, Fig.~\ref{fig3}(b) shows that for $\eta=0$, $\lambda(\varepsilon)$ is flat at small scale and decreases for $\varepsilon\approx 1$. For nonzero noise, $\lambda(\varepsilon)$ decreases, reaches a plateau and decreases again as the scale $\epsilon$ increases. As noise increases, the curves $\lambda(\varepsilon)$ permit distinguishing regions in the phase plane $(\eta,\beta)$ of Fig.~\ref{fig3}(a) where chaos is either mostly deterministic, substantially altered or even induced by noise (noisy chaos), and swamped by noise; see Appendix \ref{ap:b} and \cite{gao06}. The noise level used in the numerical simulations of Refs.~\onlinecite{att14plos,att14,cav17} is 5.65 in our units. Thus, it is fully inside the noise region of Fig.~\ref{fig3}(a), far from the noisy chaos parameter values we consider here.

When two lagged coordinates are sufficient to reconstruct the chaotic attractor from CM data, the value of the SDLE $\lambda(\epsilon)$ at the plateau coincides with the LLE calculated directly from the equations of the model. This occurs for the Lorenz or Rossler attractors, as explained in Ref.~\onlinecite{gao06}. However, to reconstruct safely a chaotic attractor, the dimension of the lagged vectors should surpass twice the fractal dimension $D_0$ \cite{ott93,cen10}. For the VM, we have found that properly reconstructing the chaotic attractor requires at least 6 lagged coordinates. Six-dimensional CM trajectories contain self-intersections in dimension 2. Fig.~\ref{fig3}(c) shows that the SDLE $\lambda(\epsilon)$ with $m=6$ displays oscillations for different noise values, not a single plateau as in Fig.~\ref{fig3}(b). Thus, we need a different algorithm to calculate the LLE from data. We have used the Gao-Zheng algorithm \cite{gao94} that requires constructing a quantity $\Lambda(k)$ whose slope near the origin produces the LLE, see Appendix \ref{ap:b}. These LLE yield the horizontal lines in Fig.~\ref{fig3}(c), which coincide with the average values of the SDLE oscillations. The latter coincide with LLE calculated from Eq.~\eqref{eq1} and  increase with noise. Thus, noise enhances chaos in the noisy chaos region of Fig.~\ref{fig3}(a), which includes critical lines of scale-free-chaos phase transitions, $\beta=\beta_c(N,\eta)$, to be explained in Section \ref{sec:4}. Numerical evidence for $100\leq N\leq 5000$ suggests that these lines move to $\beta=0$ as $N\to\infty$. Without confinement, the LLE vanishes and chaos disappears. This is corroborated by a different argument \cite{boh90}. The correlation length $\xi$ is bound by the finite velocity of propagation $c$ multiplied by the time it takes two points to move exponentially far from each other, i.e., $1/\lambda_1$:
\begin{eqnarray}
\xi\leq \frac{c}{\lambda_1}. \label{eq2}
\end{eqnarray}
Thus, a phase transition with infinite correlation length can only occur for $\lambda_1=0$ \cite{boh90,cro93}.

\section{Phase transitions within regions of chaos}\label{sec:4}
Insect swarms have a small polarization order parameter and exhibit strong correlations \cite{att14plos,att14}. To previous authors, this suggests that insect swarms may be on the disordered side, close to the ordering transition of the standard VM with periodic boundary conditions for sufficiently small box size \cite{att14plos,att14}. Beyond a certain box size, the ordering transition changes from continuous to discontinuous \cite{gre04,cha19,cha20}. The reason is that the uniform ordered phase issuing continuously from the uniform disordered phase \cite{ber06,ihl11,BT18} is unstable for an wavenumbers below a certain value \cite{ber09}. Consideration of the standard VM for small boxes implies an almost uniform disordered `gas' phase \cite{cha19,cha20}, despite experimental evidence that real insect swarms form a `condensed nucleus' far from walls surrounded by a `vapor' phase \cite{sin17}. We shall show later that swarms described by the confined VM near scale-free-chaos phase transitions also have a small polarization, have a condensed nucleus surrounded by a particle `vapor', and exhibit strong correlations. 

Cavagna {\em et al} have used data extracted from observations of natural swarms to calculate static and dynamic correlation functions and found power law behavior for susceptibility, correlation length and the dynamic correlation function \cite{att14plos,att14,cav17}. Their work indicates that the Fourier transform of the dynamic connected correlation function (DCCF) is the key tool to find power laws and critical exponents from experimental data \cite{cav17,cav18}: 
\begin{eqnarray}
\hat{C}(k,t)\!=\!\left\langle\! \frac{1}{N}\sum_{i,j =1}^{N}\!\frac{\sin(kr_{ij}(t_0,t))}{kr_{ij}(t_0,t)}\delta\hat{\mathbf{v}}_i(t_0)\!\cdot\!\delta\hat{\mathbf{v}}_j(t_0+t) \!\right\rangle_{t_0}\,\,\label{eq3}
\end{eqnarray}
Here $k$, $r_{ij}(t_0,t)$, $\mathbf{V}$, $\delta\mathbf{v}_i=\mathbf{v}_i-\mathbf{V}$, $\delta\hat{\mathbf{v}}_i=\delta\mathbf{v}_i/\sqrt{\frac{1}{N}\sum_j|\delta\mathbf{v}_j|^2}$ are the wavenumber, the distance between particles $i$ and $j$ at different times (particle positions are calculated in the center of mass reference frame), the center of mass velocity, the relative velocity, and the dimensionless velocity fluctuation of the $i$th particle, respectively. The brackets in \eqref{eq3} indicate an average over the earlier time $t_0$ and an ensemble average over random initial conditions \cite{cav17}. See Appendix \ref{ap:c} for details.

For natural swarms and their numerical simulations, conservation of the number of particles requires adapting the statistical mechanics definitions of equilibrium correlation functions, correlation length and susceptibility, \cite{hua87,ami05}; see Ref.~\onlinecite{cav18}. The equilibrium static connected correlation function (SCCF) $\hat{C}(k,0)$ reaches a maximum at $k=0$, which is the susceptibility \cite{ami05}. However, Eq.~\eqref{eq3} yields $\hat{C}(0,0)=0$. For finite $N$ and near a phase transition, $\hat{C}(k,0)$ reaches a maximum at a critical wavenumber $k_c>0$. This maximum is the susceptibility $\chi$, which tends to infinity as $N\to\infty$ at the critical point \cite{cav18}. The FSS hypothesis implies that $k_c\xi$ ($\xi$ is the correlation length) is a number of order 1 and a possible choice is $\xi=1/k_c$ \cite{cav18}.  

How do we obtain the critical confinement $\beta_c(N,\eta)$? From the theory of equilibrium phase transitions, we would expect: (i) the maximum and the inflection point of the SCCF versus $\beta$ tend to infinity as $N\to\infty$ for fixed alignment noise; (ii) the correlation (relaxation) time of the DCCF at wavenumber $k_c$ tends to infinity as $N\to\infty$ (critical slowing down). See two solvable examples illustrating the relation between susceptibility and correlation time in Appendix \ref{ap:d}. We will use criteria (i) and (ii) to identify lines of transitions in the chaotic phases of the confined VM. As $N\to\infty$, these lines all go to $\beta=0$ at the same rate, thereby characterizing a unique scale-free-chaos phase transition at $N=\infty$. By an abuse of notation, we also denote $\beta_c(N,\eta)$ and the other critical lines (see below) at finite $N$ as ``lines of phase transitions''.

\subsection{Critical confinement from correlation time}
\subsubsection{Correlation time}
\begin{center}
\begin{figure}[ht]
\begin{center}
\includegraphics[width=4.25cm]{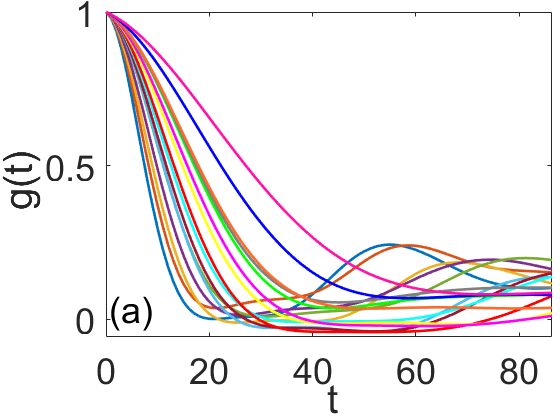}
\includegraphics[width=4.25cm]{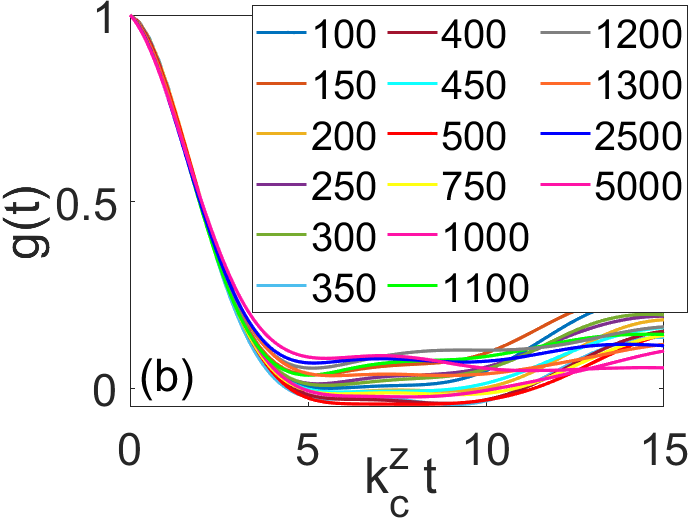}
\end{center}
\caption{Dynamic scaling of the NDCCF. $g(t)$ versus {\bf (a)} $t$, and {\bf (b)} $k_c^zt$, for $\beta=\beta_c$ and the different values of $N$ listed in the inset and $z\approx 1$. Here $\eta=0.5$. 
}
 \label{fig4}
\end{figure}
\end{center}

For the DCCF, the dynamic scaling hypothesis implies 
\begin{eqnarray}
\frac{\hat{C}(k,t)}{\hat{C}(k,0)}= f\!\left(\frac{t}{\tau_k},k\xi\right)\!= g(k^zt,k\xi); \quad g(t)=\frac{\hat{C}(k_c,t)}{\hat{C}(k_c,0)},\quad\label{eq4}
\end{eqnarray}
with $k_c=$ argmax$_k\hat{C}(k,0)$. Here $z$ is the dynamic critical exponent and $\tau_k=k^{-z}\phi(k\xi)$ is the correlation time of the normalized DCCF (NDCCF) \eqref{eq4} at wavenumber $k$ obtained by solving the equation: \cite{cav17,hal69}
\begin{eqnarray}
\sum_{t=0}^{t_{max}} \frac{1}{t}\,\sin\!\left(\frac{t}{\tau_k}\right) f\!\left(\frac{t}{\tau_k},k\xi\right)\! = \frac{\pi}{4}.  \label{eq5}
\end{eqnarray}
In Eq.~\eqref{eq5}, $t_{max}$ is the maximum time in experiments or in VM simulations \cite{cav17}. For pure exponential relaxation near an  equilibrium phase transition, $\tau_k$ obtained by solving Eq.~\eqref{eq5} equals the relaxation time \cite{cav17,hal69}. The NDCCF of a chaotic attractor first relaxes rapidly and then it exhibits damped oscillations as time elapses, cf. Fig.~\ref{fig4}(a). The rapid relaxation of $g(t)$ at short times is reminiscent of behavior near equilibrium phase transitions captured by Eq.~\eqref{eq5}. 

In simple models such as a damped harmonic oscillator forced by white noise, the DCCF has poles whose real parts also appear as reciprocal relaxation times in time dependent exponentials; see Appendix \ref{ap:d}. The SCCF contains the same poles and one of them has zero real part at the beginning of instabilities. In second order equilibrium phase transitions, one vanishing pole corresponds to a diverging susceptibility maximum and marks the critical temperature in the thermodynamic limit. Similarly, the reciprocal relaxation-correlation time of the DCCF vanishes at the critical temperature and can also be used to find it. Thus, in principle we could find the critical value of the confinement by finding the susceptibility maximum or, equivalently, the maximum correlation time. At the thermodynamic limit, the susceptibility becomes infinite and so does the correlation time (critical slowing down). We now use the same ideas to find the equivalent correlation time for the confined VM. A caveat is in order. Due to conservation of particles, $\hat{C}(0,0)=0$ and the definition of susceptibility used in equilibrium statistical mechanics has to be changed for the VM; see Appendix \ref{ap:c}. That the real parts of the poles of the susceptibility are proportional to reciprocal correlation times is no longer guaranteed; see Appendix \ref{ap:d}.

Let us endeavor to give a physical interpretation of $\tau_{k_c}$, the correlation time at $k_c\sim 1/\xi$. At fixed noise $\eta$, Fig.~\ref{fig5}(a) displays the smallest time $t_m(\beta,N)$ at which $\hat{C}(k_c,t)=0$ and the correlation time $\tau_{k_c}$ as functions of $\beta$ for $N=100, 200, 400$. The time $t_m(\beta,N)$ seems a reasonable choice for the correlation time but it varies with $\beta$. The maximal possible correlation time $t_m(\beta,N)$ would correspond to the largest negative real part of the eigenvalue posed by our hypothetical linear stability criterion. It turns out that $t_m(\beta,N)$ increases abruptly for a certain value $\beta_c$ at which $\tau_{k_c}$ is essentially minimum; see Fig.~\ref{fig5}(a). Thus, $\beta=\beta_c$ marks the largest possible correlation time based on the extension of $t_m(\beta,N)$ for $\beta\leq\beta_c$. Alternatively, the minimum value of $\tau_{k_c}$ given by Eq.~\eqref{eq5} and reached for $\beta=\beta_c$, marks the same correlation time. Fig.~\ref{fig5}(b) shows that the minimum of $\tau_{k_c}$ follows a power law $\tau_{k_c}\sim k^{-z}$, with exponent $z=1.01\pm 0.01$. For $N=100$, Fig.~\ref{fig5}(c) shows that, at $\beta$ slightly larger than $\beta_c$, the  first local minimum of $\hat{C}(k_c,t)$ becomes positive and the minimum $t_m$ having $\hat{C}(k_c,t)=0$ jumps to a much larger value. This explains the abrupt jump of $t_m(\beta,N)$ at $\beta=\beta_c$, which corresponds to the dashed line in Fig.~\ref{fig5}(a). As $N\to\infty$, $\beta_c\to 0$ and the characteristic timescale tends to infinity (critical slowing down). Fig.~\ref{fig5}(c) also shows that the correlation length decreases and the time averaged polarization $\langle W\rangle_t$, where $W$ is defined by
\begin{eqnarray}
W(t;\eta,\beta)= \left|\frac{1}{N}\sum_{j=1}^N \frac{\mathbf{v}_j(t)}{|\mathbf{v}_j(t)|}\right|\!, \label{eq6}
\end{eqnarray}
(cf Appendix \ref{ap:a}), decreases as the confinement decreases.

\begin{widetext}
\begin{center}
\begin{figure}[ht]
\begin{center}
\includegraphics[width=5.9cm]{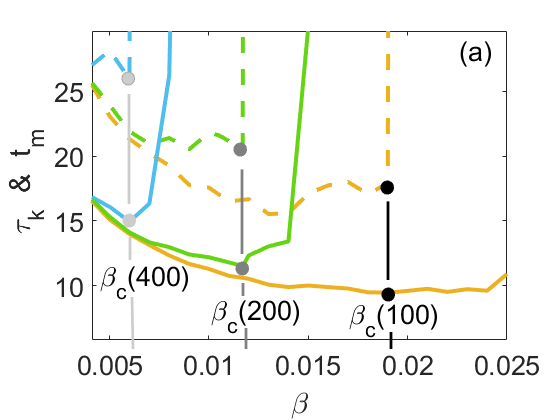}
\includegraphics[width=5.9cm]{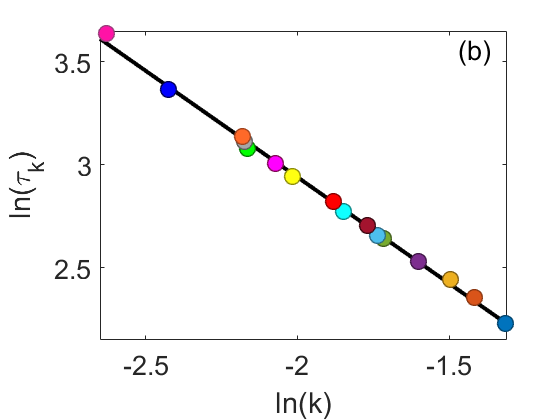}
\includegraphics[width=5.9cm]{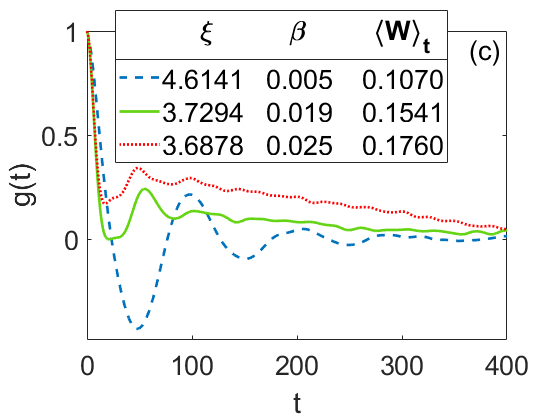}
\end{center}
\caption{{\bf (a)} Smallest time $t_m(\beta;N)$ such that  $g(t_m)=0$ (dashed curves) and characteristic timescale $\tau_{k_c}(\beta;N)$ (continuous curves) as functions of $\beta$ for $N=100, 200, 400$. The minimum characteristic timescale is close to the abrupt growth of $t_m(\beta;N)$ and marks the scale-free-chaos phase transition. {\bf (b)} Characteristic timescale, $\tau_k$, computed at $k_c=1/\xi$ for different $N$, as a function of $k$ (log-log scale): $\tau_{k_c}\sim k_c^{-z}$ with $z=1.01\pm 0.01$. {\bf (c)} Normalized DCCF vs nondimensional time for different confinement values marked and $N=100$. The inset lists the values of $\beta$, correlation length $\xi$ and time averaged polarization $\langle W\rangle_t$ for the three curves $g(t)$. In this figure, $\eta=0.5$.} \label{fig5}
\end{figure}
\end{center}\end{widetext}

\subsubsection{Collapse of the NDCCF}
\begin{center}
\begin{figure}[ht]
\begin{center}
\includegraphics[width=5.9cm]{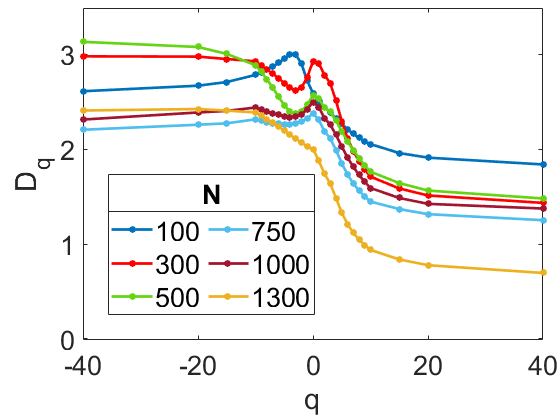} 
\end{center}
\caption{ Multifractal dimension \cite{bul99} $D_q$ vs $q$ at $\beta_c(N;\eta=0.5)$.}
 \label{fig6}
\end{figure}\end{center}

We have obtained a power law $\tau_{k_c}\sim k_c^{-z}$ with $z=1.01\pm 0.01$ for $k_c\xi=1$, as shown in Fig.~\ref{fig5}(b). For this value of $z$, Fig.~\ref{fig4}(b) illustrates how NDCCF curves $g(t)$ collapse to a single one in terms of $k_c^zt$ at the scale-free chaos line for $0<k_c^zt<4$. Moreover, NDCCF curves drop to values close to zero for $k^zt>5$ but they do not collapse for those larger times unlike critical behavior near equilibrium phase transitions. {\em What happens?} We surmize that some regions of the chaotic attractors are much more frequently visited than others, which indicates that different length and time scales coexist within the attractor. This can be ascertained by finding the multifractal dimension $D_q$. After a long transient (30000 time steps), a set of $M$ values of the CM position $\vec{x}_i=\mathbf{X}(t_i)$, $i=1,\ldots, M$, form a Poincar\'e map of the attractor. Then we define the multifractal dimension $D_q$ \cite{gra83,paw87,bul99}, 
\begin{eqnarray}
&&D_q = \frac{1}{q-1} \lim_{r \rightarrow 0} \frac{\ln [C_q(r)]}{\ln (r)}, \label{eq7}\\
&&C_q(r)=  \frac{1}{M} \sum_{i=1}^M  \left[ \frac{1}{M} \sum_{j=1}^M \theta(r-|\vec{x}_i-\vec{x}_j|) \right]^{q-1},  \label{eq8}
\end{eqnarray}
where $\theta(x)$ is the Heaviside unit step function, $M\approx 70000$, and $C_q(r)$ is the generalized correlation function. $D_0$, $D_1$ and $D_2$ are the box counting (capacity) dimension, the information dimension and the correlation dimension, respectively. As we vary $q$, different regions of the attractor will determine $D_q$. $D_\infty$  corresponds to the region where the points are mostly concentrated, while $D_{-\infty}$ is determined by the region where the points have the least probability to be found. If $D_q$ is a constant for all $q$, the CM trajectory will visit different parts of the attractor with the same probability and the point density is uniform in the Poincar\'e map. This type of attractor is called trivial, whereas a non constant $D_q$ characterizes a nontrivial attractor with multifractal structure. Fig.~\ref{fig6} shows that the box-counting dimension $D_0$ and $D_q$ for $q>0$ undergo a downward trend with increasing $N$ (decreasing $\beta_c$). Then the dimension of the more commonly visited sites of the attractor decreases. Furthermore, we shall see below that the positive LLE tends to zero and chaos disappears as $\beta\to 0$. However, the chaotic attractor remains multifractal: different time scales persist \cite{cen10}. Thus, a single rescaling of time as in Fig.~\ref{fig4}(b) cannot collapse the full NDCCF in our simulations. Curiously, the same collapse of the NDCCF as a function of $k_c^zt$ only for $0<k_c^zt<4$ occurs using data from natural swarms, as shown in Figures 2a and 2b of Ref.~\onlinecite{cav17} for $z=1.2$ (experimental data yield $z=1.12 \pm 0.16$, using the power law $\tau_{k_c}\sim k_c^{-z}$ \cite{cav17}). 

\begin{center}
\begin{figure}[h]
\begin{center}
\includegraphics[width=4.25cm]{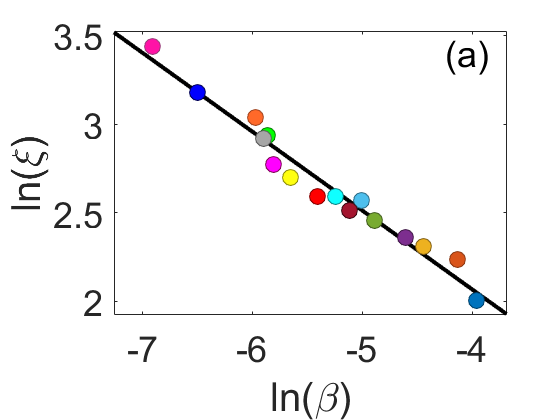}
\includegraphics[width=4.25cm]{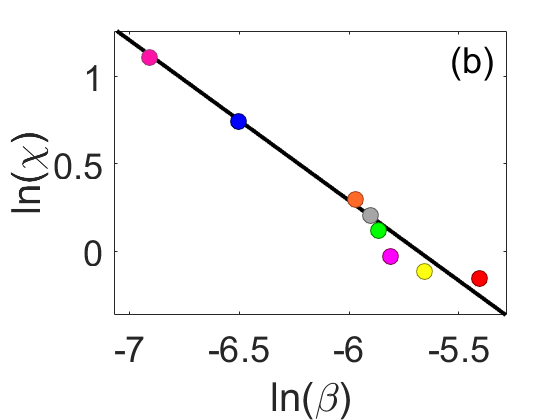}
\\
\includegraphics[width=4.25cm]{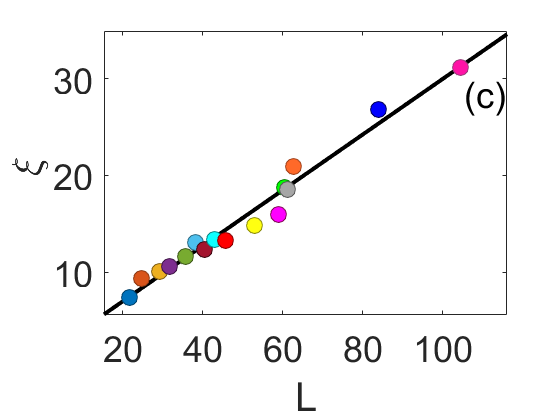}
\includegraphics[width=4.25cm]{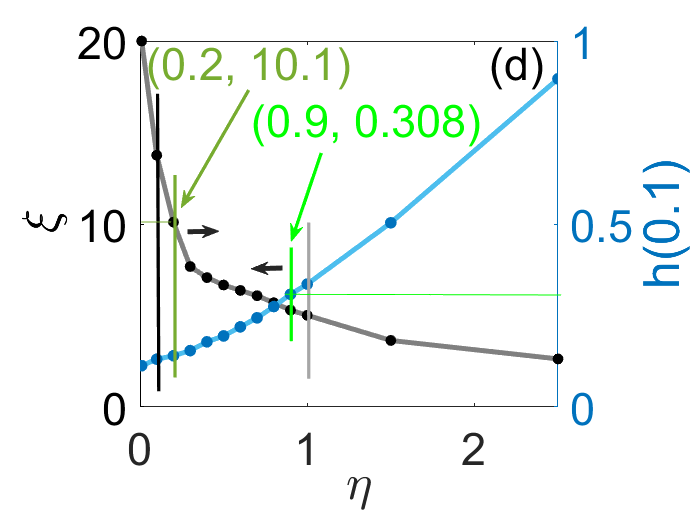}\\
\includegraphics[width=4.25cm]{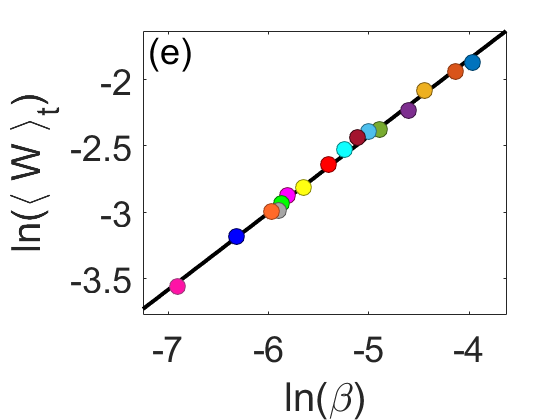}
\includegraphics[width=4.15cm]{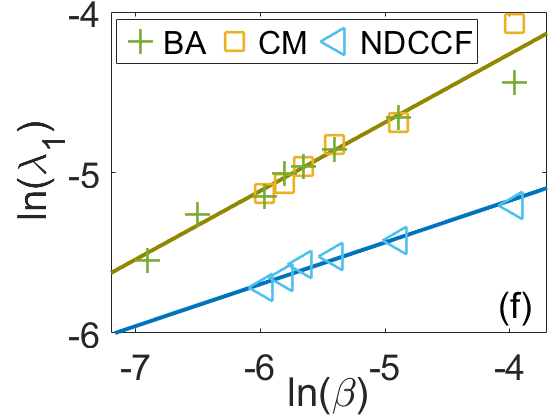}
\end{center}
\caption{{\bf (a)} Scaling of the dimensionless correlation length with $\beta$: $\xi\sim\beta^{-\nu}$, $\nu=0.436\pm 0.009$. {\bf (b)} Scaling of the real-space susceptibility with $\beta$: $\chi\sim\beta^{-\gamma}$, $\gamma=0.92\pm 0.05$ for $N=500$, 750, 1000, 1100, 1200, 1300, 2500, 5000. {\bf (c)} The correlation length increases linearly with $L$. {\bf (d)} Correlation length $\xi=r_0$ and NDCCF flatness $h(0.1)$ vs noise. Black vertical bars delimit the region of noisy chaos, occurring for smaller noise values than the ordering transition  \cite{att14plos,att14,cav17}. Green vertical bars are compatible with observations of natural swarms: the leftmost bar marks the largest observed correlation length and the rightmost bar marks when NDCCF flatness ends. {\bf (e)} Time averaged polarization versus $\beta$: $\langle W\rangle_t\sim\beta^b$, $b = 0.58 \pm 0.01$. {\bf (f)} LLE vs $\beta$ for different $N$, $\lambda_1\sim \beta^{\varphi}$, calculated by the Benettin algorithm \cite{ben80} for the complete system (crosses), from CM motion (squares) and from the NDCCF (triangles). We get $\varphi=0.43 \pm 0.03$ (crosses and squares), and $\varphi=0.24\pm 0.02$ (triangles). In Panels (a), (c) and (e), $N$ values are as in Fig.~\ref{fig4}(b), $\eta=0.5$.}  \label{fig7}
\end{figure}
\end{center}

\subsubsection{Critical exponents}
Having found the critical confinement $\beta_c(N;\eta)$ as the value of $\beta$ for which $\tau_{k_c}$ is minimum, we can find the power laws and the critical exponents for the correlation length, susceptibility, time-averaged polarization order parameter, and the LLE $\lambda_1$ in terms of $\beta=\beta_c(N;\eta)$:
\begin{eqnarray}
&&\chi(\beta,N,\eta)\equiv\max_r Q(r)\sim \beta^{-\gamma},\,\, \xi\equiv\mbox{argmax}_rQ(r)\sim\beta^{-\nu},\quad \label{eq9}\\
&&\lambda_1\sim \beta^\varphi\sim N^{-\frac{\varphi}{3\nu}}, \quad\langle W\rangle_t\sim\beta^b, \label{eq10}
\end{eqnarray}
as $\beta=\beta_c(N;\eta)\to 0$ with $N\gg 1$. Here $Q(r)$ is the cumulative real-space correlation function, which we define below.

\subsubsection{Real space susceptibility}
To calculate the susceptibility, we have used the maximum of the cumulative real-space correlation function (corresponding to the first zero $r_0$ of the real-space correlation function, which is now the correlation length) at $\beta_c(N;\eta)$ \cite{att14plos,att14}:
 \begin{eqnarray}
Q(r)= \frac{1}{N}\sum_{i=1}^{N}\sum_{j\neq i}^{N} \delta\hat{\mathbf{v}}_i\!\cdot\!\delta\hat{\mathbf{v}}_j\theta(r-r_{ij}). \label{eq11}
\end{eqnarray}
As shown in Appendix \ref{ap:c}, selecting $\hat{C}(k_c,0)$ as the susceptibility does not produce a monotonic function of $\beta_c$ or of $N$. Thus, $\hat{C}(k_c,0)$ cannot be used to fit a power law over an extended range. However, $1/k_c$ and $r_0$ are linearly related, and using either one as correlation length yields the same critical exponent $\nu$; see Appendix \ref{ap:c}. A similar relation between $1/k_c$ and $r_0$ also holds for midge data; see Fig.~SF1 of Ref.~\onlinecite{cav17}. To calculate the LLE we can use the Benettin algorithm for the VM of Eq.~\eqref{eq1} or a convenient time series obtained from the simulations, e.g., the CM evolution or the NDCCF; see Appendix \ref{ap:b}. 
 
 Figs.~\ref{fig7}(a) and \ref{fig7}(b) depict how correlation length and real-space susceptibility scale with $\beta$ and Fig.~\ref{fig7}(c) confirms that the correlation length is proportional to the size of the swarm. For $\eta=0.5$, we obtain the critical exponents $\nu=0.436\pm 0.009$ and $\gamma=0.92\pm 0.05$, respectively. Fig.~\ref{fig7}(d) shows that the correlation length decreases with alignment noise at critical confinement. Correlation length values in the region of noisy chaos are $\xi$ times 4.68 cm, and they are compatible with observations of natural swarms \cite{att14plos,att14,cav17}. 
 
 Another feature of swarm data compatible with our numerical simulations of the confined VM is that the NDCCF is flat at the origin. Referred to Eq.~\eqref{eq4}, we define the `flatness' function as 
 \begin{eqnarray}
h(x)=-\frac{1}{x}\ln f(x,1), \quad  x=\frac{t}{\tau_k},  \label{eq12}
\end{eqnarray} 
for a fixed value of the nolise $\eta$. A perfectly flat NDCCF implies that $h(0)=0$. However, $h(x)$ from experiments changes abruptly below $x=0.1$ as shown in Fig.~3b of Ref.~\onlinecite{cav17}. The same figure yields an upper value 0.3 of $h(0.1)$ for natural swarms, which we arbitrarily select as the transition value from flat to non-flat NDCCF.  For the confined VM, the transition value occurs at $\eta=0.9$ in Fig.~\ref{fig7}(d), which is close to the change to noise from noisy chaos in Fig.~\ref{fig3}(a).  Fig.~\ref{fig7}(d) shows that the correlation length decreases and $h(0.1)$ increases with increasing $\eta$. Thus, observed correlation lengths and flat NDCCFs occur in the region of noisy chaos of the confined VM that contains the scale-free-chaos phase transitions. Fig.~\ref{fig7}(e) depicts the power law of the time-averaged polar order parameter $\langle W\rangle_t$ versus $\beta$, which shows the scale-free-chaos phase transition to be of second order with critical exponent $b = 0.58 \pm 0.01$. 
 
 \subsubsection{Bound for the LLE critical exponent}
 The LLE $\lambda_1$ decreases as $\beta_c(N;\eta)$ does according to the power law \eqref{eq10} with critical exponent $\varphi= 0.43\pm 0.03$ provided the LLE is calculated using the Benettin algorithm on Eq.~\eqref{eq1} or time traces of the CM as explained in Appendix \ref{ap:b}. See Fig.~\ref{fig7}(f). For chaotic systems with short range interactions such as the confined VM, Eq.~\eqref{eq2} together with Eqs.~\eqref{eq9} and \eqref{eq10} imply that $\beta^{\varphi-\nu}\leq c$. To be consistent as $\beta\to 0$, this relation then implies 
 \begin{eqnarray}
 \varphi\geq\nu.  \label{eq13}
 \end{eqnarray} 
 Were the dynamic scaling of \eqref{eq4} to hold for all time, $e^{\lambda_1t}$ would be a function of $k^zt$; therefore $\lambda_1 \sim k_c^z\sim \beta^{z\nu}$, and $\varphi=z\nu$. Eq.~\eqref{eq13} then produces $z\geq 1$, which agrees with all our simulations carried out with the Benettin algorithm or reconstructing the chaotic attractor from center of mass data. Thus, $\varphi\approx z\nu\geq\nu$ approximately holds for one-time functions such as the center of mass trajectory with $\eta=0.5$. However, this relation fails for the two-time NDCCF, which has a smaller $\varphi$; see Fig.~\ref{fig7}(f). 

\begin{center}
\begin{figure}[h]
\begin{center}
\includegraphics[width=4.25cm]{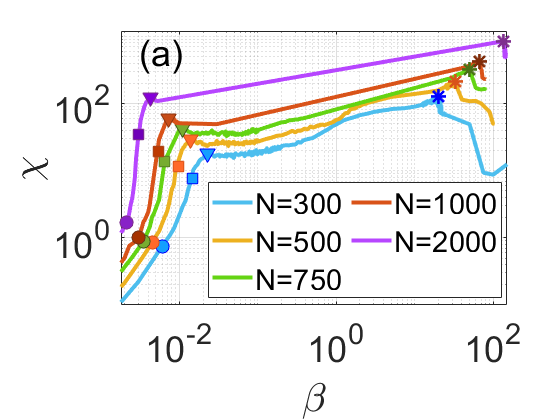}
\includegraphics[width=4.25cm]{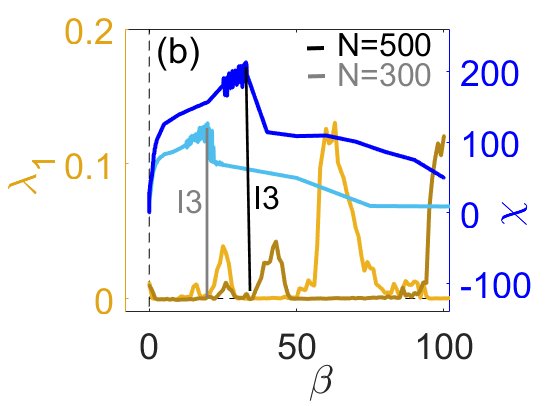}\\
\includegraphics[width=4.25cm]{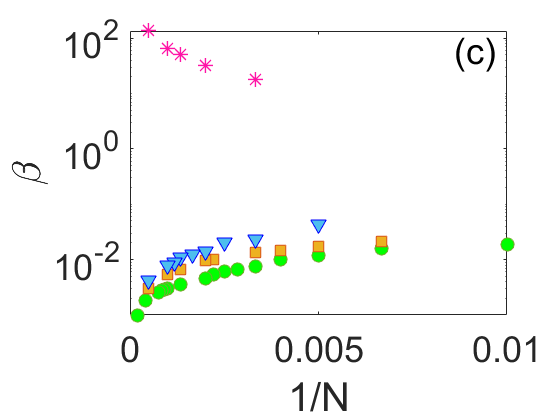}
\includegraphics[width=4.25cm]{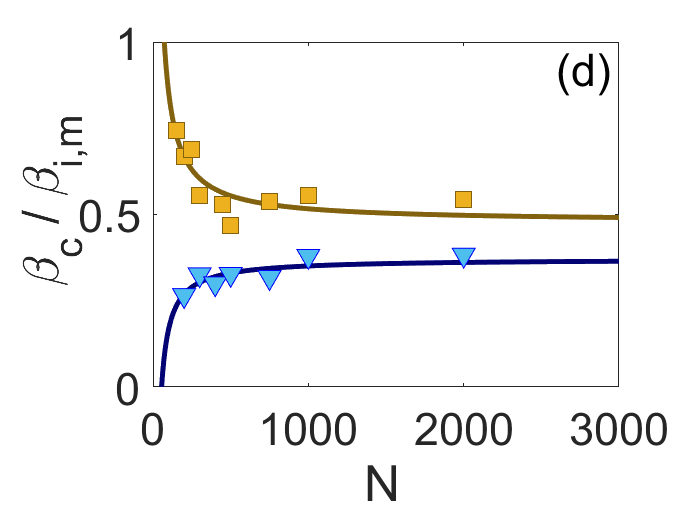}\\
\includegraphics[width=4.25cm]{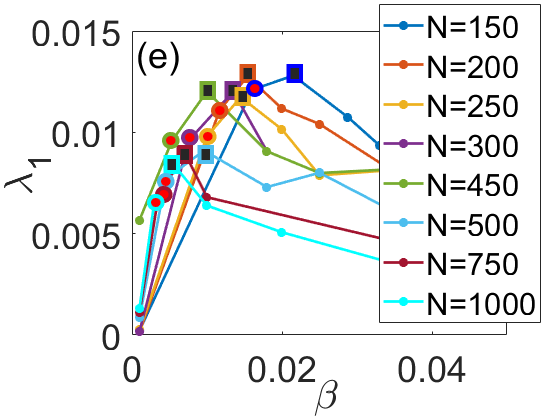}
\includegraphics[width=4.25cm]{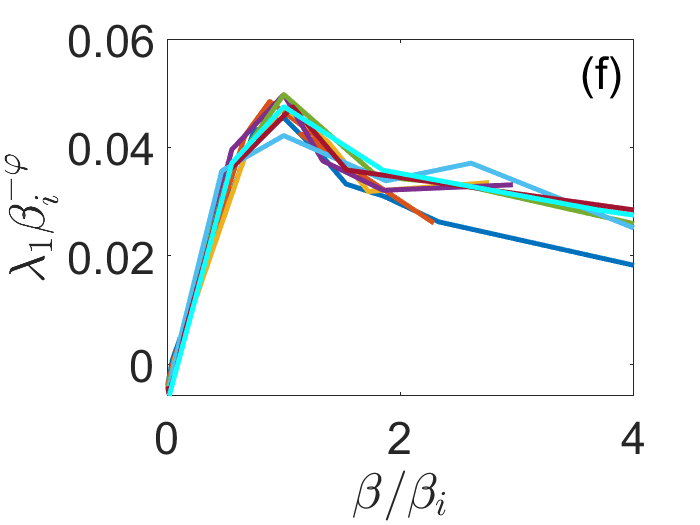}\\
\includegraphics[width=4.25cm]{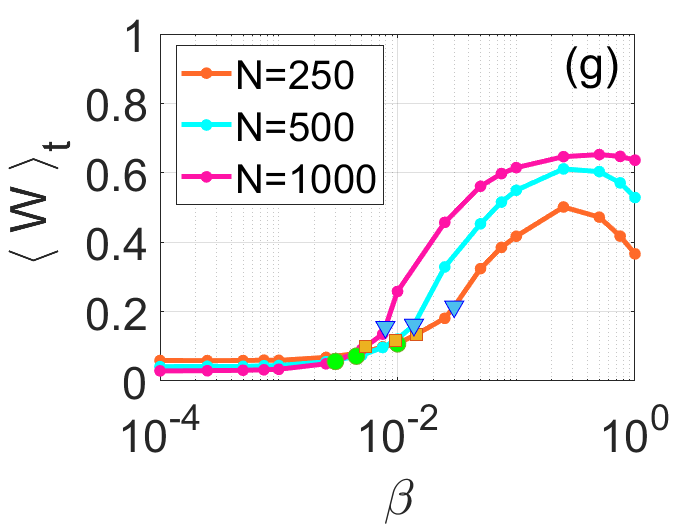}
\end{center}
\caption{{\bf (a)} Real-space susceptibility (log-log scale) for $N=300, 500, 750, 1000, 2000$.  {\bf (b)} LLE and susceptibility versus $\beta$ for $N=300, 500$. Circles, squares, triangles and asterisks mark $\beta_c$, $\beta_i$, $\beta_m$ (local $\chi$ maximum), and $\beta_M$ (global $\chi$ maximum of the susceptibility, at the beginning of the third chaotic window, marked by I3 in Panel b), respectively. {\bf (c)} $\beta_c$, $\beta_i$, $\beta_m$, and $\beta_M$ versus $1/N$. {\bf (d)} Ratios $\beta_c/\beta_i= 0.48+37.76/N$ and $\beta_c/\beta_m= 0.37-20.48/N$ as $N\gg 1$. {\bf (e)} LLE versus $\beta$ for $N$ marked in the inset. {\bf (f)} Same as Panel (e) with axes $\lambda_1/\beta_i^{\varphi_i}$ and $\beta/\beta_i$, $\varphi_i=0.33$, showing collapse of curves. {\bf (g)} Time averaged polarization versus $\beta$ for $N$ marked in the inset. Circles correspond to the critical confinement $\beta_c(N;\eta)$, squares correspond to the inflection point of the susceptibility $\beta_i(N;\eta)$. Here $\eta=0.5$. }
\label{fig8}
\end{figure}
\end{center}

\subsection{Critical confinement from the static correlation} 
	At critical confinement, the susceptibility $\chi=$max$_r Q(r)$, given by Eqs.~\eqref{eq9} and \eqref{eq11} for fixed $\beta_c(N;\eta)$, $\eta$ and $N$, becomes infinity as $N\to\infty$. For given values of the alignment noise $\eta$, we can find other values of $\beta$, e.g., the local maximum and the inflection point of $\chi=\chi(\beta,N;\eta)$ as a function of $\beta$, which also tend to infinity as $N\to\infty$. At finite $N$, these confinement values, $\beta_i(N;\eta)$ (inflection) and $\beta_m(N;\eta)$ (local maximum), are different from $\beta_c(N;\eta)$, as shown in Fig.~\ref{fig8}(a). Fig.~\ref{fig8}(b) shows that there are different regions of positive LLE separated by non-chaotic regions. The first chaotic window starts at very small positive values of $\beta$ that cannot be appreciated at the scale of the figure. The global maximum of the susceptibility is reached at large values of $\beta$ corresponding to the beginning of the third chaotic window in Fig.~\ref{fig8}(b), which is different from the scale-free chaos window of $\beta_c$, $\beta_i$, $\beta_m$. Unlike the isolated $\beta_M$, the values $\beta_c$, $\beta_i$ and $\beta_m$ tend to 0 as $N\to\infty$, as observed in Fig.~\ref{fig8}(c). Fig.~\ref{fig8}(d) shows that the ratios $\beta_c/\beta_i$ and $\beta_c/\beta_m$ tend to constant values (about 0.48 and 0.37, respectively) as $N\to\infty$. Thus, for sufficiently large $N$, the critical exponents are the same for the lines $\beta_c$, $\beta_i$, $\beta_m$ and therefore they correspond to the same phase transition.
			
Figures~\ref{fig8}(e) and \ref{fig8}(f) show that the LLE versus $\beta$ curve reaches a local maximum at $\beta_i$. Thus, maximum `chaoticity' is reached at the line of susceptibility inflection points. This qualitative feature of the line of inflection points is one of the reasons we have decided to use them to characterize the transition at finite $N$, at the same level as the line of local maxima usually selected in the statistical physics literature. %The LLE power law of Eq.~\eqref{eq10}, calculated from Eq.~\eqref{eq1}, has a critical exponent $\varphi_i= 0.33 \pm 0.04$ at $\beta=\beta_i$, and $z_i\nu_i\approx 0.66>\varphi_i$. 
Fig.~\ref{fig8}(g) shows the average polarization as a function of $\ln\beta$ for $N=250, 500, 1000$. As $N$ increases, $\beta_c$, $\beta_i$, $\beta_m$ simultaneously decrease to zero and so do the corresponding polarization order parameters, which suggest that these lines represent a second order phase transition at $N=\infty$.

\begin{center}
\begin{figure}[h]
\begin{center}
\includegraphics[width=4.25cm]{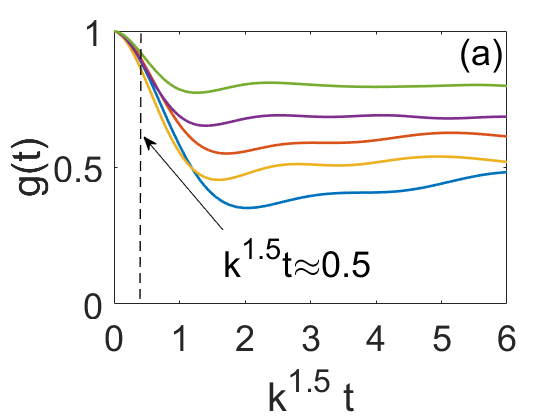}
\includegraphics[width=4.25cm]{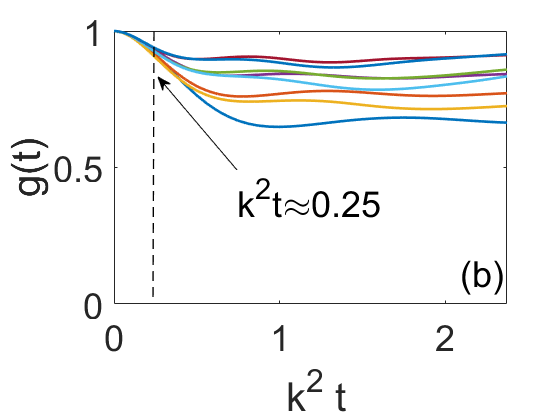}
\end{center}
\caption{ Collapse of NDCCF data as function of {\bf (a)} $k^{1.5}t$ at the inflection point and of {\bf (b)} $k^2t$ at the local maximum of the susceptibility on the indicated narrow interval near $t=0$. Here $\eta=0.5$.}  \label{fig9}
\end{figure}
\end{center}

Figs.~\ref{fig9} (a) and \ref{fig9}(b) show NDCCF collapse of the NDCCF at $\beta_i(N;\eta)$ and $\beta_m(N;\eta)$, respectively. The respective dynamical critical exponents are $z_i= 1.5$ on $0<k^zt<0.5$ and $z_m=2$ on $0<k^zt<0.25$. These exponents have been visually fitted because the correlation times obtained using Eq.~\eqref{eq5} were unable to collapse data, unlike what happened for the transition from single cluster to muticluster chaos at $\beta_c$. Furthermore, the minima of $g(t)$ are all larger than 0.3, hence they are no longer close to zero as in Fig.~\ref{fig4}.   The different dynamical critical exponents at the different critical lines could be associated to different length scales in the multifractal chaotic dynamics at the three critical lines. The connection between dynamics and susceptibility in nonequilibrium phase transitions needs further study.

\section{Subtracting rotation and dilation from CM velocity and flocking black hole phase transition}\label{sec:5}
The confined VM is said to experience an ordering transition at high values of noise. In Refs.~\onlinecite{att14,cav17}, the noise value chosen for numerical simulations is $\eta=0.45 \times 4\pi=5.65$ in our units. This value is well inside the noise region of Fig.~\ref{fig3}(a). We can expect some remnants of coherent structures exhibiting rotation and dilation there. In equilibrium phase transitions, the order parameter is independent of time and space, and it is defined using ensemble averages. To mimic these transitions, in our definitions of DCCF and SCCF for large noise, it is convenient to subtract overall rotation and dilation from the CM velocity at each time step when defining fluctuations of the velocity, as explained in Appendix \ref{ap:c}; see also Refs.~\onlinecite{att14plos,att14}.  What is the effect of these operations?

\begin{center}
\begin{figure}[ht]
\begin{center}
\includegraphics[width=4.25cm]{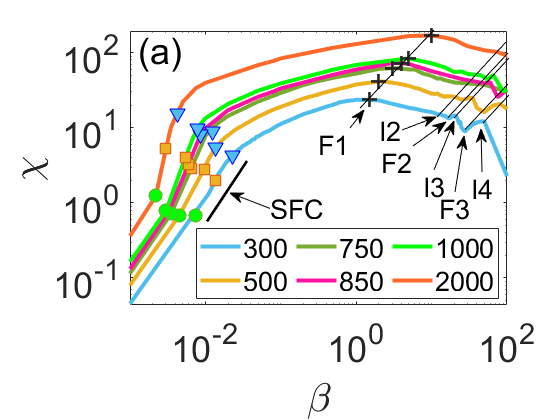}
\includegraphics[width=4.25cm]{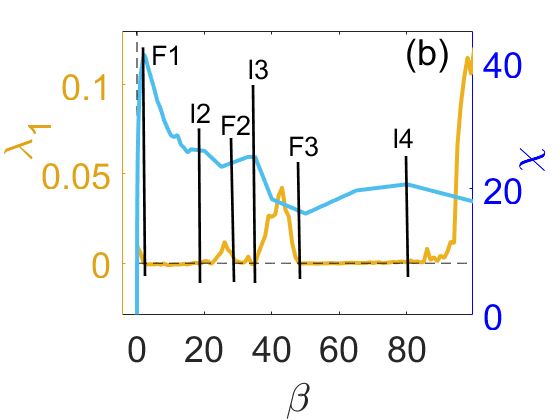}
\end{center}
\caption{ Real-space susceptibility (log-log scale) versus $\beta$ for $N=300, 500, 750, 850, 1000, 2000$. Circles and squares mark $\beta_c(N;\eta)$ and $\beta_i(N;\eta)$, which are the same with or without subtraction of rotation and dilation from CM motion. Triangles are the local maxima of susceptibility, $\beta_m$, without subtractions. Lines Ii, Fi ($i=1,\ldots, 4$) mark the initial and final $\beta$ value of the $i$th chaotic window. Crosses mark $\beta_M$ (global $\chi$ maximum of the susceptibility), which coincides with the line F1. {\bf (b)} LLE and susceptibility versus $\beta$ for $N=500$. Note the different chaotic windows and the lines Ii, Fi.}  \label{fig10}
\end{figure}
\end{center}

Below the critical line $\beta_c(N;\eta)$ but close to it, the swarm occupies a single cluster and it is disperse. Fig.~\ref{fig10}(a) shows that  $\beta_c(N;\eta)$ and $\beta_i(N;\eta)$ do not change upon subtracting rotation and dilation from the CM velocity. However, the points on $\beta_m(N;\eta)$ are not local maxima of $\chi$ vs $\beta$ in Fig.~\ref{fig10}(a). The line of global maxima of susceptibility versus $\beta$ moves to the end of the first chaotic window of Fig.~\ref{fig8}(b). The different chaotic windows for $N=500$ are shown in Fig.~\ref{fig10}(b). When the swarm splits into several clusters, they rotate and move with respect to each other. These effects are small at $\beta=\beta_i(N;\eta)$ (where the LLE reaches a local maximum) but the local maxima of the susceptibility versus $\beta$ disappear. Thus, the lines $\beta_c(N;\eta)$ and $\beta_i(N;\eta)$ move to $\beta=0$ as $N\to\infty$ at the same rate, thereby providing a finite size approximation of the scale-free-chaos phase transition. Using $\beta=\beta_c(N;\eta)$, the critical exponents $\nu=0.43\pm 0.01$ and $z=1.00\pm 0.03$ do not change when rotation and dilation are subtracted from CM motion in velocity fluctuations. We now have $\gamma=0.85\pm 0.04$. See Appendix \ref{ap:c}. What about the line $\beta_M(N;\eta)$ of global maxima?

The correlation length is finite at $\beta_M(N;\eta)$, $\xi\approx 2.5$ for $\eta=0.5$ and $N$ values up to 2000. We have checked that the end of the first chaotic window (at which the LLE becomes zero again) and the $\beta$ values of all successive chaotic windows increase with $N$. What happens? At the end of the chaotic window the clusters in the chaotic swarms are confined to regions of size $\xi\approx 2.5$ or smaller (recall that large $\beta$ values confine all particles to a sphere of diameter 2 with period 2 motion in the deterministic case, $\eta=0$). As $N$, and therefore $\beta_M(N;\eta)$, increase, more and more particles enter these regions, which implies that the average minimal distance between neighbors, $x$, decreases to zero as $N\to\infty$. Thus, confinement becomes infinitely strong, the average distance between neighbors tends to zero, and the particle density inside clusters becomes infinite. We call these clusters flocking black holes.  Qualitatively, this situation resembles gravitational collapse of self-gravitating particles \cite{alb20,chavanis20}. In particular, type II gravitational collapse to a zero mass black hole is analogous to second-order phase transitions with $x\to 0$ instead of $\xi\to\infty$ \cite{gun07}. By using $1/x$ instead of $\xi$, we can define critical exponents analogous to $\nu$ and $\gamma$ for this flocking black hole phase transition: 
\begin{eqnarray}
  x\sim  \beta^{-\nu}, \quad  \chi \sim \beta^{\gamma},\quad\tau_{k_c}\sim x^{-z}, \label{eq14}
\end{eqnarray}
with $\beta=\beta_M(N;\eta)\to\infty$ as $N\to\infty$. Subtracting rotation and dilation from CM motion, we find the critical exponents, $ \nu = 0.35 \pm 0.08$, $\gamma = 0.97 \pm 0.08$. Using fluctuations about the CM velocity without subtractions, $\beta_M(N;\eta)$ is larger and we can define similar critical exponents with it. We find $\nu = 0.33 \pm 0.02$, $\gamma =1.03\pm 0.03$. In both cases, the finite correlation length takes on the value 2.5, and the dynamical critical exponent is $z=0$. The NDCCF decays at short times and it later oscillates. All curves for different $N$ exhibit the same initial decay but the successive oscillations are irregular and displaced from one another. 

\subsection{Critical exponents from confined VM simulations}
We have found a line of phase transitions $\beta_c(N;\eta)$ representing the change from scale-free single to multicluster chaos. See Section \ref{sec:6}. The critical line is inside the confinement region where the NDCCF is flat at the origin, which corresponds to underdamped relaxation dynamics. For $\eta=0.5$ (middle of the noisy chaos region), as $\beta=\beta_c(N;\eta)\to 0$, $N\to\infty$, we have obtained $\nu=0.436\pm 0.009$ (correlation length), $\gamma=0.92\pm 0.05$ (real-space susceptibility), and $z= 1.01 \pm 0.01$ (dynamic exponent). The critical exponent for the LLE law is approximately $\varphi=z \nu$. The critical exponents change little for $0.1<\eta<1$.  Other critical lines based on the inflection points or local maxima of susceptibility versus $\beta$ collapse to $\beta=0$ at the same rate as $N\to\infty$ and therefore represent the same phase transition; cf Fig.~\ref{fig8}(d). 

While $\beta_c(N;\eta)$ and $\beta_i(N;\eta)$ do not change after subtracting rotation and dilation from the CM velocity, the values of the correlation length and time change, while still growing with $N$. For the critical line  $\beta_c(N;\eta)$, we have found the critical exponents $\nu = 0.43 \pm 0.01$, $\gamma = 0.85 \pm 0.04$ and $z=1.00\pm 0.03$, which change but little with respect to the previous values. 

The line of maxima of the susceptibility versus confinement curve, $\beta_M(N;\eta)$, is near the end of the first chaotic window and the LLE are small there.  $\beta_M(N;\eta)$ goes to infinity with $N$ and the correlation length goes to a finite constant. This phase transition at infinite confinement strength produces a collapse of particles inside bounded and infinitely dense clusters with vanishing LLE and critical exponents given by Eq.~\eqref{eq14}. For this flocking black hole phase transition, we have found $\nu=0.35\pm 0.08$, $\gamma=0.97\pm 0.08$ for $\eta=0.5$ and $N$ values up to 2000.  

\begin{center}
\begin{figure}[h]
\begin{center}
\includegraphics[width=4.25cm]{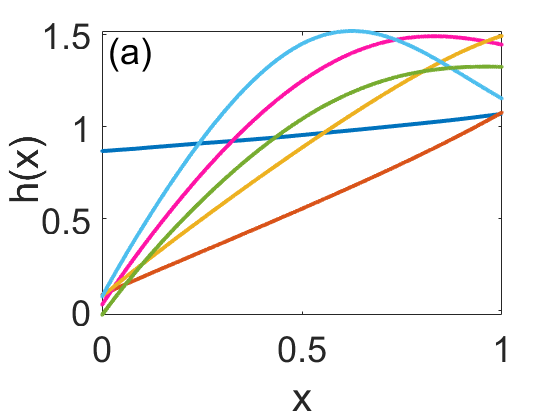} 
\includegraphics[width=4.25cm]{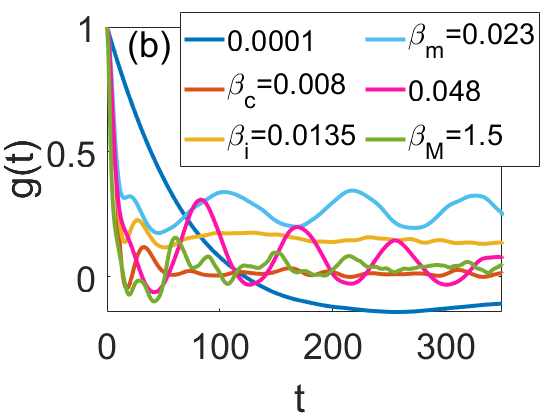} 
\end{center}
\caption{{\bf (a)} Flatness function $h(x)$ and {\bf (b)} NDCCF $g(t)$ for $N=300$, $\eta=0.5$ and $\beta$ values indicated in Panel (b) inset.}
 \label{fig11}
\end{figure}\end{center}

\subsection{Flatness function}
A feature shared by swarm data and the scale-free-chaos phase transition is that the NDCCF is flat at small times. Fig.~\ref{fig11}(a) depicts the flatness function $h(x)$ of Eq.~\eqref{eq12} for a range of $\beta\in (0,\beta_M)$, where (after subtracting rotation and dilation) $\beta_M$ is at the end of the first chaotic window of Fig.~\ref{fig8}(b). For $0<\beta\ll\beta_c(N;\eta)$, $h(x)\sim 1$ (as $x\to 0$), which implies exponential relaxation of the NDCCF with time, typical of overdamped dynamics. For these small values of $\beta$, noise overwhelms coherent dynamics induced by confinement and the LLE is negative. As $\beta$ increases towards $\beta_c(N;\eta)$, $h(x)$ decreases until it is $\approx 0$ for $x\to 0$. Then, VM dynamics is chaotic and underdamped. Fig.~\ref{fig11}(a) shows the change in $h(x)$ for different values of $\beta$. See Fig.~SF4 of Ref.~\cite{cav17} for examples of overdamped and underdamped dynamics in the stochastic oscillator. Fig.~\ref{fig11}(b) shows that the NDCCF exhibits oscillations for underdamped dynamics and a slower decay for overdamped noisy dynamics. The lines NDCCF oscillate at positive values of $g(t)$ for $\beta_i(N;\eta)$ and $\beta_m(N;\eta)$ (this later line corresponds to the local maximum without subtracting rotation and dilation from CM motion), whereas the oscillations have larger amplitude for larger values of $\beta$ and $g(t)$ may take on negative values. Thus, the flatness function indicates that the confined VM displays underdamped dynamics in the critical region about $\beta_c(N;\eta)$ for the scale-free-chaos phase transition. 

Inside the noise region of Fig.~\ref{fig3}(a), the NDCCF decays exponentially at short times and it is non-flat according to the definition \eqref{eq12}. This is the case for the noise value $\eta=5.65$ (in our units) used in Refs.~\cite{att14,att14plos,cav17}. This exponential decay led to the hasty conclusion that the confined VM displays overdamped dynamics \cite{cav17}, and to a subsequent search for convenient underdamped dynamics producing a flat NDCCF near $t=0$ \cite{cav21arxiv}. However, equations with discrete time dynamics, such as the harmonically confined VM, may exhibit overdamped and underdamped dynamics on different parameter ranges. 

In experiments, the smallest measured value of $h(x)$ occurs at $x=0.1$ and $h(0.1)<0.3$ for natural swarms \cite{cav17}. At the VM order-disorder phase transition, $h(0.1)\approx 1>0.3$ (exponential decay, clearly non-flat NDCCF) \cite{cav17}. For the confined VM, the transition value occurs at $\eta=0.9$ in Fig.~\ref{fig7}(d) and in Fig.~\ref{fig3}(a), which is close to the change to noise from noisy chaos at $\eta=1$ (much lower than the noise for the order-disorder phase transition of the VM in a box with periodic boundary conditions \cite{att14plos,att14,cav17}). As noise increases, Fig.~\ref{fig7}(d) shows that the correlation length decreases and $h(0.1)$ increases with increasing $\eta$. Thus, observed correlation lengths and flat NDCCFs occur in the region of noisy chaos of the confined VM that contains the scale-free-chaos phase transitions. 

\section{Phase transition and topological data analysis}\label{sec:6}

\begin{center}
\begin{figure}[ht]
\begin{center}
\includegraphics[width=4.25cm]{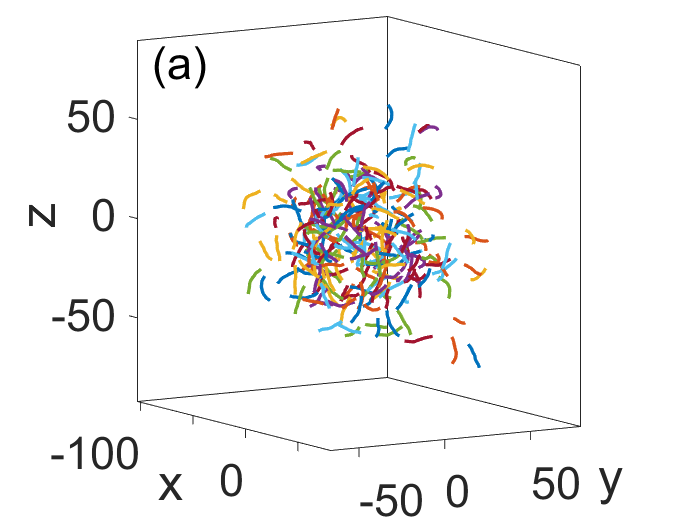}
\includegraphics[width=4.25cm]{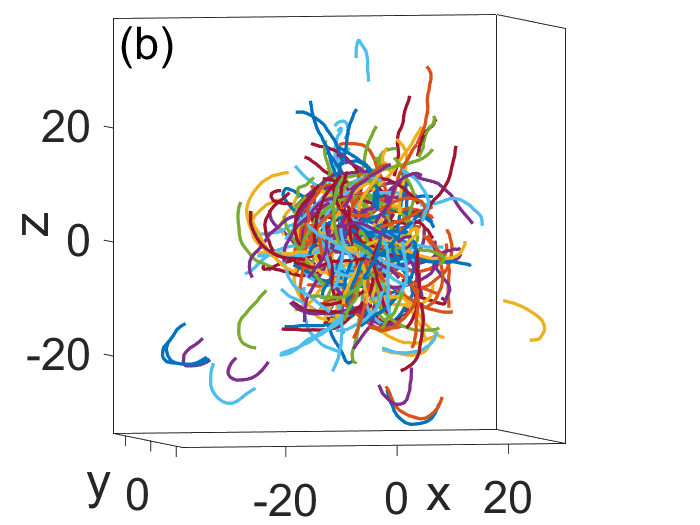}
\includegraphics[width=4.25cm]{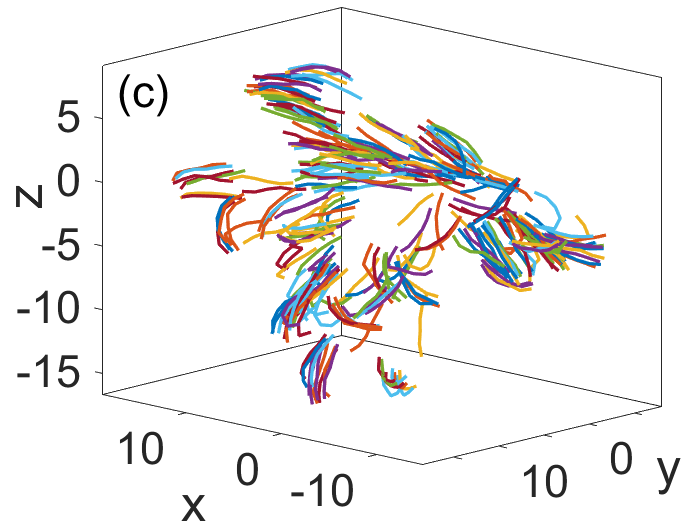}
\includegraphics[width=4.25cm]{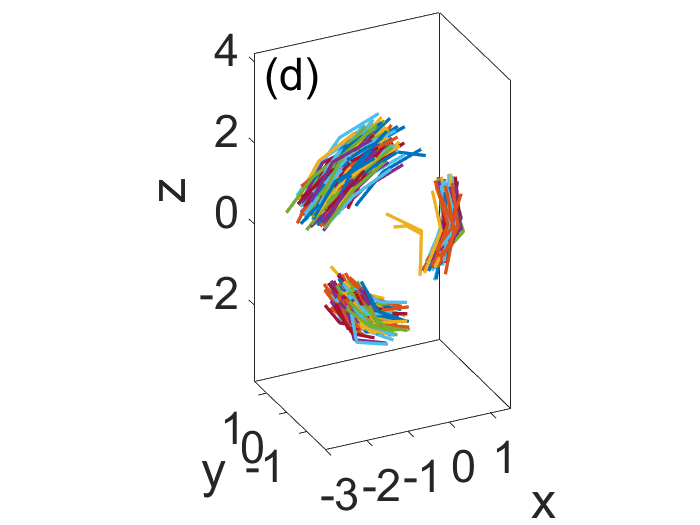}
\includegraphics[width=4.25cm]{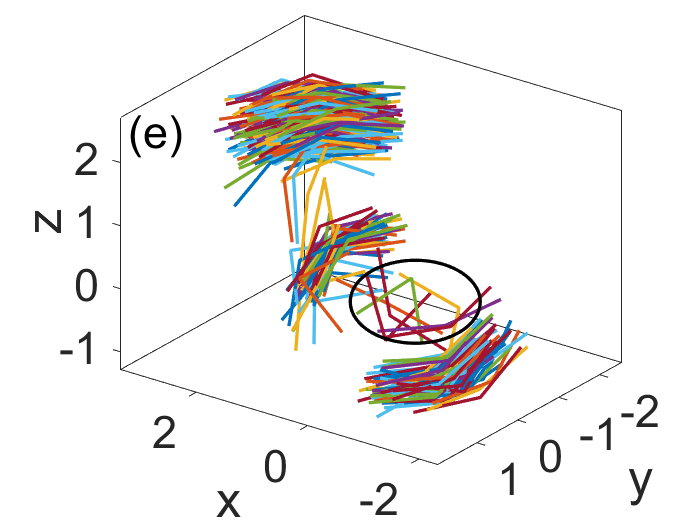}
\includegraphics[width=4.25cm]{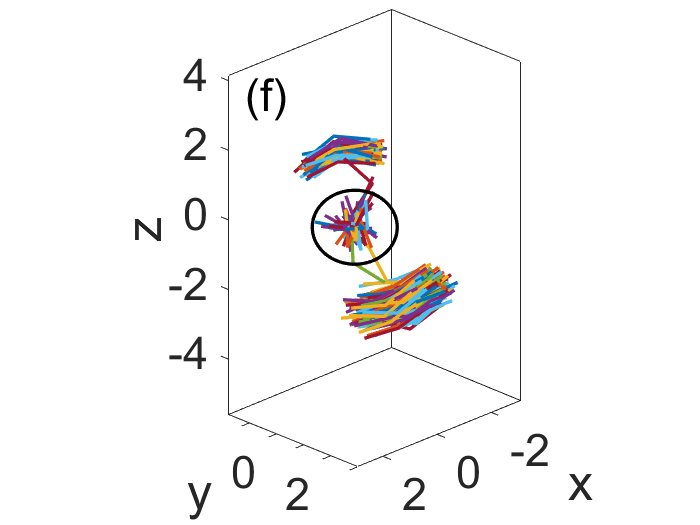}
\end{center}
\caption{Chaotic swarms of $N = 300$ particles showing short trajectories of the particles for confinements near its critical value, $\beta=\beta_c(N;\eta)$ and $\eta=0.5$: {\bf (a)} Sparse single cluster chaos for $\beta<\beta_c(N;\eta)$, {\bf (b)} compact single cluster chaos at $\beta=\beta_c(N;\eta)$, multicluster chaos for {\bf (c)} $\beta>\beta_c(N;\eta)$ and for {\bf (d)} $\beta=$ argmax $\chi(\beta)=19.8$ (global maximum without subtracting rotation and dilation from CM motion). {\bf (e)} For $\beta=21$, three chaotic clusters move around a central sphere (located by a black circle) where other particles are confined; and {\bf (f)} for $\beta=25$, only two chaotic clusters remain and more particles are trapped in the central sphere. }  \label{fig12}
\end{figure}
\end{center}

The structure of clusters changes as $\beta$ surpasses $\beta_c$, the critical confinement calculated from relaxation time. Fig.~\ref{fig12}(a) shows the swarm particles and their short time trajectories for $\beta<\beta_c(N;\eta)$: the particles form a single cluster. Figs.~\ref{fig12}(b) and \ref{fig12}(c) correspond to $\beta=\beta_c(N;\eta)$ and $\beta>\beta_c(N;\eta)$, respectively. For $\beta=\beta_c(N;\eta)$, the particles form a single cluster and fill a smaller volume, whereas for $\beta>\beta_c(N;\eta)$, the swarm has split in several clusters. Fig.~\ref{fig8}(f) shows that the average polarization is very small for sparse single-cluster chaos, $\beta<\beta_c(N;\eta)$, and it increases with $\beta$ in the multicluster chaotic region, $\beta>\beta_c(N;\eta)$. Multicluster behavior is even clearer when $\beta$ gives the global maximum of the susceptibility as in Fig.~\ref{fig12}(d). For larger values of $\beta$, some particles start being confined in a sphere centered at the origin and their number increases with $\beta$, as shown in Figs.~\ref{fig12}(e) and \ref{fig12}(f).

These findings can be rendered more precise by {\em topological data analysis} (TDA) \cite{zom02,ede10,sin17}. TDA borrows ideas from persistent homology, traditionally used to distinguish structures in low dimensional topological spaces (e.g., circle, annulus, sphere, torus, etc) by quantifying their connected components, topological circles, trapped volumes. For instance, given a point cloud $\mathbf x_1$, \ldots, $\mathbf x_N$ in $\mathbb R^3$, we can infer whether it represents a sphere or a torus by calculating the  homologies $\mathsf{H}_0$, $\mathsf{H}_1$, $\mathsf{H}_2$, and the corresponding Betti numbers $\mathsf{b}_0$, $\mathsf{b}_1$, $\mathsf{b}_2$. The different homologies can be calculated regardless of the dimension of the underlying space, as long as a distance or metric is defined \cite{zom02}.
 
\begin{figure}
\centering
\includegraphics[trim=300 200 300 200,clip=true,height=3cm]{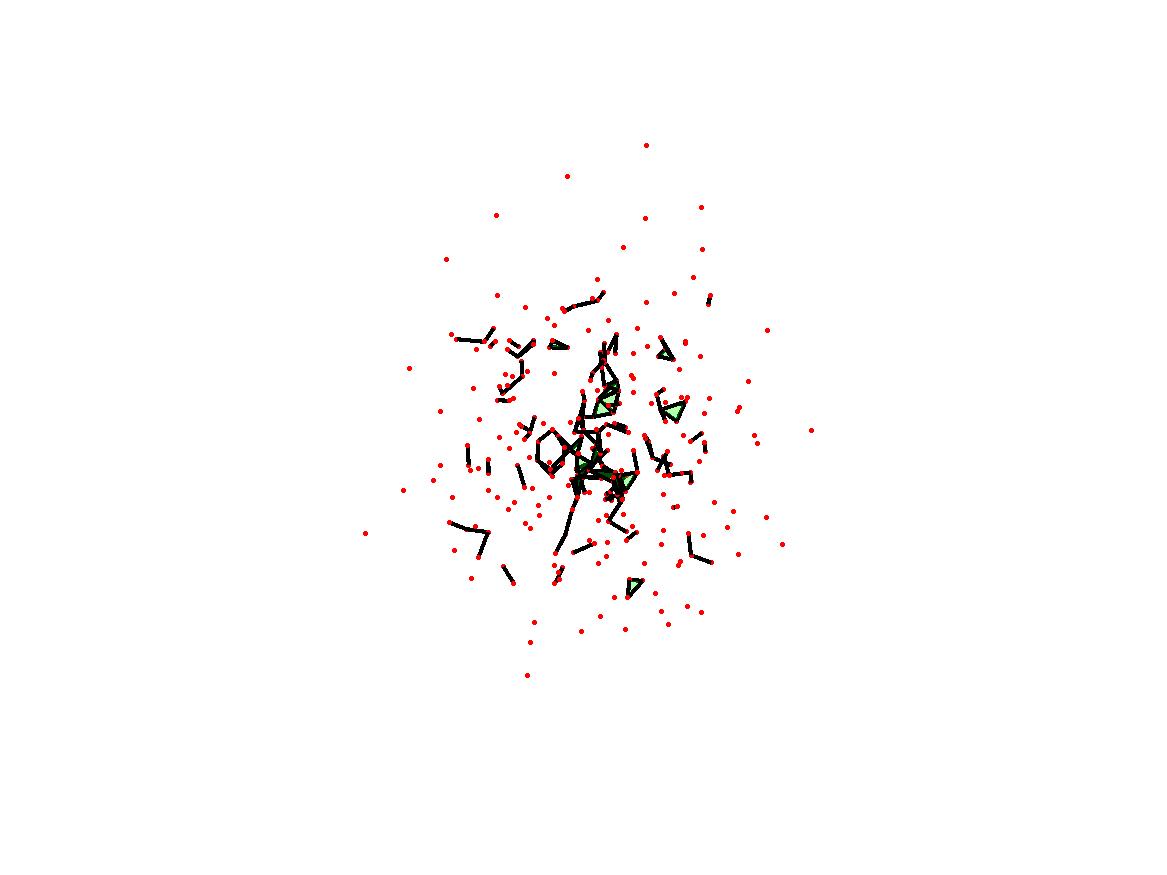} 
\includegraphics[trim=300 200 300 200,clip=true,height=3cm]{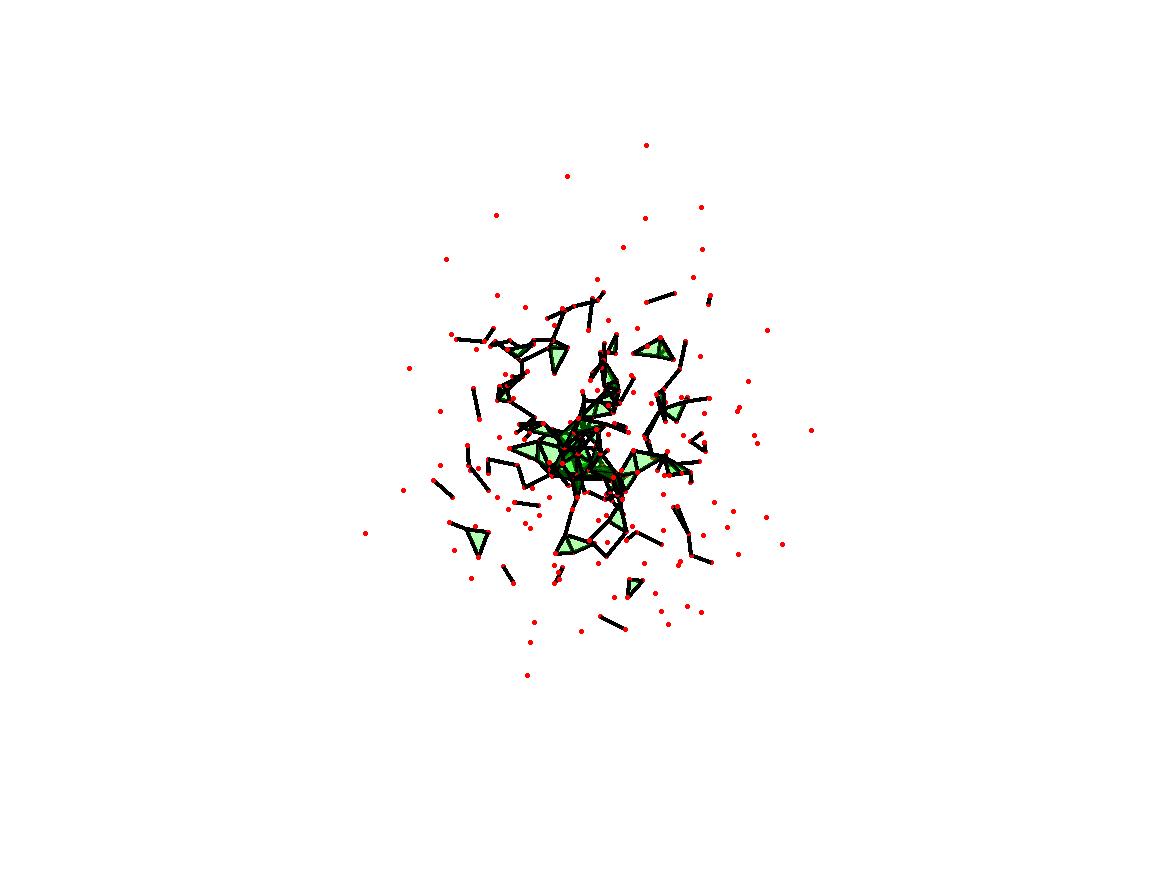}  
\includegraphics[trim=300 200 300 200,clip=true,height=3cm]{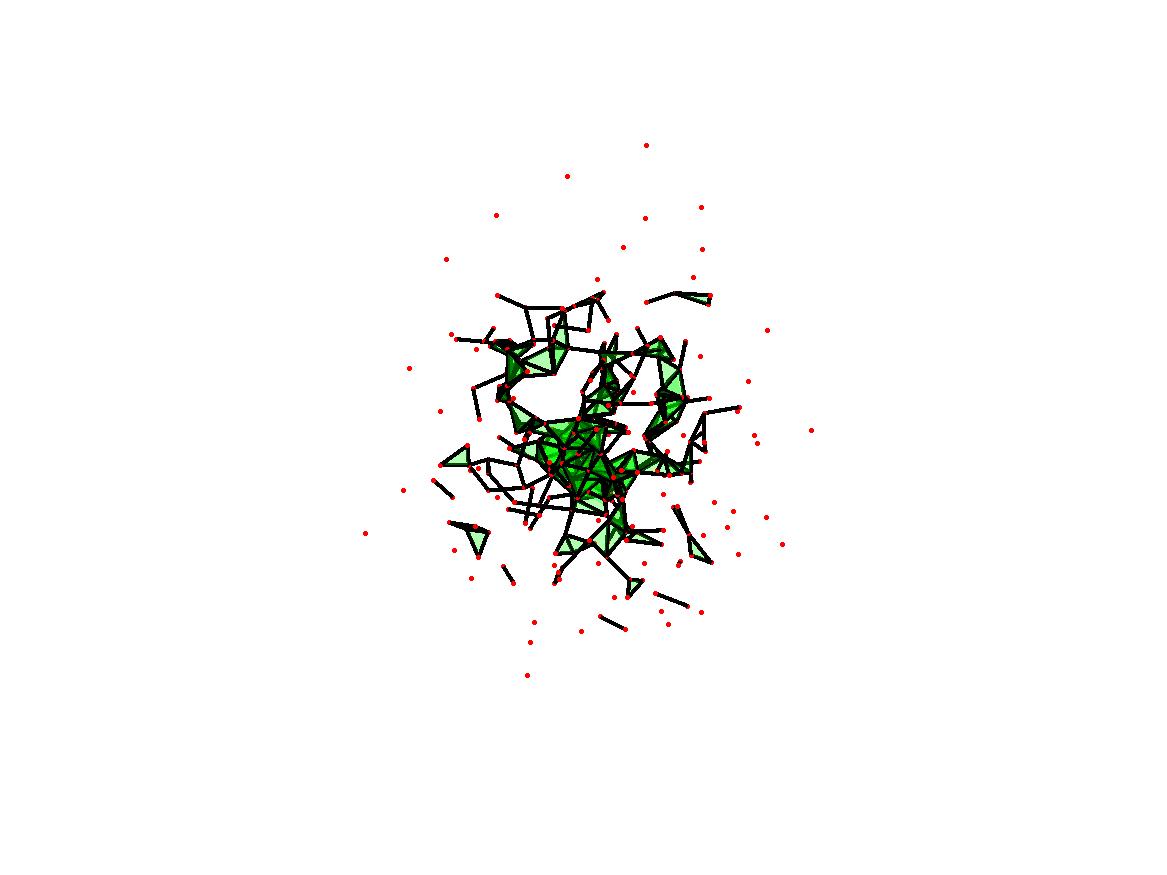} 
\includegraphics[trim=300 200 300 200,clip=true,height=3cm]{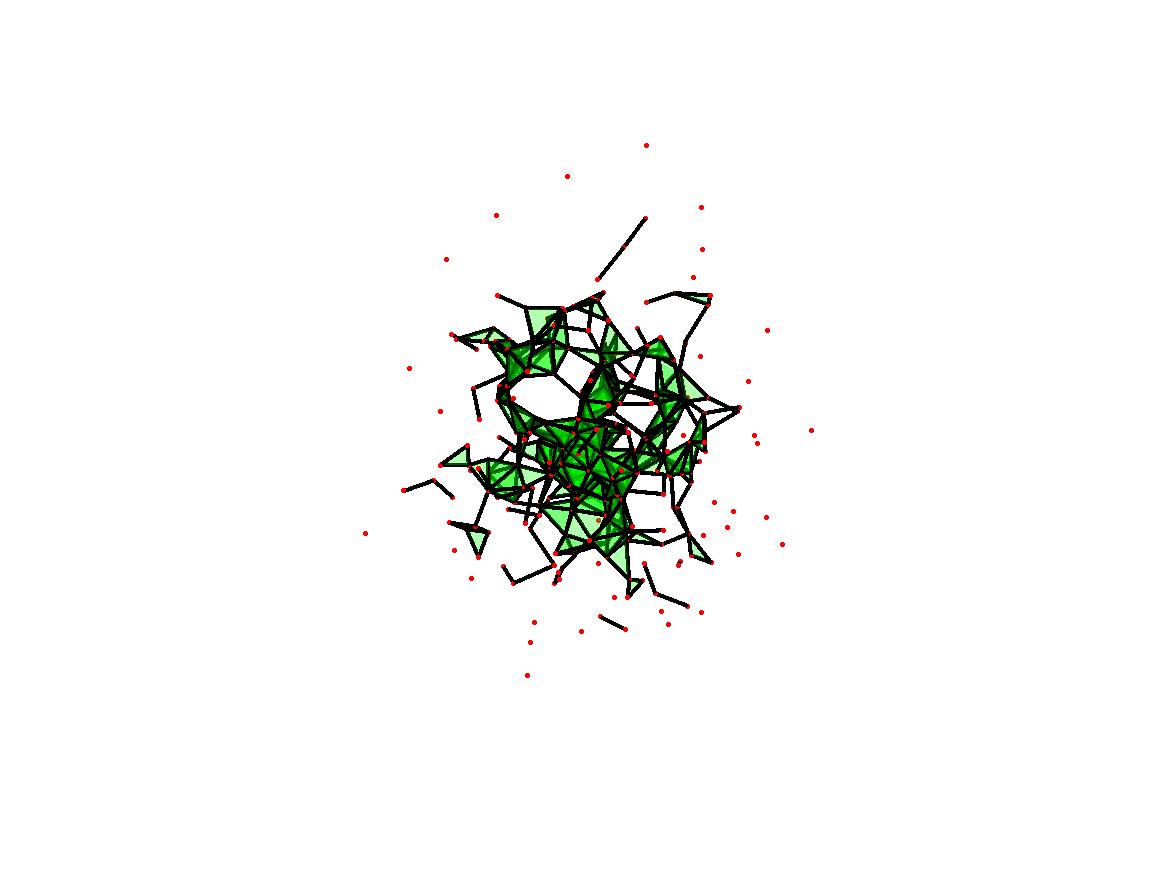}  \\
\includegraphics[trim=300 200 300 200,clip=true,height=3cm]{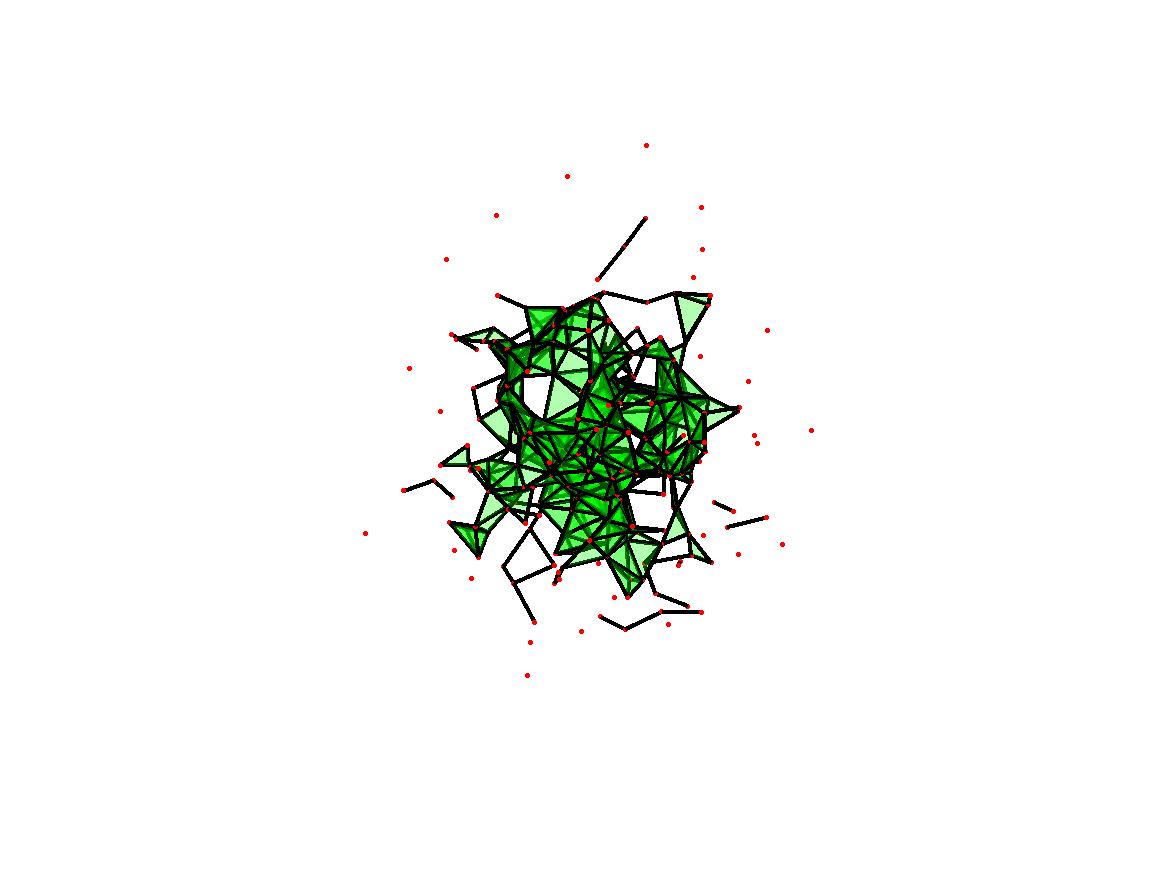} 
\includegraphics[trim=300 200 300 200,clip=true,height=3cm]{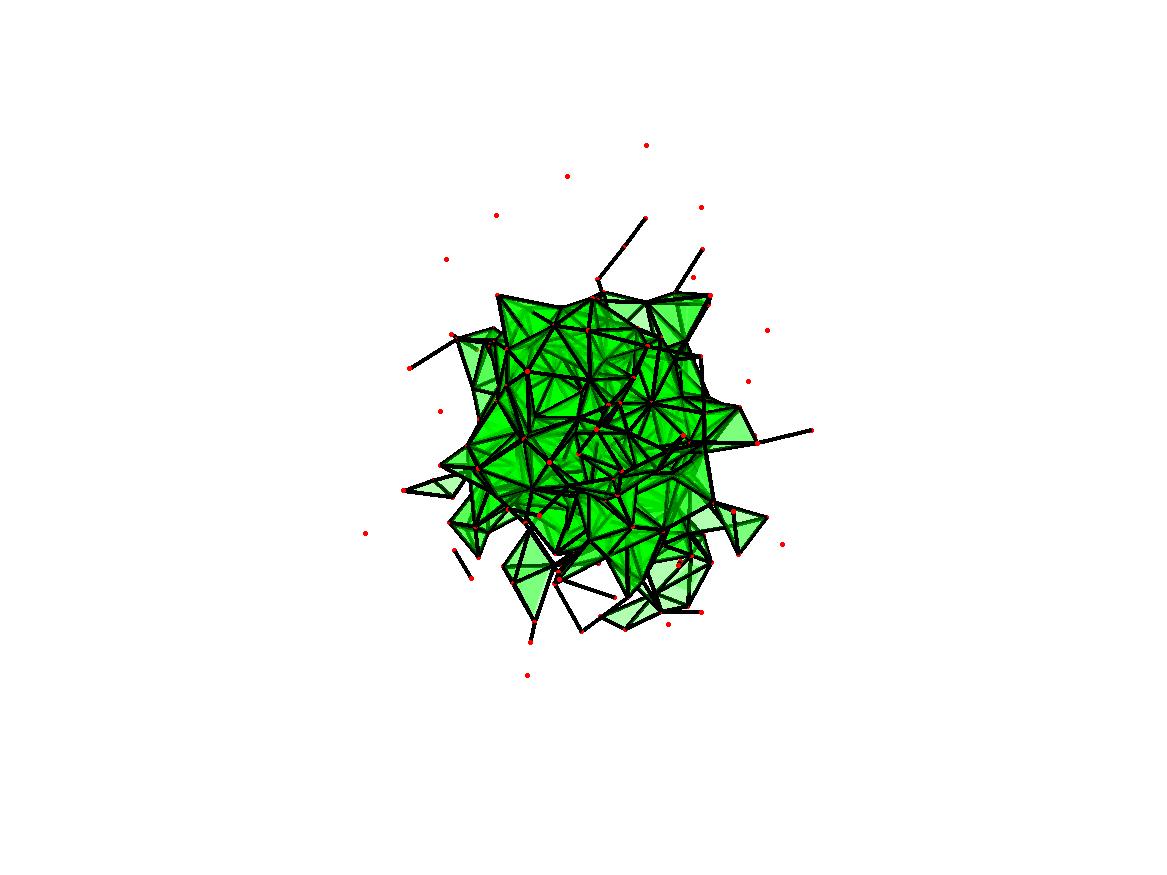}  
\includegraphics[trim=300 200 300 200,clip=true,height=3cm]{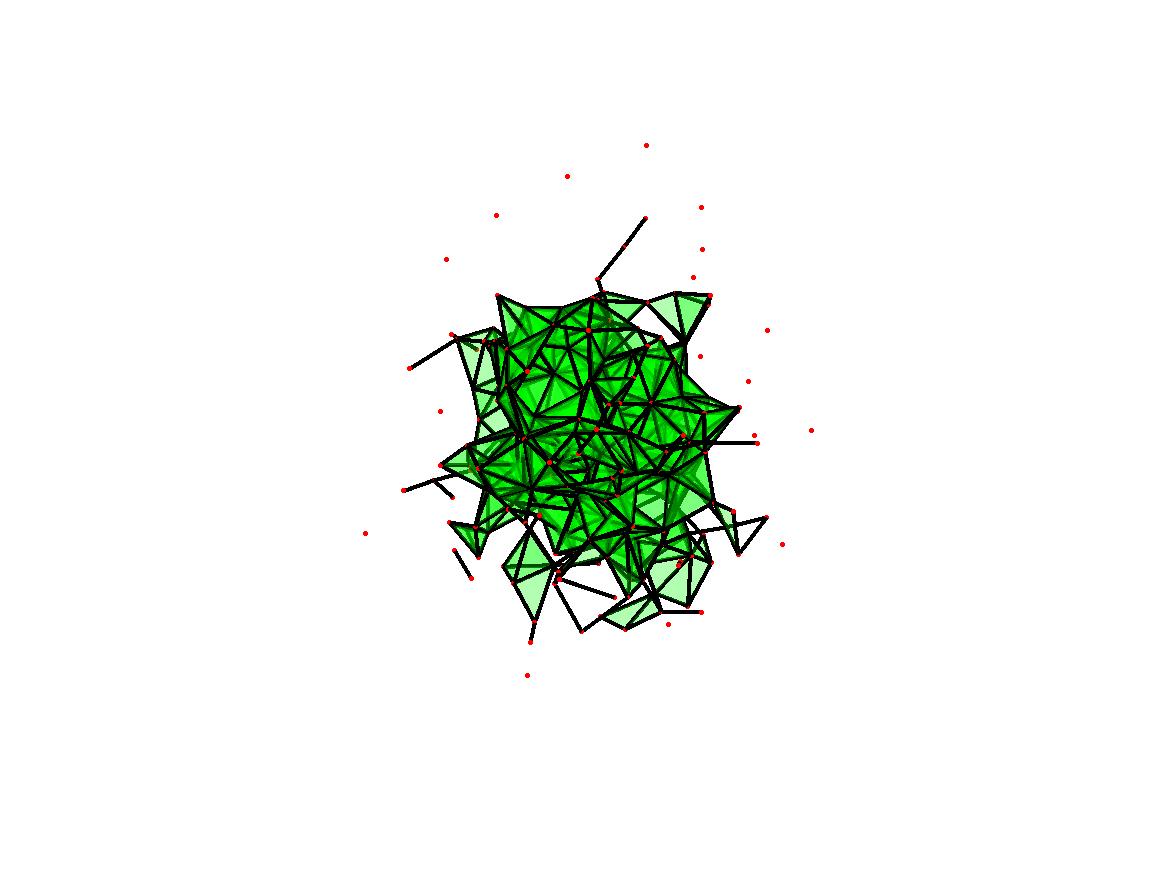}  
\includegraphics[trim=300 200 300 200,clip=true,height=3cm]{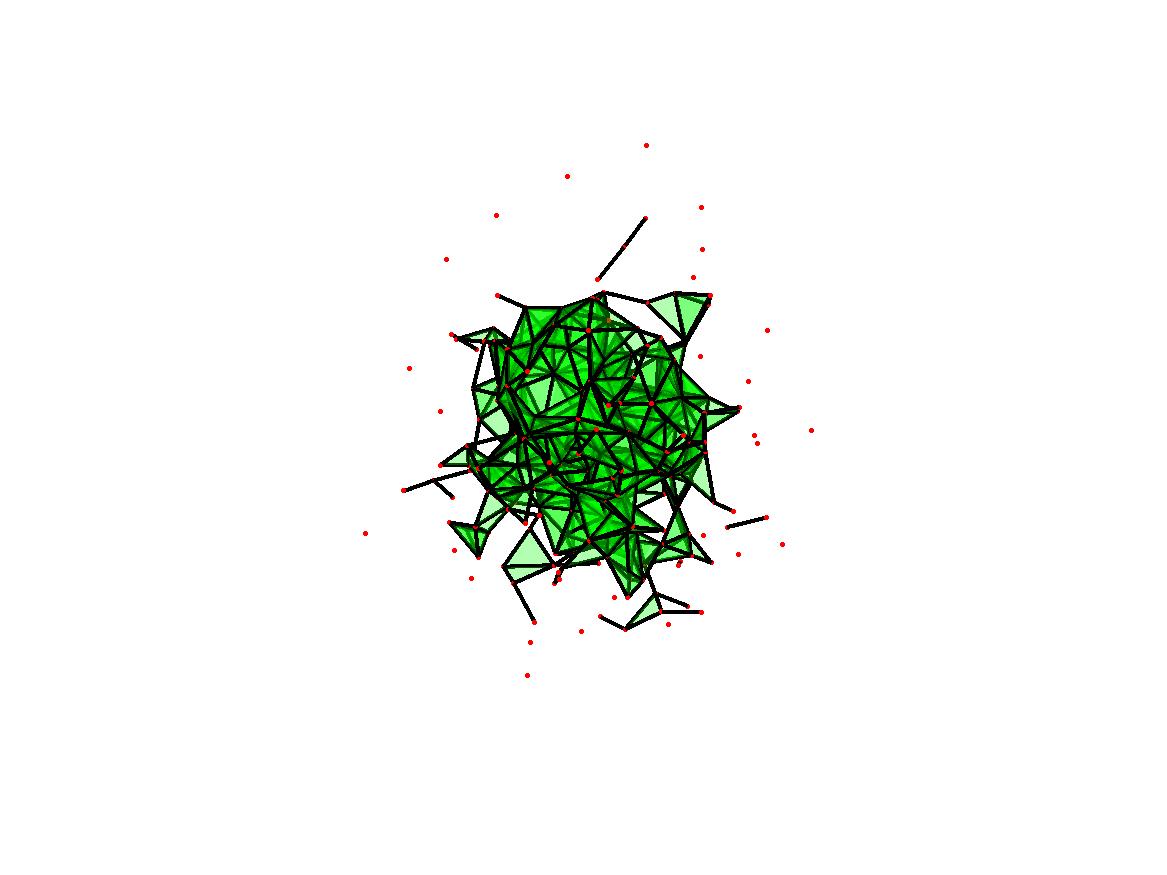}
\caption{Simplices for filtration values $r=\frac{r_M}{2}\tilde{r}$, $\tilde{r}=0.05, 0.06, 0.07, 0.08, 0.09, 0.1, 0.11, 0.12$ at a representative time of the swarm evolution. Here $r_M=150.22$ is the maximum distance between two points in the cloud, $\beta=0.001<\beta_c(300)=0.0075$. As $r$ increases, a single dominant cluster absorbs neighboring points and small components becoming a large `compact' component.}\label{fig13}
\end{figure}

\begin{figure}
\centering
\includegraphics[trim=300 200 300 200,clip=true,height=3cm]{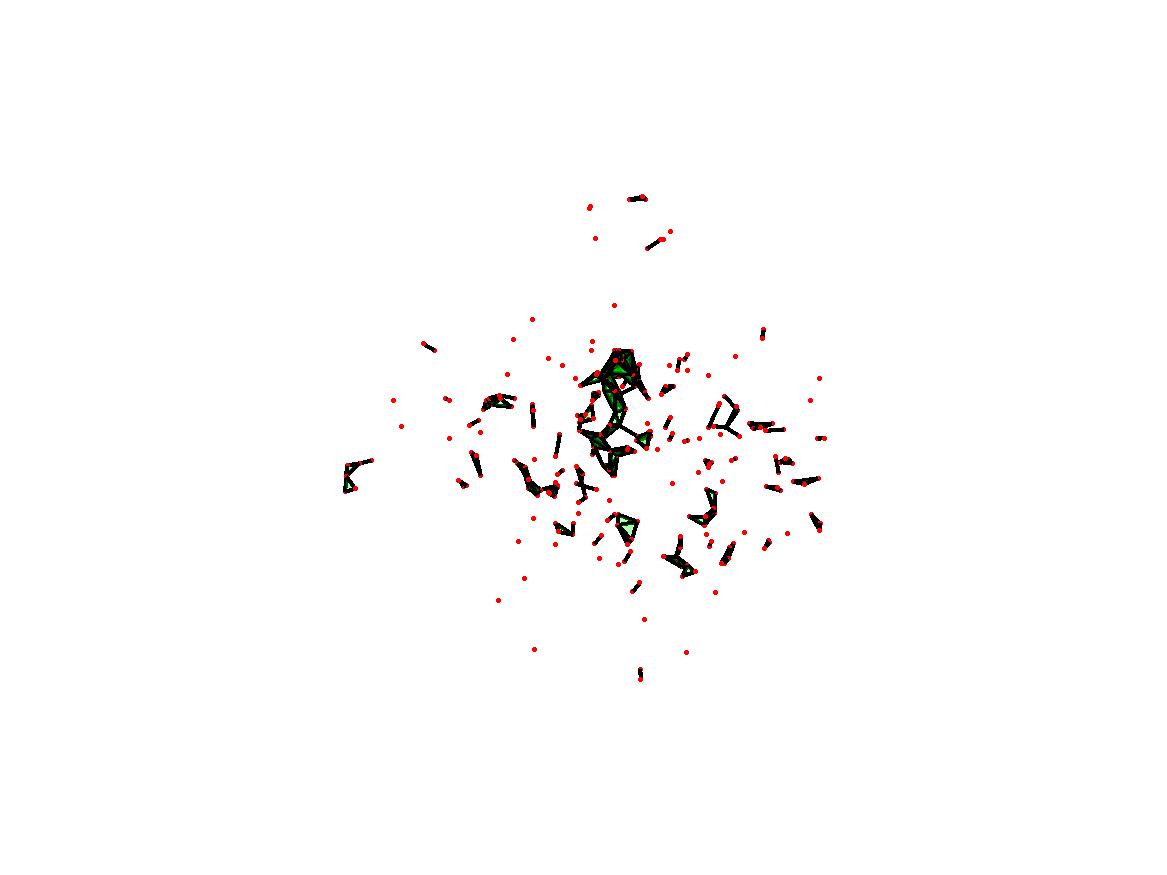} 
\includegraphics[trim=300 200 300 200,clip=true,height=3cm]{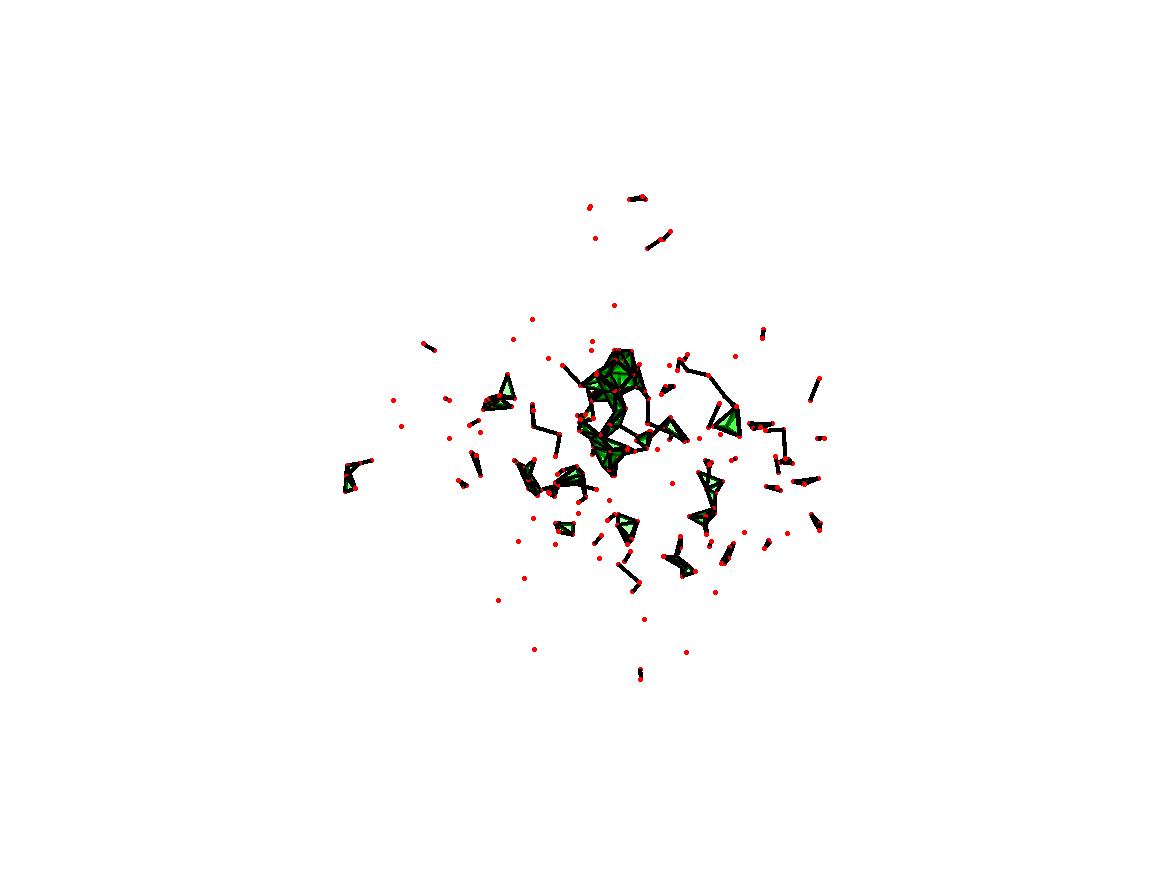}  
\includegraphics[trim=300 200 300 200,clip=true,height=3cm]{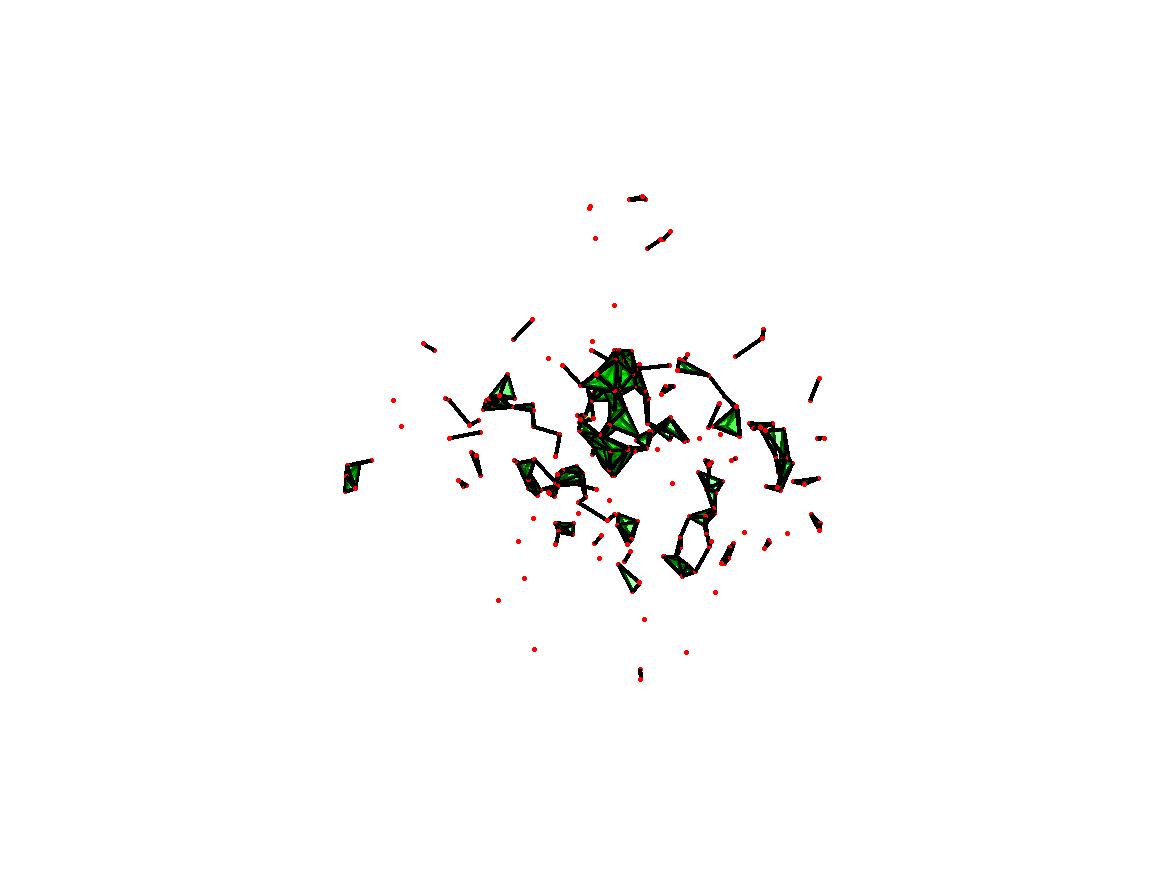} 
\includegraphics[trim=300 200 300 200,clip=true,height=3cm]{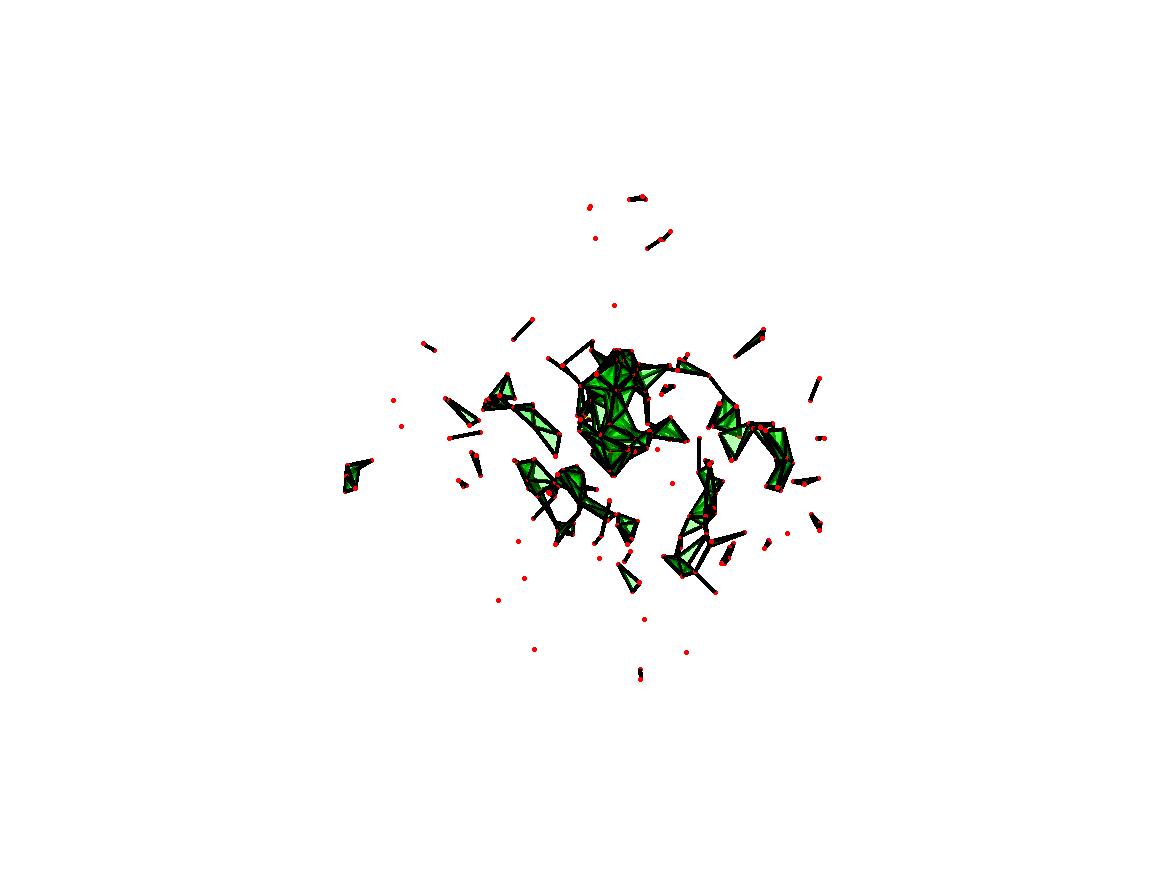}  
\includegraphics[trim=300 200 300 200,clip=true,height=3cm]{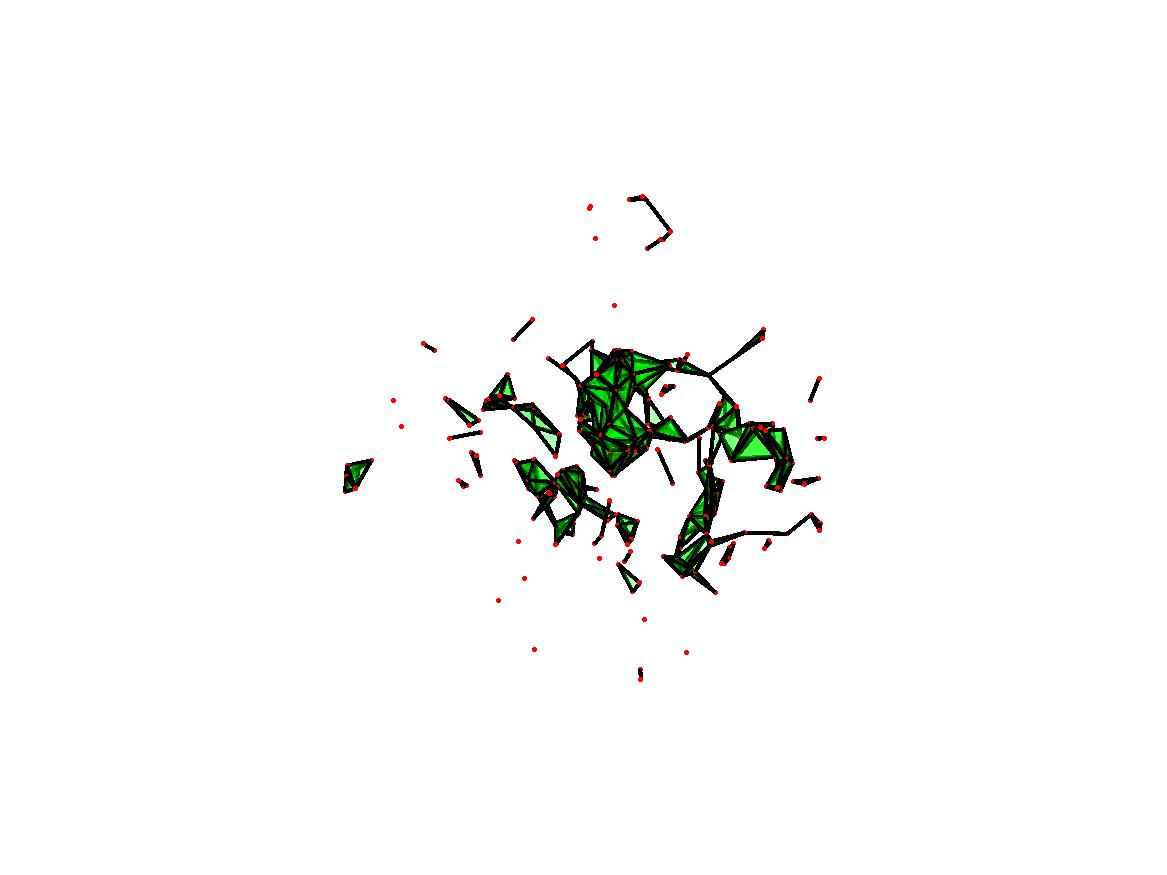} 
\includegraphics[trim=300 200 300 200,clip=true,height=3cm]{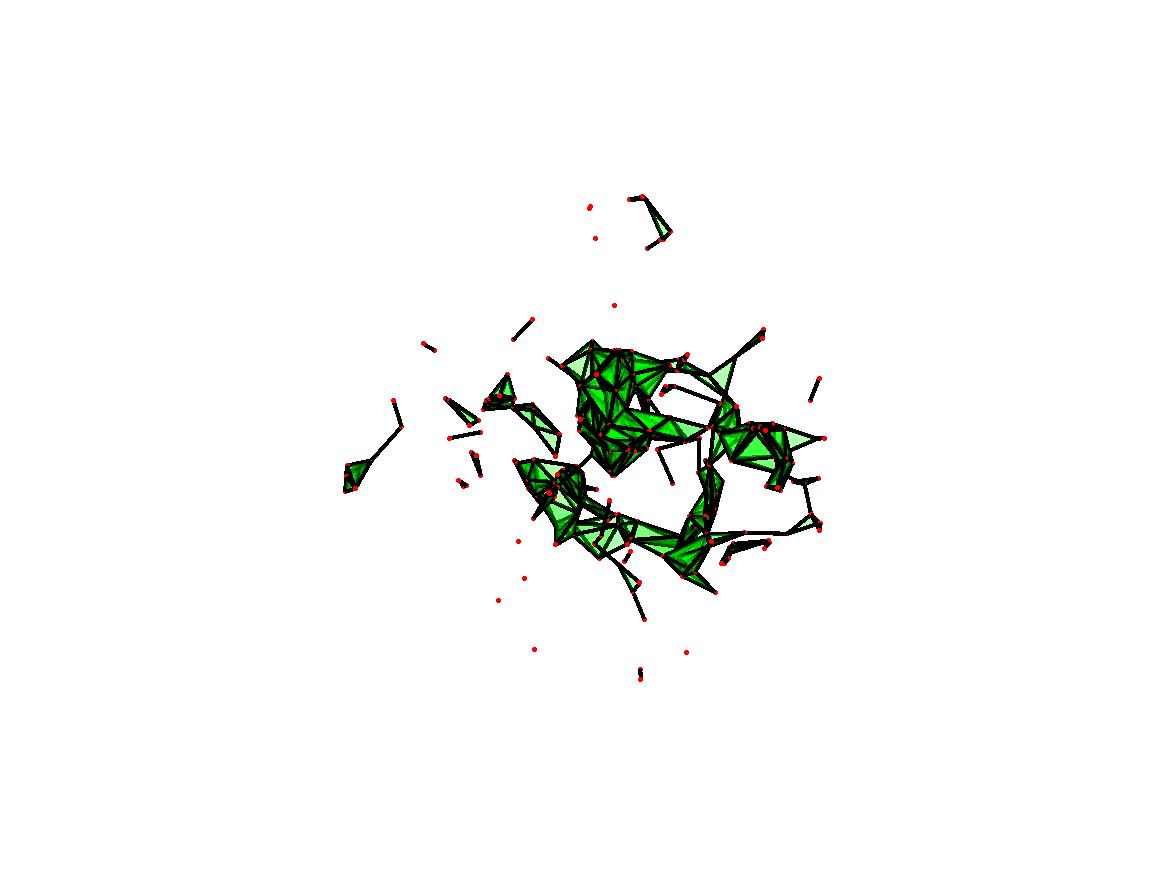}  
\includegraphics[trim=300 200 300 200,clip=true,height=3cm]{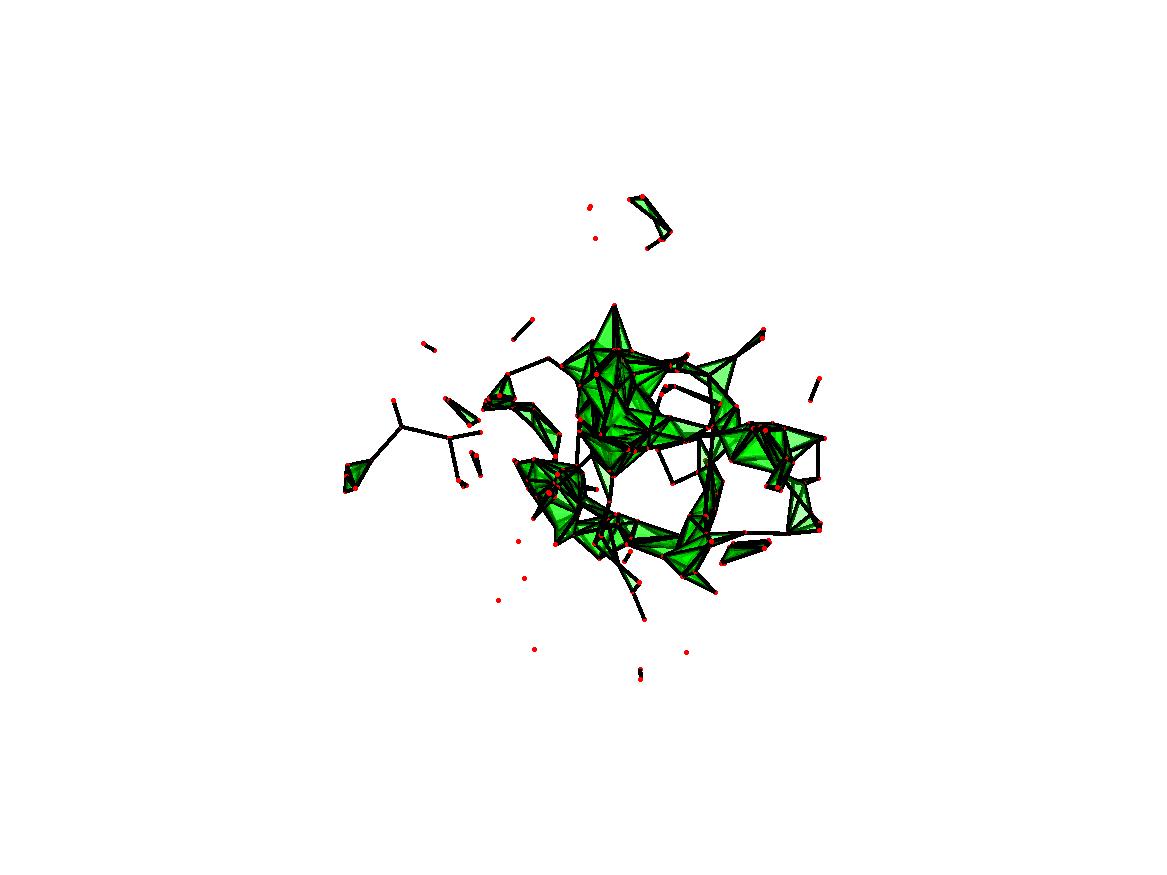}  
\includegraphics[trim=300 200 300 200,clip=true,height=3cm]{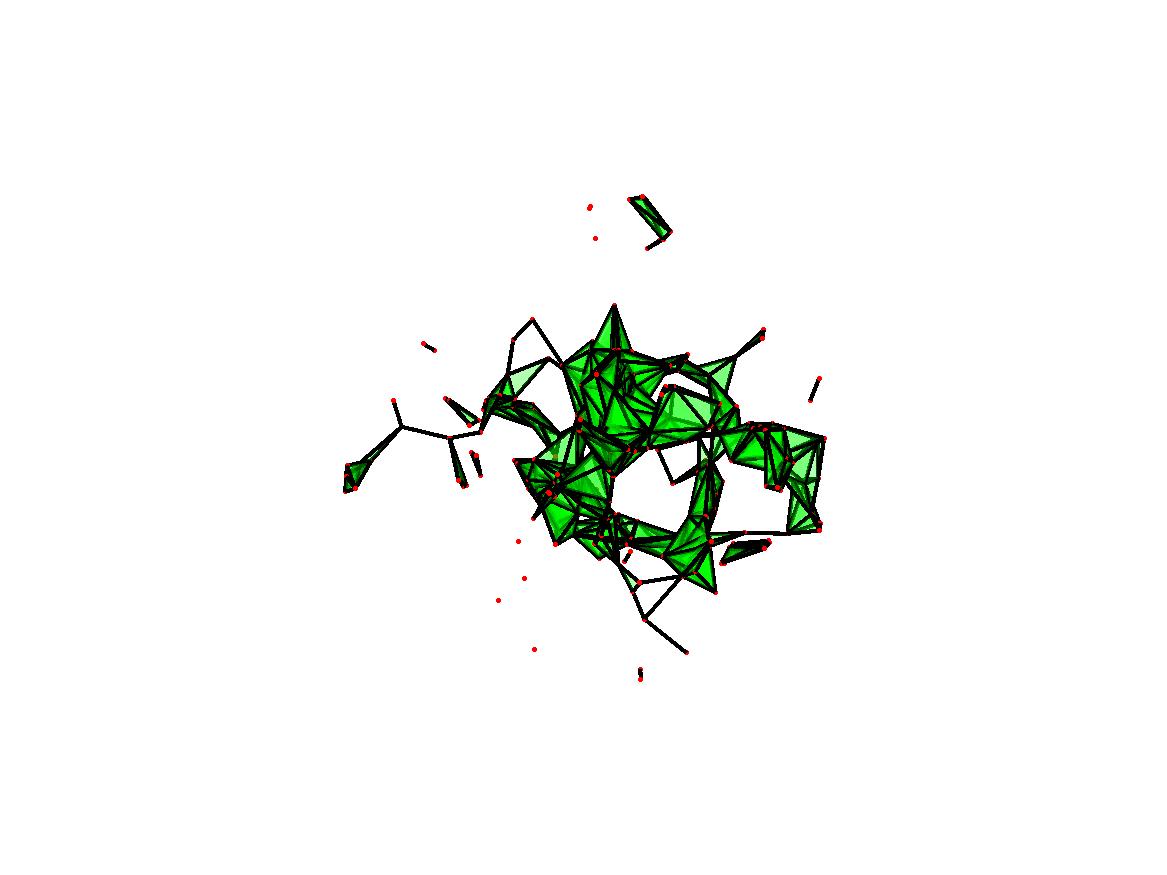}  
\caption{Same as Fig.~\ref{fig13} for $\beta=0.025>\beta_c(300)$ with $r_M=33.48$. As $r$ increases, small separated components form and eventually connect leaving large holes.} \label{fig14}
\end{figure}

We consider midges (or particles) as data points from a sampling of the underlying topological space of the swarm. Thus, we have a finite set of data points from a sampling of the underlying topological space. We measure data homology by creating connections between nearby data points, varying the scale over which these connections are made (as given by the {\em filtration parameter}), and  looking for features that persist across scales \cite{zom02,ede10}. This can be achieved  by building the {\em Vietoris-Rips complex} from all pairwise distances between points in the dataset. Assume spheres of diameter $r$ circle each particle. For each value of the filtration parameter $r>0$, we form a simplicial complex $S_r$ by finding all gatherings of $k+1$ points such that all pairwise distances between these points are smaller than $r$. Each such gathering is a $k$-simplex. The simplicial complex $S_r$ comprises finitely many simplices such that (i) every nonempty subset of a simplex in $S_r$ is also in $S_r$, and (ii) two $k$-simplices in $S_r$ are either disjoint or intersect in a lower dimensional simplex. In $S_r$, 0-simplices are the data points, 1-simplices are edges, connections between two data points, 2-simplices are triangles formed by joining 3 data points through their edges, 3-simplices are tetrahedra, and so on. See Figs.~\ref{fig13} and \ref{fig14}, which are the counterparts of Figs.~\ref{fig12}(a) and \ref{fig12}(c), respectively. These figures  illustrate how TDA automatically characterizes the formation of a loose single swarm for $\beta<\beta_c$ and of several tight smaller clusters for $\beta> \beta_c$. In the latter case, the single cluster resulting for sufficiently large filtration parameter contains large holes. 

\begin{center}
\begin{figure}[h]
\begin{center}
\includegraphics[trim={0.2cm 0.2cm 1.4cm 0.6cm},clip,width=4.25cm]{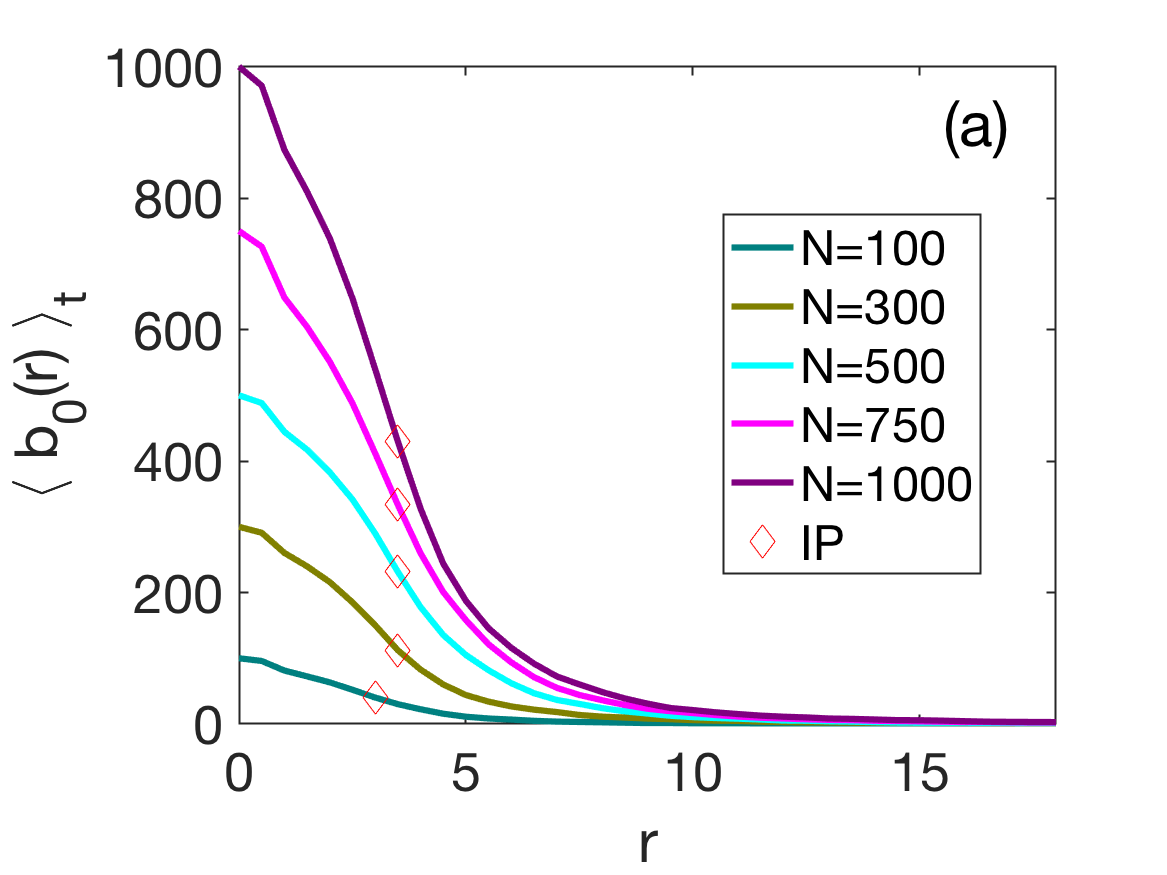}
\includegraphics[trim={0.2cm 0.2cm 1.4cm 0.6cm},clip,width=4.25cm]{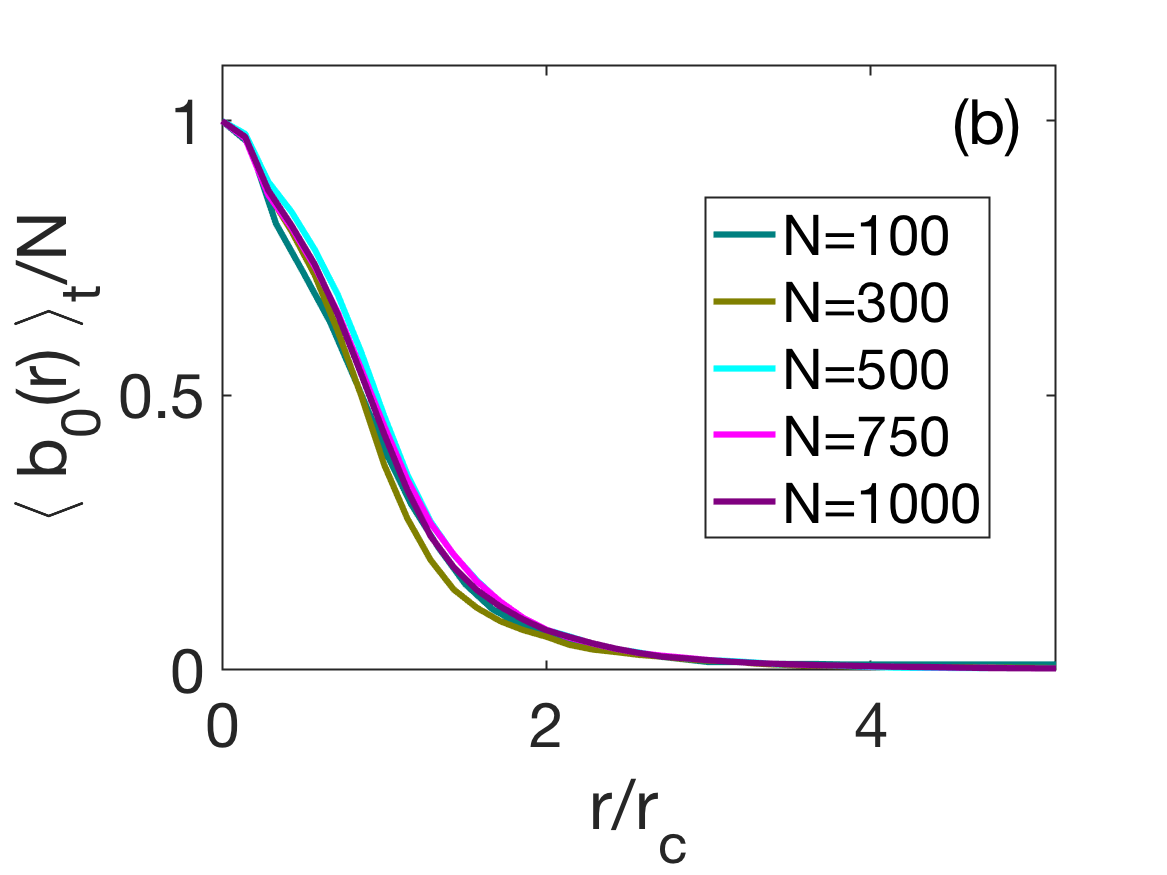}
\end{center}
\caption{{\bf (a)} Time averaged Betti number $\langle\mathsf{b}_0\rangle_t$ versus filtration parameter  $r$ for $\beta_c(N;\eta)$ and different $N$; {\bf (b)} Same for scaled averaged Betti number $\langle\mathsf{b}_0\rangle_t/N$ versus scaled filtration parameter  $r/r_c$ where $r_c(N)$ is the inflection point of each curve marked with diamonds in Panel (a). Here $\eta=0.5$.}  \label{fig15}
\end{figure}
\end{center}

\begin{center}
\begin{figure}[ht]
\begin{center}
\includegraphics[height=5.0cm]{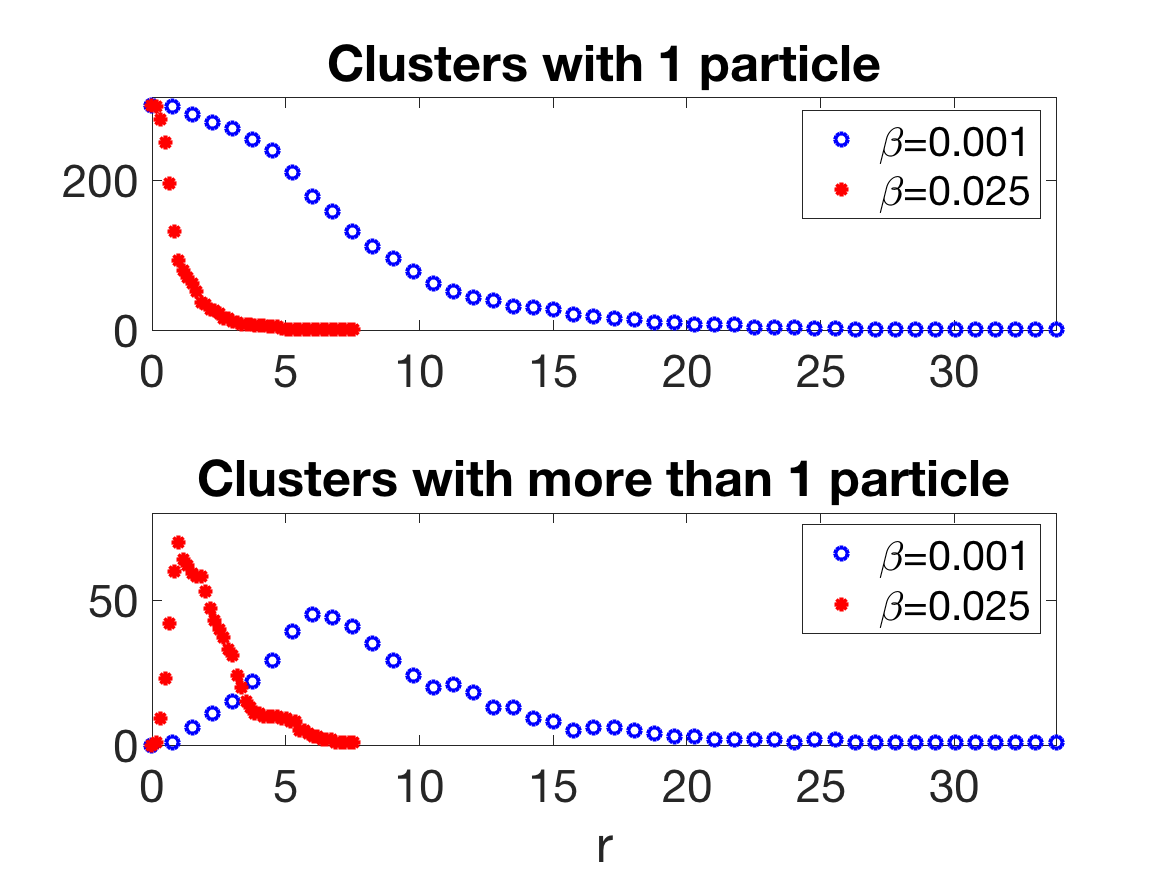} \\
{\bf (a)}\\
\includegraphics[height=5.0cm]{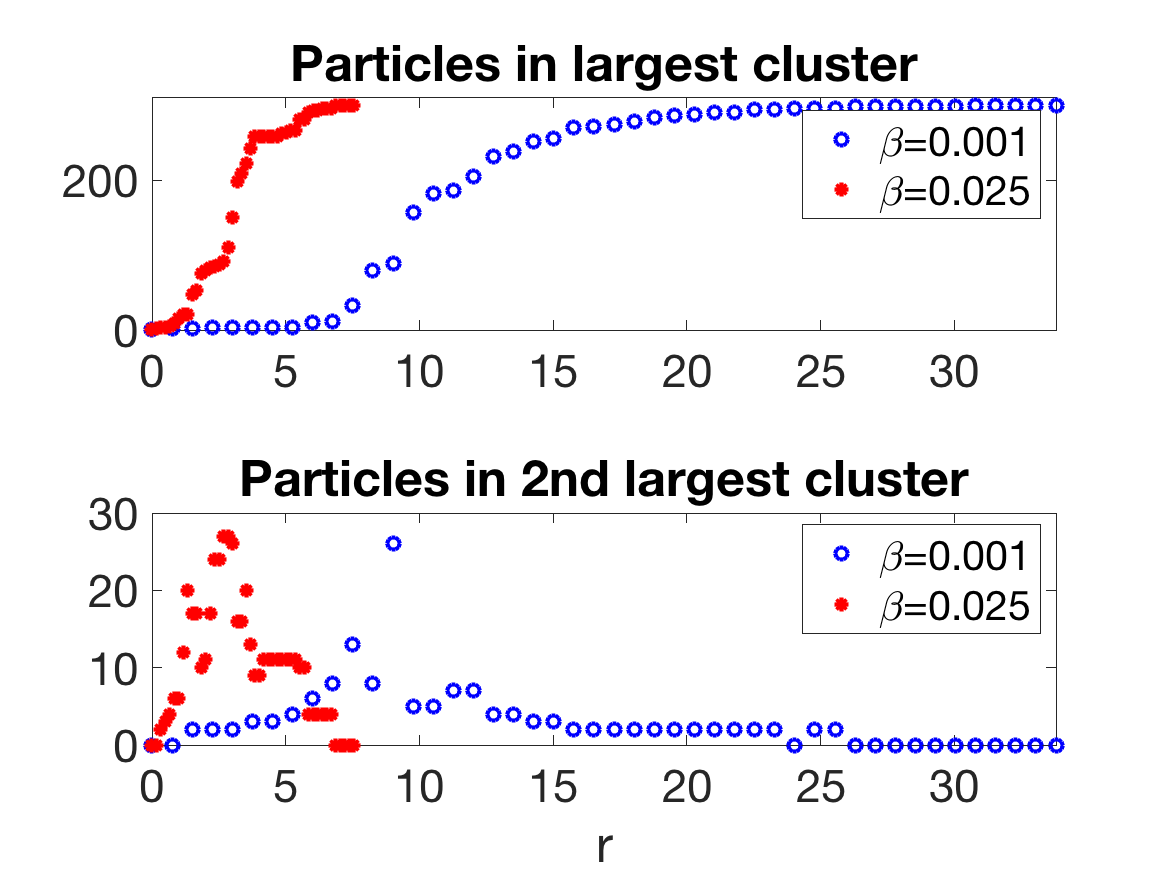} \\
{\bf (b)}\\
\includegraphics[height=5.0cm]{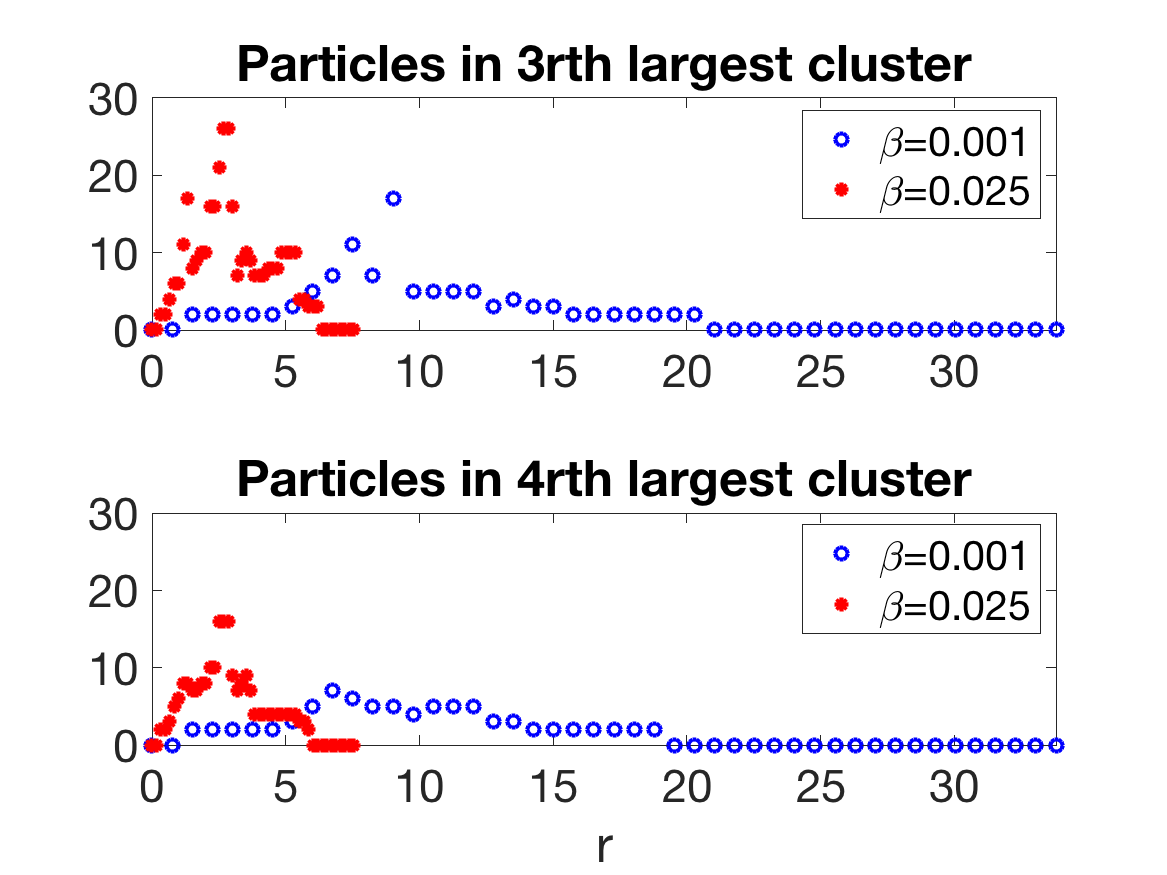}\\
{\bf (c)}
\end{center}
\caption{Hierarchical TDA clustering. {\bf (a)} Number of clusters with 1 particle (up) and with more than 1 particle (down) for $\beta=0.025$ (left red points) and $\beta=0.001$ (right blue points) vs filtration parameter $r$ at a single time. {\bf (b)} Number of particles in (up) the largest cluster and (down) the second largest cluster vs $r$. {\bf (c)} Number of particles in (up) the third largest cluster and (down) the fourth largest cluster vs $r$. Here $N=300$ and $\eta=0.5$. \label{fig16} }
\end{figure}
\end{center}

To quantify the topological structure of the swarm data points, the Betti numbers depicted in Fig.~\ref{fig15} are useful. Within the set of all $k$-simplices in $S_r$, we can distinguish closed submanifolds called $k$-\emph{cycles}, and cycles called \emph{boundaries} because they are also the boundary of a submanifold. A \emph{homology class} is an equivalence class of cycles modulo boundaries. A homology class $\mathsf{H}_k$ is the set of independent topological holes of dimension $k$, represented by cycles which are not the boundary of any submanifold. The dimension of $\mathsf{H}_k$ is the $k$th {\em Betti number} $\mathsf{b}_k$. For instance, $\mathsf{b}_0$ is the number of connected components shown in Fig.~\ref{fig15}, $\mathsf{b}_1$ is the number of topological circles, $\mathsf{b}_2$ is the number of trapped volumes, and so on. See Refs.~\onlinecite{zom02,ede10} for precise definitions. At critical confinement, we can depict average Betti numbers, $\langle \mathsf{b}_0\rangle_t$ (number of connected clusters averaged over several time snapshots of the swarm), versus $r$ for different $N$. These Betti numbers collapse when we rescale them using the inflection point of each curve \cite{sin17}, $r_c(N)$; see Figs.~\ref{fig15}(a) and \ref{fig15}(b).

 Fig.~\ref{fig16} illustrates the trend to a more compact single swarm and to swarm splitting as $\beta$ increases past its critical value. As $r$ increases, the number of clusters with a single particle decrease monotonically, as seen in the upper panel of Fig.~\ref{fig16}(a). However, the upper panel of Fig.~\ref{fig16}(b) shows that the number of particles in the largest cluster increase monotonically for $\beta<\beta_c$ but it increases with plateaus and abrupt jumps for $\beta>\beta_c$. These abrupt features indicate that the largest cluster absorbs single particles as $r$ increases if $\beta<\beta_c$, whereas several large clusters form and are abruptly absorbed by the largest cluster at particular values of $r$ for $\beta>\beta_c$. The lower panels in Figs.~\ref{fig16}(a)-\ref{fig16}(b) confirm these observations. Clusters with more than one particle form gradually if $\beta<\beta_c$ and abruptly if $\beta>\beta_c$. The plateaus and jumps in the number of particles within the second, third and fourth largest clusters in Figs.~\ref{fig16}(b) and \ref{fig16}(c) indicate absorptions thereof by the largest cluster.  These figures also illustrate the different cluster structure below and above the critical confinement $\beta_c$. When $\beta>\beta_c$, we observe the presence of several relevant clusters with a large number of particles. These clusters persist as the filtration parameter increases. Note that it is possible to have more than one cluster with the same number of particles. 
\bigskip

\section{Discussion and conclusions}\label{sec:7}
{\em Purpose.} Here we discuss a hitherto unsuspected and unexplored phase transition of free scale chaos in the harmonically confined 3D Vicsek model. The same model exhibits a different phase transition to clusters of finite size containing infinitely many particles. This work is motivated by observations of natural midge swarms, which comprise at most hundreds of insects and form about a marker \cite{att14,att14plos,cav17,kel13,gor16,sin17}. The validation of the scale-free-chaos scenario by experimental data is outside the scope of this paper.

{\em Experiments and methodology.} Cavagna {\em et al}'s observations unveiled finite size scaling and power laws in swarms of male midges. They adapted definitions from statistical physics to define correlation functions, correlation lengths and calculate critical exponents from data \cite{att14,att14plos,cav17}. To interpret data, they used the `gas' phase of the 3D VM confined in a finite box with periodic boundary conditions and ideas about universality. Contrastingly, midge swarms in an enclosure form a `condensed' nucleus far from enclosure walls surrounded by a `vapor' of insects that exit from, return to, and hover about the nucleus \cite{sin17}; see Fig.~\ref{fig12}(b) for a similar configuration of the scale-free-chaos phase in our simulations. While this is different from an gas filling a box with uniform density, definitions from correlation functions, finite size and dynamical scaling apply to both numerical simulations of the model and experimental data \cite{att14,att14plos,cav17}. We apply the methodology based on correlation functions to our simulations of the confined VM to unveil the scale-free-chaos phase transition.

{\em Dynamical systems tools.} As the confinement strength decreases, the VM with fixed number of particles $N$ displays a variety of  periodic, quasiperiodic and chaotic attractors, which may be strongly modified by alignment noise. To distinguish chaos, we have calculated the largest Lyapunov exponent directly from the VM using the Benettin algorithm \cite{ben80}. This is particularly well adapted to the discrete time dynamics of the VM. We have also calculated the LLE by reconstructing the attractor from time traces of the center of mass motion using lagged coordinates. Using only two lagged coordinates, scale-dependent Lyapunov exponents help distinguishing deterministic and noisy chaos from parameter regions where noise is dominant \cite{gao06}. This is important because the phase transitions exist within the noisy chaos region. While scale-dependent Lyapunov exponents give qualitative information, we need six lagged coordinates to faithfully reconstruct the chaotic attractor and obtain (by the Gao-Zheng algorithm \cite{gao94}) the same LLEs as provided by the Benettin algorithm. This methodology will be important to ascertain whether a real system in nature exhibits scale-free-chaos phase transitions.

{\em Statistical physics tools.} It is instructive to compare the scale-free-chaos phase transitions in the confined VM to the canonical para-ferromagnetic equilibrium phase transition whose universality class comprises the Ising  and $\phi^4$ models. A phase is an ergodic measure that exists in the thermodynamic limit (infinite volume and number of particles, finite density) and a phase transition corresponds to a discontinuous change from one to more than one phase as a parameter changes, i.e., to a bifurcation of the measure; see precise definitions and proofs in Ref.~\onlinecite{gli87}. Pure phases have different values of the magnetization order parameter. At the critical point that ends a line of first order phase transitions at zero external field, the correlation length becomes infinity in the thermodynamic limit \cite{gli87}. The magnetization order parameter undergoes a pitchfork bifurcation at the critical temperature with critical exponent 0.327 instead of 0.5 \cite{ami05,yeo92}.

The main objects to characterize critical points of second-order equilibrium phase transitions are static and dynamic correlation functions. To study flocking and other nonequilibrium phase transitions, we need to adapt the definitions of correlation functions, correlation length, susceptibility, and so on, to models such as Vicsek's. Averages over the number of particles, time averages and averages over realizations replace the ensemble averages of equilibrium statistical mechanics \cite{cav18}. Since it is important that correlation functions reflect underlying dynamic attractors, velocity fluctuations are about center of mass velocities (which may be chaotic); see Ref.~\onlinecite{cav18} for extended discussion. Subtracting an overall rotation and/or dilation at each time step \cite{att14plos,att14,cav17} does not change the critical lines $\beta_c(N;\eta)$ and $\beta_i(N;\eta)$ but the local maxima of the susceptibility versus $\beta$ curve disappear. We still have a critical line separating single from multicluster chaos followed by a narrow criticality region, both of which tend at the same rate to zero confinement as $N\to\infty$ and therefore represent the same phase transition; see Appendix \ref{ap:c}.

The chaotic phases in scale-free-chaos transitions are ergodic \cite{ott93,cen10}. The transitions are second order: as $N\gg 1$, the order parameter is close to zero in the sparse single-cluster chaotic phase and the polarization is positive in the multicluster chaotic phase. Let us discuss now the different critical lines at finite $N$ that characterize the scale-free-chaos phase transition at $N=\infty$. 

As discussed in Section \ref{sec:4}, it would be ideal if we had a relation between the poles of the susceptibility and the reciprocal correlation time, as it happens in simple models. Then vanishing of the pole would be the same as the correlation time going to infinity (critical slowing down) and this would locate the critical point. In the absence of such a relation, we have first used the correlation time that solves Eq.~\eqref{eq5} for $k_c=1/\xi$ as a reasonable substitute. The critical line $\beta_c(N,\eta)$ is the value of $\beta$ that minimizes $\tau_{k_c}$ for fixed $N$ and noise $\eta$.  Equivalently, it is the maximum value of the continuous extension of a correlation time defined as the first zero of the NDCCF. We have obtained a dynamic scaling exponent $z= 1.01\pm 0.01$ and critical exponents $\nu=0.436\pm 0.009$, $\gamma=0.92\pm 0.05$, with $\varphi\approx z\nu$ (critical exponent corresponding to the decay of the largest Lyapunov exponent). For fixed $N$, $\beta_c(N,\eta)$ is a line on the plane $(\beta,\eta)$ within the region of noisy chaos in Fig.~\ref{fig3}(a). We have checked that the correlation length is proportional to the size of the swarm for all the simulated values of $N$, and therefore the system is scale free on this critical line. The critical line is inside an interval of confinement values for which the NDCCF is flat and relaxation dynamics is underdamped. Outside this interval, the confined VM exhibits overdamped dynamics. By using topological data analysis, we lend support to the  numerical observation that chaos is single-cluster below this critical line and multicluster above it. The phase of single-cluster chaos has the smallest polarization order parameter and is therefore the most symmetric. Multicluster chaos has larger polar order. As $N\to\infty$, $\beta_c(N,\eta)$ tends to 0 and so does the LLE on that line: chaos disappears, as required by the correlation length becoming infinity and Eq.~\eqref{eq2} for finite velocity of propagation. Further study involving the invariant measure of the chaotic attractors would be desirable to explore analogies with the phase ergodic measures of equilibrium thermodynamics.

Using the susceptibility of the real-space static correlation function and finite size scaling, we have found other lines $\beta_i(N;\eta)$ and $\beta_m(N;\eta)$, with $\beta_c<\beta_i<\beta_m$, that go to zero at the same rate as $\beta_c(N;\eta)$ for $N\gg 1$, cf Fig.~\ref{fig8}(d). Thus, they represent the same phase transition and produce the same critical exponents as $N\to\infty$. For the $N$ values in our simulations, we have checked that the correlation length is proportional to swarm size (therefore they are scale free) and $\nu$ is the same. The chaotic attractors are multicluster on the lines $\beta_i(N;\eta)$ and $\beta_m(N;\eta)$ based on the inflection point and the local maximum of the susceptibility, respectively; see Fig.~\ref{fig12}. This indicates that the swarm center of mass  experiences more important rotation and dilation effects than on the single-to-multicluster line $\beta_c(N;\eta)$. 

The critical exponent $z$ is different on the three critical lines, which may simply point to the multiple time and length scale involved in the mulifractal chaotic attractors of the phase transition, cf Fig.~\ref{fig6}. That different time scales are involved in the same transition is a common occurrence in codimension two bifurcations of dynamical systems \cite{kuz04}; see e.g., the scaled normal form in Ref.~\onlinecite{dan87}. The mean field version of the standard two-dimensional Vicsek model with periodic boundary conditions also involves two time scales near the order-disorder transition. The mean field VM (in a box with periodic boundary conditions) can be described by a discrete-time Enskog-type kinetic equation which preserves the overall number of particles \cite{ihl11}. The order-disorder phase transition appears as a supercritical bifurcation of the kinetic equation when one multiplier crosses the unit circle in the complex plane; another multiplier corresponding to particle conservation is always one \cite{BT18}. On the ordered side, the scaled bifurcation equations contain two time scales, one with $z=1$ (hyperbolic scaling), the other with $z=2$ (parabolic scaling). At the hyperbolic short time scale, undamped wave propagation and resonance phenomena arise \cite{BT18}, whereas different band patterns appear at the parabolic time scale further from the bifurcation point \cite{tre22}. These patterns exist on the ordered side of the ordering phase transition. They can be found in direct simulations of the standard Vicsek model and include bands \cite{sol15} and crossbands \cite{kur20}.

In Section \ref{sec:5} and Appendix \ref{ap:c}, we show that subtracting overall rotation and dilation from CM motion in velocity fluctuations does not change the lines $\beta_c(N;\eta)$ and $\beta_i(N;\eta)$ but the line of local maxima $\beta_m(N;\eta)$ disappears. The line of global maxima, $\beta_M(N;\eta)$, appears now at the end of the first chaotic window. On $\beta_M(N;\eta)$, the correlation length is finite and does not change appreciably with $N$, the average minimal distance between particles tends to zero and the density inside these clusters of finite extension tends to infinity. This transition has $\beta_M(N;\eta)\to\infty$ as $N\to\infty$ and therefore it is no longer scale free. Instead, it is analogous to type II gravitational collapse \cite{gun07}, and it has its own critical exponents, cf Eq.~\eqref{eq14}. In conclusion, subtracting rotation and dilation from CM motion leaves only two critical lines where the system is scale free, namely, $\beta_c(N;\eta)$ and $\beta_i(N;\eta)$. Only these two critical lines need to be taken into consideration when describing the phase transition based on subtracting overall translation, rotation and dilation from particle velocities to define velocity fluctuations. These two lines illustrate the existence of a narrow criticality region following $\beta_c(N;\eta)$ that also collapses to $\beta=0$ as $N\to\infty$. Numerical simulations produce the same critical exponent $\nu$ as obtained without rotation and dilation but $\gamma$ changes as explained in Appendix \ref{ap:c}. 

{\em Critical exponents from experiments and theory.} In observations of natural swarms, the measured critical exponents are $\nu= 0.35\pm 0.10$, $\gamma=0.9\pm 0.2$ (Ref.~\onlinecite{att14plos,att14}), and $z=1.12\pm 0.16$ (Ref.~\onlinecite{cav17}), while the real-space susceptibility is between 0.32 and 5.57 for the measured swarms \cite{att14plos,att14}. More recent observations give an interval $0.93\leq z\leq 1.42$ of possible values of the dynamical exponent based on a resampling procedure; see Fig. 3 of Ref.~\onlinecite{cav21arxiv}. 

Here we have discussed the scale-free-chaos phase transition of the harmonically confined Vicsek model. For each adequate noise value within the interval of noisy chaos, cf Fig.~\ref{fig3}(a), three critical lines coalesce at the same rate to $\beta=0$ as $N\to \infty$. Thus, they represent the same phase transition. For $\beta_c(N,0.5)$, we have found $\nu=0.436\pm 0.009$ (correlation length), $\gamma=0.92\pm 0.05$ (real-space susceptibility), and $z= 1.01 \pm 0.01$ (dynamic exponent). The critical exponent for the LLE law is approximately $\varphi=z \nu$. These critical exponents change little for $0.1<\eta<1$ and are reasonably near experimentally measured ones. 

{\em Qualitative features.} In addition to reasonable critical exponents, the scale-free-chaos phase transition produces disperse chaotic swarms below $\beta_c(N,\eta)$ that are confined to a bounded region of space with a few particles entering and leaving the nucleus of the swarm, cf. Fig.~\ref{fig12}. This is akin to the observed condensed and vapor phases of natural swarms \cite{sin17}. Furthermore, as shown in Section \ref{sec:4}, the normalized dynamic correlation function coalesce to a single curve as a function of $k_c^zt$ for an interval $0<k_c^zt<4$ (cf. Fig. \ref{fig4}), which is similar to that observed in natural swarms (Fig. 2b of Ref.~\onlinecite{cav17}). Moreover, the flatness values given by Eq.~\eqref{eq14} are compatible with those observed in natural swarms \cite{cav17}. These similitudes to experimental observations and the involved theoretical challenges make worthwhile exploring more fully the confined Vicsek model and the phase transition we have discovered.  

{\em Critical exponents from models in the literature.} The ordering transition of the VM confined in a finite box with periodic boundary conditions has received much attention; see e.g, the reviews in Refs.~\onlinecite{cha19,cha20}. Near this transition, the particles form a gas and are distributed in the box with almost constant density \cite{cha19,cha20}. This contrasts with observations of natural swarms in an enclosure where most of the swarm is far from walls (condensed phase) and individual insects hover around the swarm nucleus, enter and exit from it \cite{sin17}. It is fair to say that the single-cluster chaotic phase of the confined VM resembles observations better than the ordering transition of the standard VM. Calculated critical exponents near the ordering transition of the VM in a box with periodic boundary conditions are also further away from observations:  $\gamma=1.6\pm 0.1$, $\nu= 0.75\pm 0.02$ (Ref.~\onlinecite{att14} for noise $\eta=0.45\times 4\pi= 5.65$ in our units), $z=2$ (Ref.~\onlinecite{cav17}). 

Many theoretical works study hydrodynamic equations with white noise forcing terms near a critical point which resembles that of the ordering transition of the standard VM. The idea is that all such descriptions could be analyzed using renormalization group (RG) theory and produce critical exponents compatible with experimental observations. This would then show that the appropriate hydrodynamic-type description belongs to the same universality class as the real natural swarms. These RG theories are based on weakly nonlinear expansions about a simple symmetric state. Chen {\em et al} study incompressible Toner-Tu hydrodynamic equations \cite{ton95,ton05} using RG about a unidirectional velocity and produce an exponent $z=1.72$ in 3D \cite{che15,che18}. See also Ref.~\onlinecite{cav21prr} for numerical confirmation. Cavagna {\em et al} consider incompressible Toner-Tu hydrodynamics coupled to underdamped soft spin equations under white noise forces \cite{cav21arxiv}. They study weakly nonlinear expansions about linear stochastic differential equations with constant coefficients and additive noise to obtain RG equations and calculate $z=1.3$. These values are within the range of experimental observations \cite{cav21arxiv}. 

Recently, Holubec {\em et al} have studied the VM with time delay and periodic boundary conditions. They found $\gamma\approx 1.53$, $\nu\approx 0.75$ (larger than measured in midges) and $z\approx 1$ for very long delay times using an undersampled NDCCF \cite{hol21}. Their NDCCF exhibits regular oscillations as the time delay increases, which are interpreted using a time-delayed reaction-diffusion equation (see Supplementary Information in Ref.~\onlinecite{hol21}). It is not clear whether there is a single phase transition responsible for these results. In time-delayed ordinary differential equations, oscillations often appear as Hopf bifurcations at critical delays \cite{hal77} and may evolve to relaxation oscillations as delays increase \cite{bon84}. Delayed reaction-diffusion equations can have stable relaxation-type wavetrain solutions that depend on the variable $(x+ct)$, cf Ref.~\onlinecite{bon84}. This would give a dynamic exponent $z=1$. Further study of the time-delayed VM may shed light on these connections.

{\em A universality class} comprises all physical systems that evolve to the same fixed point of the renormalization group equations under a rescaling of space and time and therefore have the same critical exponents \cite{hua87}. We have discovered a scale-free-chaos phase transition in the discrete time Vicsek model confined by a harmonic potential, which has qualitative features of natural swarms, underdamped dynamics, and compatible critical exponents. At moderate $N$, this transition is different from the well-known period-doubling, intermittency and quasiperiodic routes to chaos  (which have RGs based on maps \cite{sch05,ott93,cen10}) and from the ordering transition of the discrete time Vicsek model confined by a box with periodic boundary conditions \cite{cha19,cha20}. The scale-free-chaos phase transition encompasses phenomena at different time scales, from dynamical exponent $z\approx 1$ to larger $z$ for $\beta_i$ and $\beta_m$, which might require additional theoretical tools to understand. While there are RG calculations about Hopf bifurcations to stable oscillatory states \cite{bon88,ris05}, it would be desirable to have RG calculations about a single-cluster chaotic attractor, instead of the ordering transition of the standard VM (or related simple states of other models). Would it be possible to derive effective equations near the scale-free-chaos phase transition playing roles similar to amplitude equations in bifurcation theory \cite{bon88}? Could these effective equations exhibit new instabilities and coexistence of stable solutions and spinodal lines akin to those found in the Vicsek model with periodic boundary conditions \cite{cha19,cha20}? Time will tell.

Summarizing, we have numerically simulated the harmonically confined Vicsek model, which is an idealized description of insect swarms. Depending on confinement strength $\beta$ and noise $\eta$, the model exhibits different periodic, quasiperiodic and chaotic attractors. Our results support the existence of a line of phase transitions in a noisy chaos region of $\eta$ values as the number of particles $N$ tends to infinity and $\beta\to 0$. For finite $N$, there is a line $\beta_c(N;\eta)$ at which the correlation time is minimal and the correlation length is proportional to the system size. Topological data analysis supports the interpretation of $\beta_c(N;\eta)$ as a line separating single from multicluster chaos. The time averaged polarization acts as an order parameter: near $\beta_c(N;\eta)$, it is almost zero for $\beta<\beta_c(N;\eta)$ and positive and increasing with $\beta$ for $\beta>\beta_c(N;\eta)$. On the line of scale free chaos, the dynamic critical exponent is $z\approx 1$ and the dynamic correlation function collapses on an interval of the same length as in measured swarms. Close to the critical line $\beta_c(N;\eta)$ and for fixed $N$ and $\eta$, there are other critical lines obtained from the inflection point and local maximum of the susceptibility versus confinement curve. As $N\to\infty$, $\beta_c(N;\eta)\approx 0.48\beta_i(N;\eta)$ and $\beta_c(N;\eta)\approx 0.37\beta_m(N;\eta)$. Thus, the three lines represent a narrow criticality region, collapse at the same rate to $\beta=0$ as $N\to\infty$ and stand for the same  phase transition. Different exponent $z$ on the lines may reflect the multiplicity of time and length scales involved in the chaotic attractors. The particle swarms at the scale-free-chaos phase transition share qualitative features and similar critical exponents of insect swarms. Our simulations also point to a different phase transition reminiscent of gravitational collapse to clusters of finite size containing infinitely many particles.

This work paves the way to studies in many directions. Possible directions consist of exploring other possible transitions on chaotic and non-chaotic windows of the parameter space and the effect of anisotropic confinement on the phase transition studied here. Exploring a possible  phase transition to flocking black holes in self-gravitating models of swarms \cite{gor16} might be worth pursuing. On the theoretical side, can we find the invariant measure of the chaotic attractors and characterize scale-free-chaos phase transitions as $N\to\infty$ in terms of the invariant measure? This could bring together dynamical systems and nonequilibrium statistical mechanics studies and yield fruitful new ideas and methods.

\begin{acknowledgments}
We thank the anonymous referees for insightful and useful comments that have helped improving our paper. This work has been supported by the FEDER/Ministerio de Ciencia, Innovaci\'on y Universidades -- Agencia Estatal de Investigaci\'on grants PID2020-112796RB-C21 (RGA \& AC) and  PID2020-112796RB-C22 (LLB), by the Madrid Government (Comunidad de Madrid-Spain) under the Multiannual Agreement with UC3M in the line of Excellence of University Professors (EPUC3M23), and in the context of the V PRICIT (Regional Programme of Research and Technological Innovation) (LLB). RGA acknowledges support from the Ministerio de Econom\'\i a y Competitividad of Spain through the  Formaci\'on de Doctores program Grant PRE2018-083807 cofinanced by the European Social Fund. 
\end{acknowledgments}

\appendix
\section{Nondimensionalized equations of the confined Vicsek model}\label{ap:a}
We consider the three-dimensional confined Vicsek model: 
\begin{eqnarray}
&&\mathbf{x}_i(t+\Delta t)=\mathbf{x}_i(t)+ \Delta t\,\mathbf{v}_i(t+\Delta t), \nonumber\\
&&\mathbf{v}_i(t+\Delta t)= v\mathcal{R}_\eta\!\left[\frac{\sum_{|\mathbf{x}_j-\mathbf{x}_i|<r_1R_0}\mathbf{v}_j(t)-\beta_0\mathbf{x}_i(t)}{\left|\sum_{|\mathbf{x}_j-\mathbf{x}_i|<r_1R_0}\mathbf{v}_j(t)-\beta_0\mathbf{x}_i(t)\right|}\right]\!,\label{eqa1}
\end{eqnarray} 
where $\mathcal{R}_\eta(\mathbf{w})$ rotates the unit vector $\mathbf{w}$ randomly within a spherical cone centred at it and spanning a solid angle in $(-\frac{\eta}{2},\frac{\eta}{2})$ \cite{wan92}. Initially, the particles are randomly placed within a sphere with unit radius and the particle velocities are pointing outwards. 

We nondimensionalize the model using data from the experiments on midges reported in the supplementary material of Refs.~\onlinecite{att14plos,att14,cav17}. We select the event labeled $20120910\_A1$ in Table I \cite{cav17}. We measure times in units of $\Delta t=0.24$ s, lengths in units of the time-averaged nearest-neighbor distance of the $20120910\_A1$ swarm, which is $r_1=4.68$ cm, and velocities in units of $r_1/\Delta t$, whereas $v=0.195$ m/s. Then Eq.~\eqref{eq1} is the nondimensional version of Eq.~(\ref{eqa1}) with $\Delta t=1$ and
\begin{eqnarray}
v_0=v\, \frac{\Delta t}{r_1}, \quad \beta=\beta_0 \Delta t. \label{eqa2}
\end{eqnarray}
For the example we have selected, $v_0=1$, whereas other entries in the same table produce order-one values of $v_0$ with average 0.53. For these values, the confined Vicsek model has the same behavior as described here. Thus, the Vicsek model describing midge swarms is far from the continuum limit $v_0\ll 1$. Cavagna {\em et al} consider a much smaller speed, $v_0=0.05$, closer to the continuum limit where derivatives replace finite differences \cite{cav17}. 

Collective consensus is quantified by the polarization $W\in[0,1]$: 
\begin{eqnarray}
W(t;\eta,\beta)= \left|\frac{1}{N}\sum_{j=1}^N \frac{\mathbf{v}_j(t)}{|\mathbf{v}_j(t)|}\right|\!.\label{eqa3}
\end{eqnarray}
The time average $\langle W\rangle_t$ coincides with the ensemble average of \eqref{eqa3} by ergodicity.

{\em Effect of the boundary conditions.} In the standard VM, the particles are enclosed in a cubic box, the boundary conditions are periodic and the system is invariant under translations. On the other hand, in the confined VM, there are no boundaries, the particles are confined by a harmonic potential, and translation invariance is broken. There are many studies of the standard VM, which is not the case for the confined VM. In fact, the confined VM has time-dependent attractors that are different from those of the standard VM. Among them, chaotic attractors. Another qualitative difference between both VMs is that broken translation symmetry precludes particles filling uniformly the available space for the confined VM. Thus, the ordering transition of the periodic-box VM cannot be the same for the confined VM.

\section{Chaotic and noisy dynamics}\label{ap:b}
We calculate the LLE in different ways that are complementary to each other: (i) directly from the equations by using the Benettin {\em et al} (BA) algorithm \cite{ben80,ott93,cen10}, (ii)-(iii) using from time traces of the center-of-mass motion or the NDCCF to reconstruct the phase space of the chaotic attractor. We need model equations to use the BA whereas time traces can be obtained from numerical simulations of equations or from experiments and observations. To obtain the LLE from time traces, we have used (ii) the scale-dependent Lyapunov exponent (SDLE) algorithm \cite{gao06} and (iii) the Gao-Zheng algorithm \cite{gao94}. The SDLE algorithm is useful to separate the cases of mostly deterministic chaos from noisy chaos and mostly noise even in the presence of scarce data and a reconstruction of the attractor that is not very precise \cite{gao06} whereas the Gao-Zheng algorithm requires more data points \cite{gao94}. We now describe these different algorithms and illustrate the results they provide for the confined VM. In all cases, we eliminate the effects of initial conditions by leaving out the first 30000 time steps before processing the time traces.

\subsection{Benettin algorithm}
We have to simultaneously  solve Eqs.~\eqref{eq1} and the linearized equations
\begin{widetext}\begin{subequations}\label{eqb1}
\begin{eqnarray}
\delta\mathbf{\tilde{x}}_i(t+1)\!&=&\! \delta\mathbf{\tilde{x}}_i(t)+\delta\mathbf{\tilde{v}}_i(t+1),  \quad i=1,\ldots, N,  \label{eqb1a}\\
\delta\mathbf{\tilde{v}}_i(t+1)\!&=&\! v_0\mathcal{R}_\eta\!\left[\!\left(\mathbb{I}_3-\frac{[\sum_{|\mathbf{x}_j-\mathbf{x}_i|<R_0}\mathbf{v}_j(t)-\beta\mathbf{x}_i(t)]^T[\sum_{|\mathbf{x}_j-\mathbf{x}_i|<R_0}\mathbf{v}_j(t)-\beta\mathbf{x}_i(t)]}{|\sum_{|\mathbf{x}_j-\mathbf{x}_i|<R_0}\mathbf{v}_j(t)-\beta\mathbf{x}_i(t)|^2}\right)\cdot \frac{\sum_{|\mathbf{x}_j-\mathbf{x}_i|<R_0}\delta\mathbf{\tilde{v}}_j(t)-\beta\delta\mathbf{\tilde{x}}_i(t)}{|\sum_{|\mathbf{x}_j-\mathbf{x}_i|<R_0}\mathbf{v}_j(t)-\beta\mathbf{x}_i(t)|}\right]\!,  \label{eqb1b}
\end{eqnarray}
\end{subequations}\end{widetext}
in such a way that the random realizations $\mathcal{R}_\eta$ are exactly the same for Eqs.~\eqref{eq1} and \eqref{eqb1}. The initial conditions for the disturbances, $\delta\mathbf{\tilde{x}}_i(0)$ and $\delta\mathbf{\tilde{v}}_i(0)$, can be randomly selected so that the overall length of the vector $\delta\bm{\chi}=(\delta\mathbf{\tilde{x}}_1,\ldots,\delta\mathbf{\tilde{x}}_N,\delta\mathbf{\tilde{v}}_1,\ldots,\delta\mathbf{\tilde{v}}_N)$ equals 1. After each time step $t$, the vector $\delta\bm{\chi}(t)$ has length $\alpha_t$. At that time, we renormalize $\delta\bm{\chi}(t)$ to $\hat{\bm{\chi}}(t)=\delta\bm{\chi}(t)/\alpha_t$ and use this value as initial condition to calculate $\delta\bm{\chi}(t+1)$. With all the values $\alpha_t$ and for sufficiently large $l$, we calculate the Lyapunov exponent as
\begin{eqnarray}
&&\lambda_1= \frac{1}{l}\sum_{t=1}^l\ln\alpha_t, \label{eqb2}\\
&& \alpha_t=|\delta\bm{\chi}(t)|=|(\delta\mathbf{\tilde{x}}_1(t),\ldots,\delta\mathbf{\tilde{x}}_N(t),\delta\mathbf{\tilde{v}}_1(t),\ldots,\delta\mathbf{\tilde{v}}_N(t))|, \nonumber
\end{eqnarray}

\begin{center}
\begin{figure}[h]
\begin{center}
\vspace{-0.5cm}
\includegraphics[width=5.9cm]{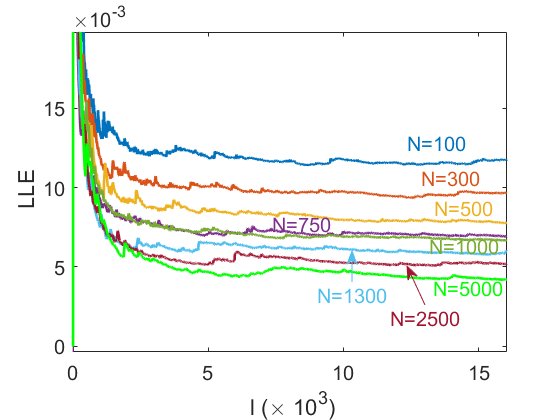}
\end{center}
\caption{Largest Lyapunov exponent as a function of $l$ as given by Eq.~\eqref{eqb2} for $\eta=0.5$, $\beta=\beta_c(N)$ and different $N$}
 \label{fig17}
\end{figure}
\end{center}

\begin{center}
\begin{figure}[h]
\begin{center}
\includegraphics[width=4.25cm]{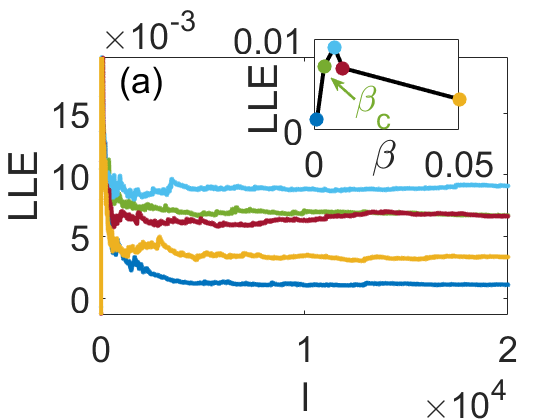}
\includegraphics[width=4.25cm]{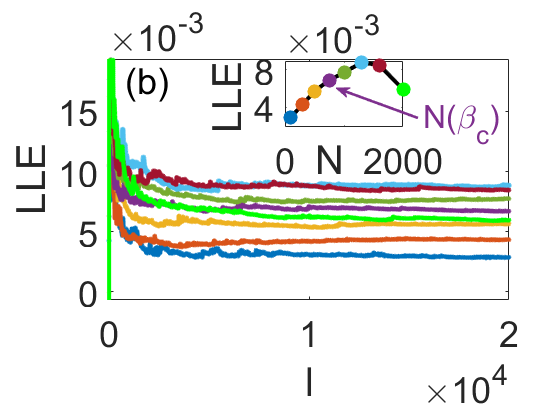}
\end{center}
\caption{ {\bf (a)} LLE versus $l$ as in Eq.~\eqref{eqb2} for $N=750$ and $\beta=0.001, 0.0035, 0.007, 0.01, 0.05$. Inset: LLE vs $\beta$; marked: $\beta_c=0.0035$. {\bf (b)} LLE vs $l$ for $\beta=0.0035$ and $N=100, 300, 500, 750, 1000, 1300, 1600, 2000$. Inset: LLE vs $N$ for $\beta_c=0.0035$; marked: $N=750$. Here $\eta=0.5$. }
 \label{fig18}
\end{figure}
\end{center}

 Fig.~\ref{fig17} plots $\lambda_1$ versus $l$ at critical confinement $\beta=\beta_c(N)$ showing convergence of the exponent for different values of $N$. For $N=750$, Fig.~\ref{fig18}(a) depicts the LLE versus $l$ for different values of $\beta$ whereas Fig.~\ref{fig18}(b) fixes $\beta=\beta_c(750)=0.0035$ and shows the LLE versus $l$ for different values of $N$, including $N=750$. The insets of these figures indicate that the LLE is not a monotonic function of either $\beta$ or $N$. See also Fig.~\ref{fig8}(e).

\subsection{Scale dependent Lyapunov exponents}
We use scale dependent Lyapunov exponents (SDLE) from the CM motion to characterize deterministic and noisy chaos as different from noise \cite{gao06}. 

Adding the components of $\mathbf{X}(t)$, we form the time series $x(t)=X_1(t)+X_2(t)+X_3(t)$. To calculate the SDLE, we construct the lagged vectors: $\mathbf{X}_\alpha=[x(\alpha),x(\alpha+\tilde{\tau}),..,x(\alpha+(m-1)\tilde{\tau})]$. The simplest choice is $m=2$ and $\tilde{\tau}=1$ (other values can be used, see below). From this dataset, we determine the maximum $\varepsilon_\text{max}$ and the minimum $\varepsilon_\text{min}$ of the distances between two vectors, $\|\mathbf{X}_\alpha-\mathbf{X}_\beta\|$. Our data is confined in $[\varepsilon_\text{min},\varepsilon_\text{max}]$. Let $\varepsilon_0$, $\varepsilon_t$ and $\varepsilon_{t+\Delta t}$ be the average separation between nearby trajectories at times 0, $t$, and $t+\Delta t$, respectively. The SDLE is
 \begin{subequations}\label{eqb3}
 \begin{eqnarray}
\ln \lambda(\varepsilon_t)=\frac{\ln\varepsilon_{t+\Delta t}-\ln\varepsilon_t}{\Delta t}. \label{eqb3a}
\end{eqnarray}
The smallest possible $\Delta t$ is of course the time step $\tilde{\tau}=1$, but $\Delta t$ may also be chosen as an integer larger than 1. Gao {\em et al} introduced the following scheme to compute the SDLE \cite{gao06}. Find all the pairs of vectors in the phase space whose distances are initially within a shell of radius $\epsilon_k$ and width $\Delta\epsilon_k$:
\begin{eqnarray}
\varepsilon_k\leq\|\mathbf{X}_\alpha-\mathbf{X}_\beta\|\leq\varepsilon_k+\Delta\varepsilon_k,\quad k=1,2,\ldots.\label{eqb3b}
\end{eqnarray}
We calculate the Lyapunov exponent as follows:  
\begin{eqnarray}
\lambda(\varepsilon_t)=\frac{\langle\ln\|X_{\alpha+t+\triangle t}-X_{\beta+t+\triangle t}\|-\ln\|X_{\alpha+t}-X_{\beta+t}\|\rangle_{k}}{\triangle t},\quad \quad \label{eqb3c}
\end{eqnarray} 
where $\langle\rangle_{k}$ is the average within the shell $(\varepsilon_k,\varepsilon_k+\triangle \varepsilon_k)$. The shell dependent SDLE $\lambda(\varepsilon)$ in Fig.~\ref{fig3}(b) displays the dynamics at different scales for $\tilde{\tau}=1$ and $m=2$. \cite{gao06} Using 2 lagged coordinates produces plateaus having a value of $\lambda(\varepsilon)$ equal to the LLE of deterministic chaos. This value differs from the LLE calculated using the BA or a more appropriate reconstruction of the phase space involving more lagged coordinates (see below). However, the SDLE with $m=2$ yields a qualitative idea of the effects of noise on chaos. In deterministic chaos, $\lambda(\varepsilon)>0$ presents a plateau with ends $\varepsilon_1< \varepsilon_2\ll 1$, in noisy chaos, this plateau is preceded and succeeded by regions in which $\lambda(\varepsilon)$ decays as $-\gamma\ln\varepsilon$, whereas it shrinks and disappears when noise swamps chaos. As $\eta$ increases, $\lambda(\varepsilon)$ first decays to a plateau for $\eta=0.1$. A criterion to distinguish (deterministic or noisy) chaos from noise is to accept the largest Lyapunov exponent as the positive value at a plateau $(\varepsilon_1,\varepsilon_2)$ satisfying 
\begin{eqnarray}
\log_{10}\frac{\varepsilon_2}{\varepsilon_1}\geq \frac{1}{2}.  \label{eqb3d}
\end{eqnarray}\end{subequations}
For $\eta=0.5$, the region where $\log_{10}(\varepsilon_2/\varepsilon_1)= 1/2$ is marked in Fig.~\ref{fig3}(b) by vertical lines. Plateaus with smaller values of $\log_{10}(\varepsilon_2/ \varepsilon_1)$ or their absence indicate noisy dynamics \cite{gao06}. This occurs for $\eta=1$. The ends of the interval $(0.1,1)$ of noisy chaos are marked by two vertical dashed lines in Fig.~\ref{fig3}(a). 

\begin{widetext}
\begin{center}
\begin{figure}[h]
\begin{center}
\includegraphics[width=5.9cm]{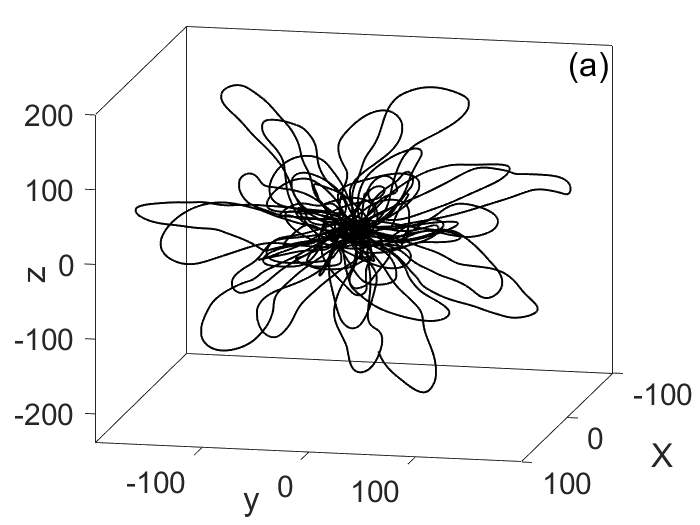}
\includegraphics[width=5.9cm]{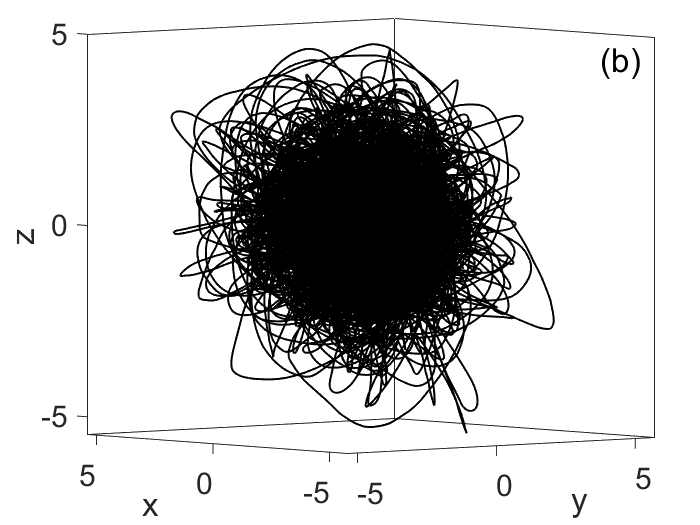}
\includegraphics[width=5.9cm]{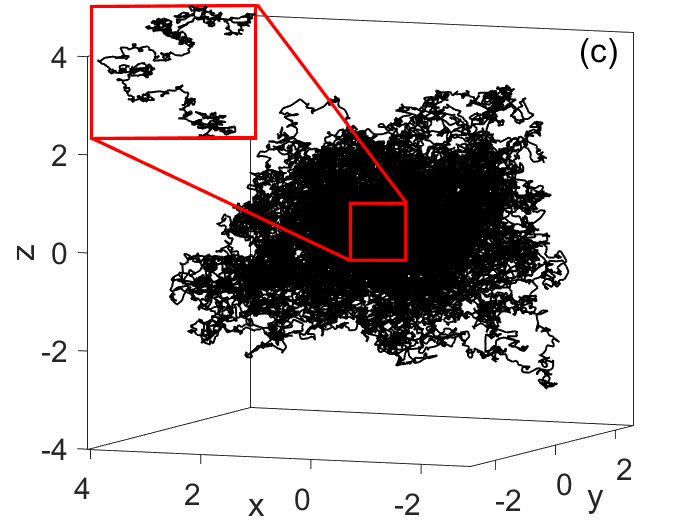}
%\\ {\bf (a)} \hskip 5.5cm {\bf (b)} \hskip 5.5cm{\bf (c)} 
\end{center}
\caption{{\bf (a)} Trajectory of the center of mass for $\eta=0.01$, which corresponds to deterministic chaos with flower shape phase portrait. {\bf (b)} Same for $\eta=0.3$, which corresponds to noisy chaos: the trajectories of the center of mass cover more densely part of the  space. {\bf (c)} Predominantly noisy motion for $\eta=5.5$. The trajectories from $t_0=1000$ to $t_f=50000$ are depicted. Here, $N=100$, $\beta=\beta_c$ for each $\eta$.  \label{fig19}}
\end{figure}
\end{center}
\end{widetext}

The chaotic dynamics of the swarm is reflected in quantities that depend on the positions and velocities of the particles. Important global quantities are the motion of the CM and the NDCCF of Eq.~\eqref{eq3}. Figs.~\ref{fig19} displays the CM trajectory, thereby visualizing the dynamics of the swarm. For increasing values of noise corresponding to the different regions in Fig~\ref{fig3}(a), the CM motion goes from deterministic chaos, Fig.~\ref{fig19}(a), to noisy chaos, Fig.~\ref{fig19}(b), to mostly noise, Fig.~\ref{fig19}(c). 

Note that all the plateaus in Fig.~\ref{fig3}(b) produce the same positive value of the LLE $\lambda(\varepsilon)$. This is not very realistic because the BA yields different values of the LLE depending on the noise strength $\eta$. Why? Recall that we have used $m=2$ (two lagged coordinates) in the reconstruction of the attractor from the time traces. However, as shown in Fig.~\ref{fig6}, the CM chaotic attractor has fractal dimension $D_0$ between 2 and 3, and we need $m\geq 2 D_0$ to faithfully reconstruct the chaotic attractor \cite{ott93,cen10}. Thus, we need at least $m=6$ to reconstruct it. Using $m=6$ and its optimal value of $\tilde{\tau}$ (Ref.~\onlinecite{gao94}) produces Fig.~\ref{fig3}(c). Now $\lambda(\epsilon)$ presents large oscillations whose averages in the plateau regions coincide with the LLE as calculated by the Gao-Zheng algorithm \cite{gao94}.

\subsection{Largest Lyapunov exponent from high dimensional reconstructions of CM motion}
\begin{center}
\begin{figure}[ht]
\begin{center}
\includegraphics[width=5.9cm]{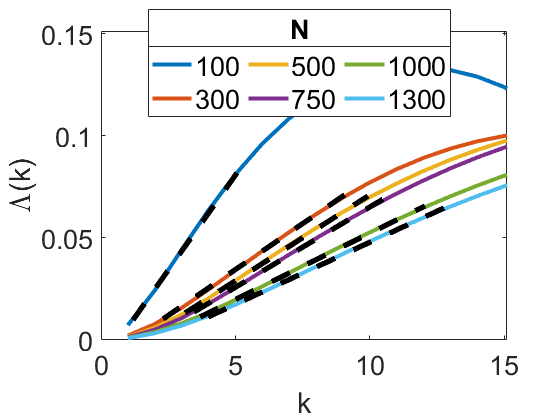}
\end{center}
\caption{Plot of $\Lambda(k)$ vs $k$ for different  values of $\beta$.  Thick dashed lines mark the slope of $\Lambda(k)$ for different values of $N$ at $\beta_c(N)$. }
 \label{fig20}
\end{figure}
\end{center}

As explained above, the previous reconstruction of the phase space for CM motion used to calculate SDLE considers 2D lagged vectors ($m=2$). This produces useful qualitative phase diagrams with flat plateaus, but the dimension of this vector space is too small to reconstruct faithfully the attractor. More realistic CM trajectories in higher dimension contain self-intersections in dimension 2. This explains the different values of the LLE found in the SDLE plateaus of Fig.~\ref{fig3}(b) as compared with those found by the BA of Eq.~\eqref{eqb2}. To reconstruct  safely a chaotic attractor, the dimension of the lagged vectors should surpass twice the fractal dimension $D_0$. \cite{ott93} For the confined VM, $m=6$ is sufficient in view of Fig.~\ref{fig6}. However, the SDLE $\lambda(\varepsilon)$ presents oscillations as indicated in Fig.~\ref{fig3}(c) and their average values replace the plateaus in Fig.~\ref{fig3}(b). In contrast with  Fig.~\ref{fig3}(b), the averaged oscillations produce LLEs that increase with noise. Averaging oscillations is not going to produce precise values of the LLE. Thus, we calculate the LLE from the lagged coordinates with $m=6$ using the Gao-Zheng algorithm \cite{gao94}. This requires constructing the quantity $\Lambda(k)$ whose slope near the origin gives the LLE \cite{gao94}
\begin{eqnarray}
\Lambda(k)= \left\langle\ln\frac{\|X_{i+k}-X_{j+k}\|}{\|X_i-X_j\|}\right\rangle.\label{eqb4}
\end{eqnarray}
Here the brackets indicate ensemble average over all vector pairs with $\|X_i-X_j\|<r^*$ for an appropriately selected small distance $r^*$. Fig.~\ref{fig20} displays the graph of $\Lambda(k)$ given by Eq.~\ref{eqb4}. The slopes of $\Lambda(k)$ for different values of $N$ at $\beta_c(N)$ equal the LLEs, increase with $\beta$ and agree with the averaged oscillations marked in Fig.~\ref{fig3}(c). 

\begin{table}[ht]
\begin{center}\begin{tabular}{|c||c|c|c|c|c|c|}
 \hline
$\mathbf{N}$& $100$ & $300$ & $500$& 750 & 1000 &1300\\ 
\hline
{\bf BA} & 0.0118& 0.0095& 0.0078& 0.0070& 0.0067& 0.0058\\ 
\hline
{\bf CM }& 0.017&  0.0092&    0.008&  0.007&  0.0063& 0.0059\\
\hline
$\mathbf{g(t)}$ & 0.0055& 0.0044& 0.0041&0.0038&0.0035 &0.0033\\
\hline
\end{tabular}
\end{center}
\caption{LLE for $\eta=0.5$ and different $N$ at $\beta_c(N;\eta)$ as calculated using the BA for the complete VM equations, Eq.~\eqref{eqb2}, and Eq.~\eqref{eqb4} for CM motion and for the NDCCF $g(t)$. Note that the LLE as calculated using the BA and Eq.~\eqref{eqb4} for CM motion are similar whereas the LLE corresponding to the NDCCF $g(t)$ is smaller.}
\label{table2}
\end{table}

For different particle numbers with $\eta=0.5$, Table \ref{table2} lists the LLEs calculated using the BA for the complete system as in Eq.~\eqref{eqb2}, and  using Eq.~\eqref{eqb4} for CM motion and for the NDCCF $g(t)$. We observe that the LLE values calculated from the CM motion are similar to those found by the BA, whereas they are noticeably smaller if calculated for the NDCCF. While the NDCCF is still chaotic, we speculate that the subtraction of the CM motion and ensemble average involved in two-time NDCCF $g(t)$ dilute chaos by lowering the LLE. We observe that the difference between LLEs calculated from BA and CM motion and those from $g(t)$ in Table \ref{table2} decreases as $N$ increases. Thus, it could happen that both sets of LLEs eventually converge to similar smaller values as $N\to\infty$ and chaos disappear.

\section{Dynamic and static connected correlation functions}\label{ap:c}
{\em Definitions.} Following Refs.~\onlinecite{att14plos,att14}, we define the dynamic connected correlation function (DCCF) as 
\begin{eqnarray}\
&&C(r,t)\!=\!\!\left\langle\! \frac{\sum_{i=1}^{N}\!\sum_{j=1,j\neq i}^{N}\delta\hat{\mathbf{v}}_i(t_0\!)\!\cdot\!\delta\hat{\mathbf{v}}_j(t_0\!+t)\delta[r\!-r_{ij}(t_0\!,t)]}{\sum_{i=1}^{N}\sum_{j=1,j\neq i}^{N}\delta[r-r_{ij}(t_0,t)]}\! \right\rangle_{t_0}\quad \label{eqc1}\\
&&C(r)=C(r,0),\nonumber\\
&&\delta\hat{\mathbf{v}}_i\!=\frac{\delta\mathbf{v}_i}{\sqrt{\frac{1}{N}\sum_k \delta\mathbf{v}_k\cdot\delta\mathbf{v}_k}},\quad \delta\mathbf{v}_i=\mathbf{v}_i - \mathbf{V},\nonumber\\
&&r_{ij}(t_0,t)=|\mathbf{r}_i(t_0)-\mathbf{r}_j(t_0+t)|,\,  \mathbf{r}_i(t_0)=\mathbf{x}_i(t_0) - \frac{1}{N}\sum_{j=1}^N\mathbf{x}_j(t),   \nonumber\\
&&\langle f\rangle_{t_0}=\frac{1}{t_{max}-t}\sum_{t_0=1}^{t_{max}-t}f(t_0,t).\nonumber 
\end{eqnarray}
In these equations, $\delta(r-r_{ij})=1$ if $r<r_{ij}<r+dr$ and zero otherwise, and $dr$ is the space binning factor. The usual dynamic correlation function and susceptibility in statistical mechanics are 
\begin{subequations} \label{eqc2}
\begin{eqnarray}
&&C(r,t)=\langle (\phi(0,0)-\langle\phi(0,0)\rangle)(\phi(\mathbf{r},t)-\langle\phi(\mathbf{r},t)\rangle)\rangle,\quad\quad \label{eqc2a}\\
&&\chi=\int C(r,0)\, d\mathbf{r}=\hat{C}(0,0),\label{eqc2b}
\end{eqnarray}\end{subequations}
respectively, where $\langle\ldots\rangle$ are averages over the appropriate ensemble average and we have set $t_0=0$.  In Appendix \ref{ap:d}, we show two solvable examples indicating the relation between correlation time and susceptibility for different dynamics, which may or may not lead to thermal equilibrium.

 In Eq.~\eqref{eqc1}, we have replaced arithmetic means instead of the ensemble averages and added a time average. Dropping the condition $j\neq i$ adds one term proportional to $\delta(r)$ to numerator and denominator of Eq.~\eqref{eqc1}, which is the choice made in Refs.~\onlinecite{cav17,cav18}.

The function $C(r,t)$   sums all the products $\delta\mathbf{v}_i(t_0)\cdot\delta\mathbf{v}_j(t_0+t)$ for those pairs $i$ and $j$ with a distance $r_{ij}(t_0,t)$ between $r$ and $r+dr,$ and then divides by the number of such pairs (denominator). It depends only on the distance $r$ at time $t$ because inter-particle interactions are local and distance dependent. The static connected  correlation function (SCCF) is the equal time connected correlation function given by Eq.~\eqref{eqc1}. As discussed in Ref.~\onlinecite{cav18}, these definitions are inspired in statistical mechanics taking into account $\sum_j\delta \mathbf{\hat{v}}_j=0$ because ensemble averages have been replaced by averages over the particles. 

{\em Susceptibility.} For a SCCF that decays exponentially, the correlation length $\xi$ is such that $C(\xi)=1/e$. In the present work, there is finite size scaling \cite{att14plos,att14,cav17} and $C(r)$ or $C(r,t)$ do not decay exponentially with $r$ and can take on negative values. Then the correlation length $\xi$ is the first zero of $C(r)$, corresponding to the first maximum of the cumulative correlation function \cite{att14plos,att14}:
\begin{eqnarray}
&&Q(r)= \frac{1}{N}\sum_{i=1}^{N}\sum_{j\neq i}^{N} \delta\hat{\mathbf{v}}_i\!\cdot\!\delta\hat{\mathbf{v}}_j\theta(r-r_{ij}), \quad\chi= Q(\xi),\label{eqc3}\\ 
&&\xi=\mbox{argmax}Q(r),\, C(\xi)=0\,\mbox{ with }\, C(r)>0,\, r\in(0,\xi), \quad\nonumber
\end{eqnarray}
where $\theta(x)$ is the Heaviside unit step function. It turns out that this correlation length $\xi$ is proportional to the swarm size $\ell$, which is the hallmark of scale free behavior. At equilibrium, the susceptibility measures the response of the order parameter to changes in an external field linearly coupled to it, and equals the integral of the SCCF \eqref{eqc2b} with $C(r)>0$. A susceptibility thus defined would be $Q(\infty)$. However,  by Eq.~\eqref{eqc3}, $Q(\infty)=Q(\ell)=-1$ because $\theta(\ell-r_{ij})= 1$ and $\sum_i \delta\hat{\mathbf{v}}_i=0$. Thus, we cannot define susceptibility by Eq.~\eqref{eqc2b}. Instead, we define susceptibility $\chi$ as the value of $Q(r)$ at its first maximum, as in Eq.~\eqref{eq9}, and Refs.~\onlinecite{att14plos,att14}. For values of $N$ corresponding to insects in measured swarms, our numerical simulations produce susceptibility values defined by Eq.~\eqref{eqc3} between 0.7 and 1.2, which are included in the measured interval $(0.32,5.57)$ \cite{att14plos,att14}.

\begin{center}
\begin{figure}[ht]
\begin{center}
\includegraphics[width=4.25cm]{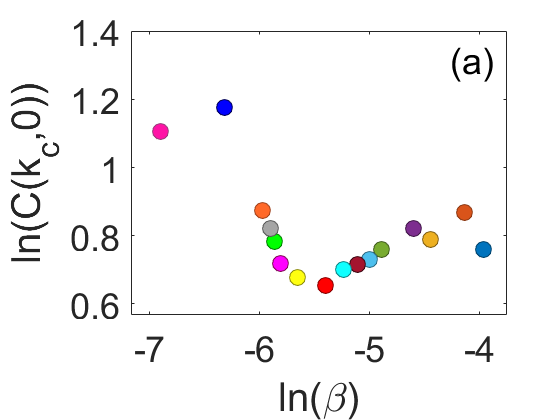}
\includegraphics[width=4.25cm]{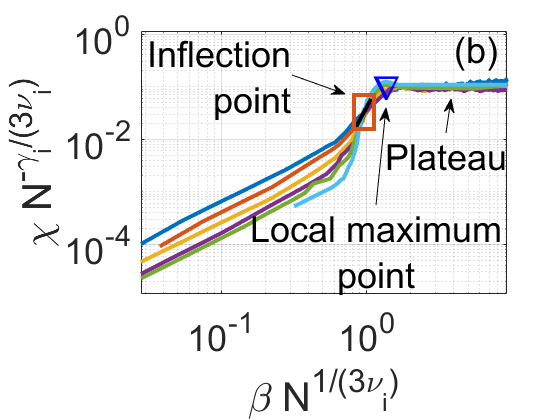}
\end{center}
\caption{{\bf (a)} Log-log scale plot of the susceptibility as given by max$_k\hat{C}(k,0)$ of Eq.~\eqref{eq3} versus $\beta_c$.  {\bf (b)} Scaled susceptibility versus scaled confinement showing data collapse at the inflection point (square) and the local maximum (triangle) of the susceptibility. The local maximum is followed by a plateau of the scaled confinement. Here $\eta=0.5$, $\nu_i=0.44$, $\gamma_i=1.2$. 
}
\label{fig21}
\end{figure}
\end{center}

\begin{center}
\begin{figure}[ht]
\begin{center}
\includegraphics[width=5.9cm]{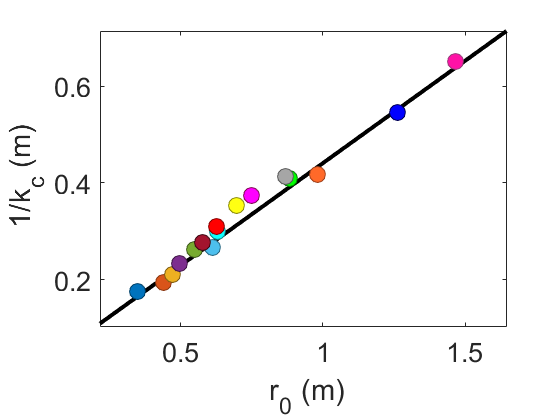}
\end{center}
\caption{The correlation length $\xi= 1/k_c$ computed from the static correlation function in Fourier space as a function of the correlation length $\xi=r_0$ computed from the static correlation function in real space. }
\label{fig22}
\end{figure}
\end{center}

At equilibrium and for $N=\infty$, the susceptibility becomes infinity at critical points and it marks a phase transition. The susceptibility scales as 
 \begin{eqnarray}
\chi(x)\sim (x-x_c)^{-\gamma}, \label{eqc4}
\end{eqnarray}
where $x$ is the control parameter and $x_c$ the value thereof for $N=\infty$. In our case, $x=\beta_c(N,\eta)$ and $x_c = \beta_c(\infty,\eta) = 0$, which produces $\gamma=0.92\pm 0.05$ as shown in Fig.~\ref{fig7}(b). Eq.~\eqref{eq3} is related to the Fourier transform of $C(r,t)$, as discussed in Ref.~\onlinecite{cav18}. Fig.~\ref{fig21}(a) shows that $\hat{C}(k,0)$ of Eq.~\eqref{eq3} oscillates with $\beta_c(N;\eta)$. Thus, $\max_k\hat{C}(k,0)$ is not a convenient  definition of susceptibility. Contrastingly, Fig.~\ref{fig7}(b) plots the real-space susceptibility max$_rQ(r)$ using many more values, $500\leq N\leq 5000$, which makes this fitting more reliable. %and similar to the measured values in Refs.~\onlinecite{att14plos,att14}. 
Fig.~\ref{fig21}(b) shows data collapse of scaled susceptibility and scaled confinement at $\beta_i$ (susceptibility inflection point) and $\beta_m$ (susceptibility local maximum). For our data, the relation between the correlation length as defined by Eq.~\eqref{eqc3} and $1/k_c$, given by $k_c=$ argmax$_k\hat{C}(k,0)$ is 
\begin{eqnarray}
\frac{1}{k_c}=0.44\, r_0+0.36;   \label{eqc5}
\end{eqnarray} 
see Fig.~\ref{fig22}. Since our unit of length is 4.68 cm, the straight line in Fig.~\ref{fig22} is quite close to that of Fig.~SF1 of Ref.~\onlinecite{cav17} (Supplementary data) obtained from measurements in natural midge swarms.

{\em Perception range.} Instead of setting $x=\beta$, we can use the rescaled average nearest neighbor distance $x=r_1/R_0$ (perception range, inversely proportional to density), as in Ref.~\onlinecite{att14plos,att14}. The perception range is calculated as the time average of the arithmetic mean of the minimal distance between each particle and its closest neighbor. We find a critical perception range $x_c=2.945\pm 0.047$, with $(x-x_c)$ proportional to $\beta$. This is larger than $x_c=0.421\pm 0.002$ at the order-disorder transition of the standard VM \cite{att14plos,att14}, indicating a less dense swarm at critical confinement. As $x_c=12.5 \pm 0.1$ in natural swarms (measured in units of the average insect size) \cite{att14plos,att14}, the critical perception range is 4.2 insect bodies for the scale-free chaotic transition of the confined Vicsek model versus 30 insect sizes at the ordering transition of the VM with periodic boundary conditions. At the phase transition to clusters of finite size containing infinitely many particles, $x\to 0$ as $x\sim \beta^{-\nu}$ and $\chi\sim\beta^\gamma$, with $\beta_M\to\infty$ for $N\to\infty$. The critical exponents for this transition are $\nu=0.33$ and $\gamma= 1.03$. 

As chaos disappears when $\beta\to 0$, it may seem surprising that an ordered chaotic phase is less dense than the disordered phase at the larger noise of the order-disorder transition for the standard VM with periodic boundary conditions. Recall that density is inversely proportional to the average nearest neighbor distance (perception range). However, the confined VM does not morph seamlessly to the standard VM as $\beta_c(N;\eta)\to 0$.  Firstly, confinement by a harmonic potential and confinement due to a large box with periodic boundary conditions are qualitatively different and they may not produce the same swarm patterns in the thermodynamic limit. Secondly, the standard VM with periodic boundary conditions experiences a crossover to a discontinuous order-disorder phase transition for $N\gg 1$ \cite{gre04,cha19,cha20}. Thirdly, the noise values ($\eta=5.65$ in our units) for which the confined VM and the standard VM with periodic boundary conditions have similar critical behaviors according to \cite{att14,att14plos,cav17} are much larger than the noisy chaos interval of Fig.~\ref{fig3}(a). Thus, we think that the scale-free-chaos phase transition of the confined VM as $\beta\to 0$ is not related to the continuous ordering transition of the standard VM. 

{\em Numerical calculation of the connected correlation functions.} Fixing the parameters $N$, $\eta$ and $\beta$, we simulate the VM for five different random initial conditions during 10000 iterations. After a sufficiently long transient period, the polarization of Eq.~\eqref{eqa3} fluctuates about a constant value. Once this regime is established, we use the last 2000 iterations to calculate the static correlation function $\hat{C}(k,0)$, whose first maximum provides the critical wave number $k_c$. Using the definition in Eq.~\eqref{eq3} and averaging over the five realizations, we obtain the time dependent correlation function. 

{\em Rotation and dilation.} In Refs.~\onlinecite{att14,att14plos}, the average swarm velocity is defined subtracting overall rotations and dilations from $\mathbf{V}$ at each time step. Note that an overall rotation does not change the distance between trajectories (which are used to calculate Lyapunov exponents) but an overall dilation does. To subtract an overall rotation, we proceed as follows. The fluctuations of the velocity are  
\begin{subequations}\label{eqc6}
\begin{eqnarray}
&&\delta\mathbf{v}_i(t+1)=\mathbf{y}_i(t+1)-\mathbf{y}_i(t),\label{eqc6a}\\
&&\mathbf{y}_i(t)=\mathbf{x}_i(t) - \mathbf{X}(t), \label{eqc6b}\\
&&\mathbf{X}(t)=\frac{1}{N}\sum_{j=1}^N\mathbf{x}_j(t),\label{eqc6c}\\
&&\mathbf{X}(t+1)-\mathbf{X}(t)=\mathbf{V}(t+1).\label{eqc6d}
\end{eqnarray}
\end{subequations}
We can define the optimal rotation matrix as the $3\times 3$ orthogonal matrix $\mathbf{U}$ that minimizes 
\begin{subequations}\label{eqc7}
\begin{eqnarray} 
\mathbf{U}=\mbox{argmin}_{\mathbf{U}^T\mathbf{U}=\mathbf{I}}\left[\frac{1}{2}\sum_{i=1}^N[\mathbf{y}_i(t+1)-\mathbf{U}\mathbf{y}_i(t)]^2\right]\!. \label{eqc7a}\end{eqnarray}
The optimal dilation is
\begin{eqnarray}
&&\Lambda=\mbox{argmin}_{\Lambda}\left[\frac{1}{2}\sum_{i=1}^N[\mathbf{y}_i(t+1)-\Lambda\mathbf{U}\mathbf{y}_i(t)]^2\right]\!. \label{eqc7b}\end{eqnarray}\end{subequations}
From Ref.~\onlinecite{kab76}, the optimal rotation matrix for Eq.~\eqref{eqc7a} is
\begin{subequations}\label{eqc8}
\begin{eqnarray}
&&U_{ij}=\sum_{k=1}^N B_{ki}A_{kj},\label{eqc8a}\\
&&B_{ki}=\sum_{n=1}^N\sum_{l=1}^3Y_{ni}(t+1)Y_{nl}(t)\frac{A_{kl}}{\sqrt{\mu_k}}, \label{eqc8b}\end{eqnarray}
where $Y_{ni}(t)$ is the $N\times 3$ matrix formed by the components of the vector $\mathbf{y}_n(t)$. The orthogonal matrix $A_{kl}$ is formed by the orthogonalized eigenvectors of the eigenvalue problem
\begin{eqnarray}
&&\sum_{lm=1}^N\sum_{n,p=1}^N [Y_{li}(t)Y_{ln}(t+1)Y_{mp}(t)Y_{mn}(t+1)]A_{kp}=\mu_kA_{ki},\quad\quad\label{eqc8c}\\ 
&& \mu_k\geq 0,\nonumber
\end{eqnarray}
for the $3\times 3$ positive semidefinite symmetric matrix within square brackets appearing in this expression. In the generic case, the three eigenvalues $\mu_k$ are positive. If one eigenvalue is zero, e.g., $\mu_3=0$, then  Eq.~\eqref{eqc8b} yields only two vectors, $B_{1i}$, $B_{2i}$, corresponding to the eigenvectors with nonzero eigenvalues, $A_{1i}$, $A_{2i}$. The other eigenvector $A_{3i}$ and $B_{3i}$ are 
\begin{eqnarray}
A_{3i}=\sum_{j,k=1}^3\epsilon_{ijk}A_{1j}A_{2k},\quad B_{3i}=\sum_{j,k=1}^3\epsilon_{ijk}B_{1j}B_{2k}, \label{eqc8d}
\end{eqnarray}
where $\epsilon_{ijk}$ is the completely antisymmetric unit tensor with $\epsilon_{123}=1$ \cite{kab76}. With these definitions of $A_{3i}$ and $B_{3i}$, Eq.~\eqref{eqc8a} holds. 

To subtract overall dilation, we use the optimal dilation matrix from Eq.~\eqref{eqc8b},  
\begin{eqnarray}
\Lambda=\frac{\sum_{j,k=1}^N\sum_{i,l=1}^N Y_{jl}(t+1)B_{ki}A_{kl}Y_{jl}(t)}{\sum_{j,k=1}^N\sum_{i,l=1}^N [B_{ki}A_{kl}Y_{jl}(t)]^2}.\label{eqc8e}
\end{eqnarray}\end{subequations}
Then the fluctuations in Eq.~\eqref{eq3} are \cite{att14plos}
\begin{eqnarray}
\delta\mathbf{v}_i(t_0)=\mathbf{y}_i(t_0+1)-\Lambda\mathbf{ Uy}_i(t_0).\label{eqc9}
\end{eqnarray}
Note that $\sum_i\delta\mathbf{v}_i(t_0)=\sum_i\mathbf{y}_i(t_0+1)-\Lambda\mathbf{ U}\sum_i\mathbf{y}_i(t_0)=0$. Subtracting only overall rotations, we would use $\delta\mathbf{v}_i(t_0)=\mathbf{y}_i(t_0+1)-\mathbf{ Uy}_i(t_0)$ instead of Eq.~\eqref{eqc9}.

\begin{center}
\begin{figure}[h]
\begin{center}
\includegraphics[width=4.25cm]{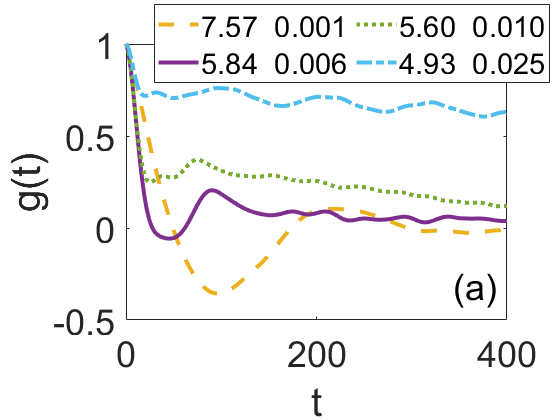}
\includegraphics[width=4.24cm]{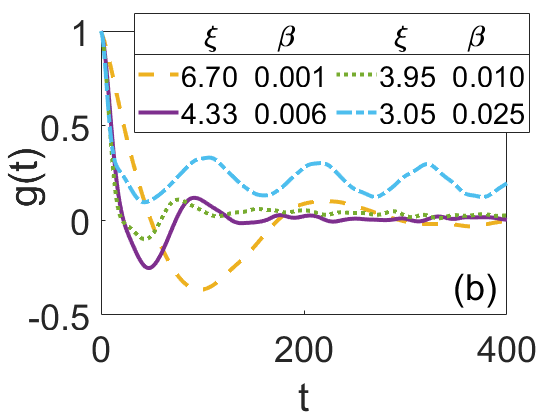}
\end{center}
\caption{Normalized dynamic correlation function with $\xi= 1/k_c$ for different values of $\beta$ calculated from (a) Eq.~\eqref{eq3} and (b) from Eq.~\eqref{eq3} subtracting rotation and dilation in the velocity fluctuations. Here, $N=300$, $\eta=0.5$. }
\label{fig23}
\end{figure}
\end{center}

\begin{center}
\begin{figure}[h]
\begin{center}
\includegraphics[width=4.25cm]{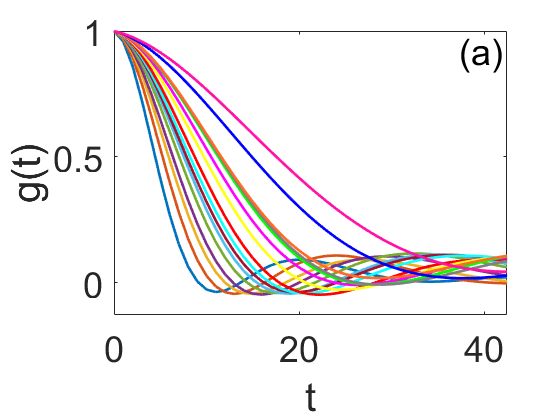}
\includegraphics[width=4.25cm]{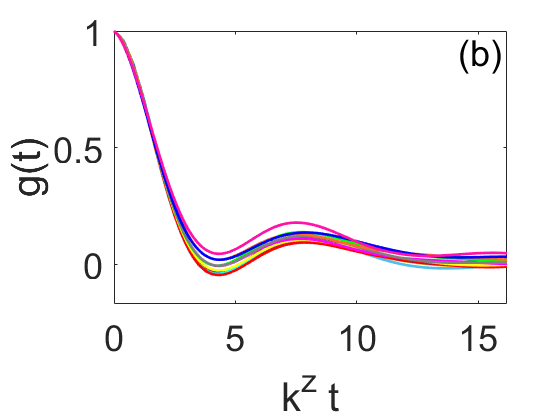}
\end{center}
\caption{Same as Figure \ref{fig4} but calculated subtracting rotation and dilation: NDCCF $g(t)$ versus (a) $t$ and (b) $k^zt$ with $z=1.00\pm 0.03$. }
\label{fig24}
\end{figure}
\end{center}

\begin{center}
\begin{figure}[h]
\begin{center}
\includegraphics[width=4.25cm]{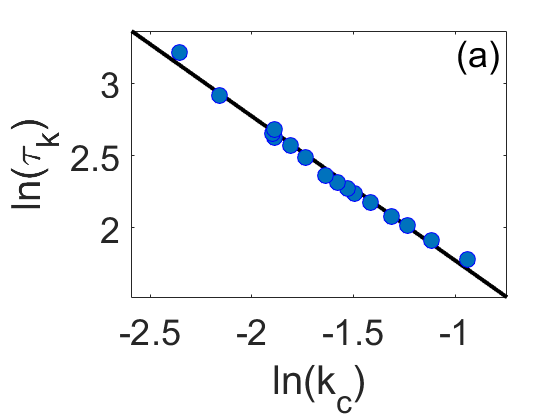}
\includegraphics[width=4.25cm]{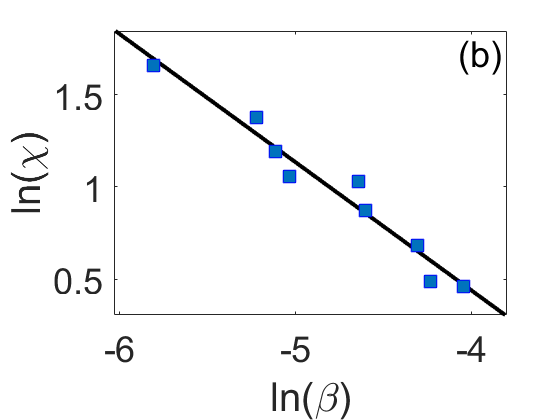}
\end{center}
\caption{(a) Correlation time versus $\beta$ at the critical line $\beta_c(N;0.5)$. (b) Susceptibility versus $\beta$ at the critical line $\beta_i(N;0.5)$. Both power laws calculated subtracting rotation and dilation from CM motion.}
\label{fig25}
\end{figure}
\end{center}

\begin{center}
\begin{figure}[h]
\begin{center}
\includegraphics[width=4.25cm]{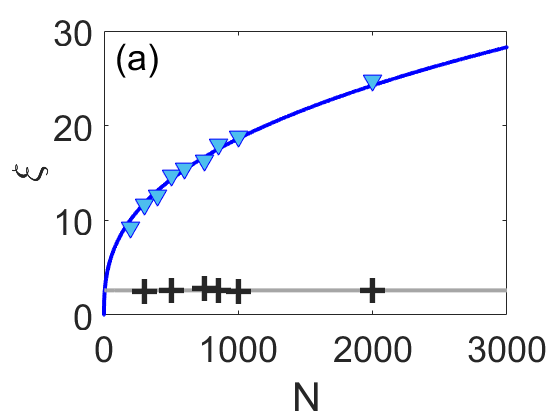}
\includegraphics[width=4.25cm]{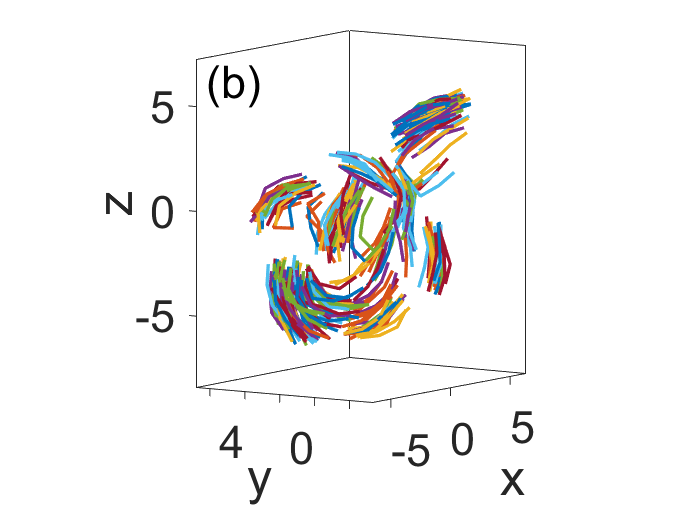}
\end{center}
\caption{(a) Correlation length versus $N$ obtained without (triangles) and with (crosses) subtraction of rotation and dilation from CM velocity at confinements corresponding to the susceptibility maximum. (b) Chaotic swarm showing short trajectories of 300 particles for $\beta_M(N;\eta)$: Subtractions shift the critical line to larger confinement well inside the multicluster chaos region. Here,  $\eta= 0.5$. }
\label{fig26}
\end{figure}
\end{center}

{\em Results.} Figs.~\ref{fig23}(a) and \ref{fig23}(b) compare the NDCCF $g(t)$ calculated as in Eq.~\eqref{eq3} and the same function subtracting rotation and dilation. We observe that both functions look alike and that subtracting rotation and dilation changes slightly the times $t_m(\beta,N)$ where $g(t_m)=0$. Then the critical line $\beta_c(N,\eta)$ is unchanged by subtractions of rotation and dilation; see also Fig.~\ref{fig24} for the collapse of the NDCCF with dynamical exponent $z$. The relation \eqref{eqc5} becomes $k_cr_0=2$ after subtractions. We expect small rotation and dilation for single-cluster chaos and larger rotation and dilation for multicluster chaos. Thus, subtracting rotation and dilation brings down $g(t)$. In in Fig.~\ref{fig23}, this effect is largest for $\beta=0.025$, well inside the region of multicluster chaos. 

The critical lines $\beta_c(N;\eta)$ and $\beta_i(N;\eta)$ do not change by subtractions from CM motion but the local maxima defining the line $\beta_m(N;\eta)$ disappear. The critical exponent $\nu=0.45\pm 0.02$ rests unchanged by subtractions. However, for $N\leq 2000$, $\gamma_i=0.70 \pm 0.06$ (on the line of inflection points of $\chi$ vs $\beta$) drops from the value 1.2 obtained without subtractions, cf Fig.~\ref{fig25}(b). Together with the larger value of the polar order parameter, this confirms that the critical line $\beta_i(N;\eta)$ lies in the multicluster chaos region where rotation and dilation effects are more prominent; see Figs.~\ref{fig12}(c)-\ref{fig12}(e) for pictures of swarms splitting into different clusters. 

The global maxima $\beta_M(N;\eta)$ of the susceptibility vs confinement curve in Fig.~\ref{fig8}(a) move to the end of the first chaotic window when subtracting rotation and dilation from CM motion and do not correspond to scale free transitions. Fig.~\ref{fig26}(a) shows that the correlation length for these states does not change with the number of particles maxima and is no longer proportional to the swarm size. At the line $\beta_M(N;\eta)$, chaos is multicluster and rotation and dilation effects are stronger. As shown in Figs.~\ref{fig12} and \ref{fig26}(b), these chaotic clusters are not connected, and their correlation length remains unchanged with $N$: it takes on a value similar to the diameter of the sphere of influence (2.5 versus 2, in nondimensional units). In conclusion, using the critical lines $\beta_c(N;\eta)$ and $\beta_i(N;\eta)$ (for which rotation and dilation effects are very minor), and not $\beta_m(N;\eta)$, is crucial to unveil the scale-free-chaos phase transition in the limit of infinitely many particles. 

\section{Susceptibility and correlation time}\label{ap:d}
Here we illustrate the relations between reciprocal correlation time and singularities of the susceptibility by solvable examples. Note that these examples could be linearizations of stochastic equations about a time-independent homogeneous state. Thus, they are still far from the phase transition about a chaotic spatially non-homogeneous state and are not directly applicable to the confined VM. 

I. Consider the diffusive and noisy overdamped oscillator
\begin{subequations} \label{eqd1}
\begin{eqnarray}
\frac{\partial\phi}{\partial t} = D\nabla^2\phi - \omega_0^2\phi + \sqrt{2T}\,\xi(x,t),\label{eqd1a}
\end{eqnarray}
where $\xi(x,t)$ is a zero mean delta correlated white noise. The equilibrium probability density corresponding to Eq.~\eqref{eqd1a} is $Z^{-1}e^{-\int H\, dx/T}$, with hamiltonian density $H=(D|\nabla\phi|^2+\omega_0^2\phi^2)/2$ and temperature $T$. The Fourier transformed solution and the Fourier transformed correlation function are
\begin{eqnarray}
&&\hat{\phi}(k,t)= \int_{-\infty}^{t} e^{-(\omega_0^2+Dk^2)(t-s)}\hat{\xi}(s,k)\, ds, \label{eqd1b}\\ 
&&\hat{C}(k,t)= 2T \int_{-\infty}^{t_0} e^{-(\omega_0^2+Dk^2)(t_0-s)}e^{-(\omega_0^2+Dk^2)(t_0+t-s)} ds\nonumber\\
&&\quad\quad\quad=\frac{T}{\omega_0^2+Dk^2}e^{-(\omega_0^2+Dk^2)t}. \label{eqd1c}
\end{eqnarray}
Here we have taken the initial time to be $-\infty$ and a zero initial condition. Instability of the trivial state $\phi=0$ is reached when $\omega_0^2=0$, $k\propto 1/L=0$, which is pole of the susceptibility, $\hat{C}(0,0)=T/\omega_0^2$, and, equivalently, infinite value of the maximal correlation time, $1/\omega_0^2$. The nonlinear version of Eq.~\eqref{eqd1a},
\begin{eqnarray}
\frac{\partial\phi}{\partial t} = D\nabla^2\phi - \omega_0^2\phi-\zeta\phi^3 + \sqrt{2T}\,\xi(x,t),\label{eqd1d}
\end{eqnarray}
produces an equibilibrium probability density corresponding to the Landau-Wilson hamiltonian density $H=(D|\nabla\phi|^2+\omega_0^2\phi^2)/2+\zeta\phi^4/4$, which has a paradigmatic second order phase transition  provided $\omega_0^2$ may become negative \cite{ami05}.
\end{subequations}

II. To ascertain the influence of dynamics, consider the underdamped version of Eq.~\eqref{eqd1a}:
\begin{subequations} \label{eqd2}
\begin{eqnarray}
\frac{\partial^2\phi}{\partial t^2}\!&+&\!2 \omega_d\frac{\partial\phi}{\partial t} = D\nabla^2\phi - \omega_0^2\phi  +
\sqrt{4\omega'_dT}\,\xi(x,t).\label{eqd2a}
\end{eqnarray}
Proceeding as before, we find
\begin{eqnarray}
&&\hat{\phi}(k,t)= \int_{-\infty}^{t} G(t-s;k)\hat{\xi}(s,k)\, ds, \label{eqd2b}\\ 
&&G(t;k)=e^{-\omega_dt}\frac{\sin[\Omega(k)t]}{\Omega(k)} \,\theta(t),\label{eqd2c}\\
&&\Omega(k)=\sqrt{ \omega_0^2+Dk^2-\omega_d^2},
\label{eqd2d}
\end{eqnarray}
\begin{eqnarray}
&&\hat{C}(k,t)= 4T\omega'_d \int_0^\infty G(s;k) G(t+s;k)\, ds\nonumber\\
&&\quad\quad\quad=\frac{T\omega'_de^{-\omega_d t}}{\omega_0^2+Dk^2}\!\left(\frac{\cos[\Omega(k)t]}{\omega_d}+\frac{\sin[\Omega(k)t]}{\Omega(k)}\right)\!, \label{eqd2e}\\
&&\chi=\hat{C}(0,0)= \frac{T\omega'_d}{(\omega_0^2+Dk^2)\omega_d}. \label{eqd2f}
\end{eqnarray}
\end{subequations} 
Here we assume $\Omega(k)^2>0$. For  $\omega'_d=\omega_d$, the underdamped dynamics about thermal equilibrium yields the same susceptibility as Eq.~\eqref{eqd1c}. The pole of the susceptibility is again $\omega_0(k)^2=\omega_0^2+Dk^2$. Allowing $\omega_0^2$ to change sign and adding a nonlinearity as in Eq.~\eqref{eqd1d} leads to the same equilibrium phase transition as Eq.~\eqref{eqd1a}. However, for $0<\omega'_d\neq\omega_d$, $\omega_0^2>0$, and allowing $\omega_d$ to change sign, the system cannot reach thermal equilibrium as it did in the overdamped case. We have an additional pole of the susceptibility \eqref{eqd2f}, $\omega_d$, which coincides with the reciprocal relaxation time of Eq.~\eqref{eqd2e}. Adding a nonlinearity $-\zeta\phi^2\partial\phi/\partial t$ to the right hand side of Eq.~\eqref{eqd2a} may produce a quite different van der Pol-like instability and phase transition for $\omega_d<0$. Certainly, the van der Pol  limit cycle appears as a supercritical Hopf bifurcation \cite{kuz04} and the corresponding nonequilibrium phase transition would be similar to that analyzed in Ref.~\onlinecite{ris05} by RG calculations. For this nonequlibrium phase transition, vanishing of the reciprocal relaxation time coincides with vanishing of the pole $\omega_d=0$.

\end{document}